\documentstyle[12pt,umthesis,epsf]{report}
\newfont{\script}{eusb10}

\newcommand{\no}{\nonumber}
\newcommand{\rms}{\rm\scriptsize}

\newcommand{\n}{\noindent}
\newcommand{\p}{\hspace*{\parindent}}

\newcommand{\bel}[1]{\be\label{#1}}
\newcommand{\be}{\begin{equation}} 
\newcommand{\ee}{\end{equation}}

\newcommand{\ba}{\begin{array}{c}}
\newcommand{\bat}{\begin{array}{cc}}
\newcommand{\ea}{\end{array}}
\newcommand{\beqn}{\begin{eqnarray}}
\newcommand{\eeqn}{\end{eqnarray}}

\newcommand{\bi}{\begin{itemize}}
\newcommand{\ei}{\end{itemize}}

\def\an#1#2#3{{{\it Ann. Rev. Nucl. Part. Sci.} {\bf #1}, pg. #2, #3}}

\def\ijm#1#2#3{{{\it Int. J. Mod. Phys.} {\bf A#1}, pg. #2, #3}}

\def\nc#1#2#3{{{\it Nuovo Cimento} {\bf #1A}, pg. #2, #3}}
\def\npb#1#2#3{{{\it Nucl. Phys.} {\bf B#1}, pg. #2, #3}}
\def\plb#1#2#3{{{\it Phys. Lett.} {\bf B#1}, pg. #2, #3}}
\def\ppnp#1#2#3{{{\it Prog. Part. Nucl. Phys.} {\bf #1}, pg. #2, #3}}
\def\rpp#1#2#3{{{\it Rep. Prog. Phys.} {\bf #1}, pg. #2, #3}}
\def\prd#1#2#3{{{\it Phys. Rev.} {\bf D#1}, pg. #2, #3}}
\def\prl#1#2#3{{{\it Phys. Rev. Lett.} {\bf #1}, pg. #2, #3}}
\def\prt#1#2#3{{{\it Phys. Rep.} {\bf #1}, pg. #2, #3}}

\def\rmp{{\it Rev. Mod. Phys.}\ }

\def\zp#1#2#3{{{\it Z. Phys.} {\bf C#1}, pg. #2, #3}}

\def\ie{i.e.}

\def\etal{{\it et al.}}

\def\ri{\rightarrow}
\def\ov{\overline}

\def\gsim{\ \rlap{\raise 2pt \hbox{$>$}}{\lower 2pt \hbox{$\sim$}}\ }
\def\lsim{\ \rlap{\raise 2pt \hbox{$<$}}{\lower 2pt \hbox{$\sim$}}\ }
\def\Vtd{V_{td}^{\phantom{*}}}

\def\Vud{V_{ud}^{\phantom{*}}}
\def\Vus{V_{us}^{\phantom{*}}}

\def\Vuss{V_{us}^*}
\def\prk{p^{\rho}_K}
\def\psk{p^{\sigma}_K}
\def\pkm{p^{\phantom{l}}_{K\mu}}
\def\pnk{p^{\phantom{l}}_{K\nu}}
\def\pr0{p^{\rho}_0}
\def\ps0{p^{\sigma}_0}
\def\p0m{p^{\phantom{l}}_{0\mu}}
\def\pn0{p^{\phantom{l}}_{0\nu}}
\def\as{\arctan{\sqrt{s \over {4 m^2_{\pi}-s}}}}
\def\ak{\arctan{\sqrt{k^2_1 \over {4 m^2_{\pi}-k^2_1}}}}
\def\sa{\sqrt{4 m^2_{\pi}-s}}
\def\ka{\sqrt{4 m^2_{\pi}-k^2_1}}
\def\ds{(s-k^2_1)}
\def\ns{(8 m^4_{\pi}+2 m^2_{\pi} s -s^2)}
\def\nk{(8 m^4_{\pi}+2 m^2_{\pi} k^2_1 -k^4_1)}
\def\mp{m_{\pi}}
\def\k1{k_1}
\def\nns{(8 m^4_{\pi}-6 m^2_{\pi} s +s^2)}
\def\nnk{(8 m^4_{\pi}-6 m^2_{\pi} k^2_1 +k^4_1)}

\begin{document}
%
%%%

%%% Page styles [default is 'preliminary']
%
%\pagestyle{thesis}      % Page # upper right hand corner.
%\pagestyle{draft}       % Page # as 'thesis' with `DRAFT' written on top.
%\pagestyle{preliminary} % Page # at bottom middle
%\pagestyle{draftpreliminary}
%                        % Page # at bottom with `DRAFT'.
%%%

%%% List of Figures/Tables [default is both true]
\figurespagetrue        % We DO want a list of figures.
\tablespagetrue         % We do NOT want a list of tables.
%%%

%%% Preliminary pages (title, signature, copyright...) in file title.tex

%%%%%%%%%%%%%%%%%%%%%%%%%%%%%%%%%%%%%%%%%%%%%%%%%%%%%%%%%%%%%%%%%%%%%%%%
%%%% This makes all the required pages before con-text
%%%%
%%%%

%%%% You need to enter the appropriate information here:
%%
%
\title{Rare Kaon Decays and CP Violation}
\author{Fabrizio Gabbiani}
\headofdepartment{John F.~Donoghue}
\committeechair{John F.~Donoghue}
\firstreader{Eugene Golowich}
\secondreader{Barry R.~Holstein}
\thirdreader{Monroe S.~Z.~Rabin}
\outsidereader{Thurlow A.~Cook}
\submitdate{September 1997}      % Should be February, May or September

\copyrightyear{1997}           % I think umthesis.sty figures this
                                % one out from the previous information.
%%%%

%%%% This makes a cover page, a copyright page and a signature page.
%%   using the above information.

\beforepreface
%\pagestyle{preliminary}         %% MLS 032796
%
%%%%

%%%% If you want to dedicate your thesis to someone, do this :
%%

\dedicationsection{~}
\begin{center}
\em \bf To the Opera Singer \\
\end{center}
\pagebreak
%
% If you don't want the heading ``Dedication'' to appear on the page,
% use a tilde instead: \dedicationsection{~}
%---------------------------------------------------------
%%%%

%%%% Acknowledgements
%%

\prefacesection{Acknowledgments}
\vskip -0.3truecm

My thesis advisor, Prof. John F. Donoghue, will never be thanked
enough, but I must try anyway. His patience and availability, in spite
of all the workload, both academical and administrative, that he has to
shoulder, were invaluable.

Thanks to Prof. Barry Holstein and Prof. Eugene Golowich, who
were a source of inspiration and physical insight, as well as humor,
throughout my stay at this University.

Prof. Monroe S.~Z.~Rabin and Prof. Thurlow A.~Cook are
warmly thanked for being on my committee.

I wish to acknowledge and thank my colleagues in the physics
department with whom I shared such a large part of my life and my
thoughts while at UMASS: Gustavo Alberto Burdman, Lars Kielhorn, Alexey
Anatolievich Petrov, Jo\~{a}o Manuel Soares, Jusak Tandean, Tibor
Torma, Sundar Viswanathan, and especially Thomas Robert Hemmert,
Antonio Felipe P\'erez and Eswara Prasad Venugopal. Their support, not
restricted only to the complicated paths of theoretical physics, was
precious.

I cannot forget all the fun I had getting involved with the GSS, and
the people who taught me how to get things done in the rude world of
campus politics: Stefanie Louise (Aloisia) Ameres, Carmelita Patricia
(Rosie) Casta\~{n}eda, Colin Sean Cavell, Hussein Yousuf Ibish,
Fatimah Batul Ihsan, Shyamala Rao Ivatury, Julia Ruth Johnson, Julia
Darnell Mahoney, Sandra S.~Rose, Susan E.~Shadwick, Alison Leah Wing, and
especially Deepika Gautama.

An important aspect of the life of a graduate student is to relax in
the appropriate place, to recharge after academic fatigue. The place
is the Graduate Lounge, with its motley crowd of unforgettable
characters: Sulin D.~Allen, Sveva Besana, Katherine Henrietta Brannum,
Maria Nella Carminati, Corinna Maria Colatore, Claire Marie Darwent,
Juliana Dezavalia, Alessandra Dimaio, Robert Eaton, Touria El-Jaoual,
Shibani Ajay Ghosh, Shrikumar Hariharasubrahmanian, Elizabeth
(Doctress Neutopia) N.~Hubbard, Wenping Jin, Sundari Josyula, Barsha
Khattry, Satish Kumar Kolluri, Fausto Marincioni, Rafael
Mill{\'a}n-Gabet, Ali Husain Mir, Raza Ali Mir, Nnamudi P.~Mokwunye,
Goldie Osuri, Lucia Ponginebbi, Amit Prothi, Sunita Rajouria,
Madanmohan Rao, Sangeeta Vaman Rao, Badri N.~Toppur, Richard Ram\'on
Ure\~{n}a, Elena Piera Vittadini, Joanna Wisniewska, and Maya Kirit
Yajnik.

A special award goes to my longtime friend Th{\'e}r{\`e}se
Marie Bonin for keeping my morale high.

Finally, I want to thank all the staff in the UMASS Physics
Department who helped me in numerous ways. Thanks to Ann E.~Adams Cairl,
Margaret C.~MacDonald, Mary E.~Pelis and Mary Ann Ryan. Another ``thank
you'' is for Peter R.~deFriesse and Stephen C.~Svoboda for their
indispensable technical support, so much needed when I had to get the
computers to work.

\newpage
%
%%%%

%%%% Abstract 
%%   (yes, I know)
\prefacesection{Abstract} 
\begin{center}
%\large\sc
\large          %% MLS 032796
RARE KAON DECAYS AND CP VIOLATION \\
SEPTEMBER 1997 \\
FABRIZIO GABBIANI\\
LAUREA IN FISICA, UNIVERSIT\`A DEGLI STUDI DI PADOVA \\
Ph.D., UNIVERSITY OF MASSACHUSETTS AMHERST\\
{\large Directed by: Professor John F. Donoghue}
\end{center}

% To be changed
Rare kaon decays are an important testing ground of the electroweak
flavor theory. They can provide new signals of CP-violating phenomena
and open a window into physics beyond the Standard Model. The
interplay of long-distance QCD effects in strangeness-changing
transitions can be analyzed with Chiral Perturbation Theory
techniques. Some theoretical predictions obtained within this
framework for radiative kaon decays are reviewed, together with the
present experimental status. In particular, two rare kaon decays are
analyzed: The first decay, $K_L \rightarrow \pi^0 e^+ e^-$, is being
searched for as a signal of direct $\Delta S = 1$ CP violation. We
provide a thorough updating of the analysis of the three components of
the decay: 1) Direct CP violation, 2) CP violation through the mass
matrix and 3) CP-conserving (two-photon) contributions. First the
chiral calculation of the $K_S \rightarrow
\pi^0 e^+ e^-$ rate, due to Ecker, Pich and de Rafael, is
updated to include recent results on the nonleptonic amplitude. Then
we systematically explore the uncertainties in this method. These
appear to be so large that they will obscure the direct CP violation
unless it is possible to measure the $K_S
\rightarrow \pi^0 e^+ e^-$ rate. The CP-conserving amplitude remains
somewhat uncertain, but present indications are such that there may be a
sizable CP-violating asymmetry in the $e^+, e^-$ energies from the
interference of CP-conserving and CP-violating amplitudes and this may
potentially be useful in determining whether direct CP violation is present.
The second decay, $K_L \rightarrow \pi^0 \gamma e^+ e^-$, which occurs
at a {\it higher} rate than the nonradiative process $K_L \rightarrow
\pi^0 e^+ e^-$, can be a background to CP violation studies using
the latter reaction. It also has interest in its own right in the
context of Chiral Perturbation Theory, through its relation to the
decay $K_L \rightarrow \pi^0 \gamma\gamma$. The leading order chiral
loop contribution to $K_L \rightarrow \pi^0 \gamma e^+ e^-$, including
the $(q_{e^+}+q_{e^-})^2/m^2_{\pi}$ dependence, is completely
calculable. We present this result and also include the higher order
modifications that are required in the analysis of $K_L \rightarrow
\pi^0 \gamma \gamma$.

%%%% This makes the Table of Contents   (always!)
%%                  List of Figures     (\figurespagetrue ?) See thesis.tex
%%                  List of Tables      (\tablespagetrue  ?) See
%%                  thesis.tex
%%
%%   using a special format

\afterpreface

%%%%

% This completes all the stuff before your first chapter ....
         % Makes everything before Chapter one.
%%%

%%% And now include the text of the chapters, bibliography etc.
%   The \includeonly command in the preamble determines what actually 
%   gets included; e.g. appendix is ignored here (it doesn't exist
%   anyway).

\pagestyle{thesis}
\chapter{INTRODUCTION}                   % CAPITALIZED

\vspace{2mm}
High-precision experiments on rare kaon decays offer the
possibility of unraveling new physics beyond the Standard Model.
Searching for forbidden flavor-changing processes \cite{LW}
at the $10^{-10}$ level
[$BR(K_L\ri\mu e) < 3.3\times 10^{-11}$,
$BR(K_L\ri\pi^0\mu e) < 3.2\times 10^{-9}$,
$BR(K^+\ri\pi^+\mu e) < 2.1\times 10^{-10}$, \ldots],
one is actually exploring energy scales above the 10 TeV
region. The study of allowed (but highly suppressed) decay modes
provides, at the same time, very interesting tests of the Standard
Model itself. Electromagnetic-induced nonleptonic weak transitions
and higher-order weak processes are a useful tool to improve our
understanding of the interplay among electromagnetic, weak and strong
interactions. In addition, new signals of CP violation, which would
help to elucidate the source of CP-violating phenomena, can be looked
for.

In this chapter we shall briefly describe a few relevant rare kaon
decays, reserving the rest of this thesis for the processes $K_L
\rightarrow \pi^0 \gamma \gamma$, $K_L \rightarrow \pi^0 e^+e^-$ and
$K_L \rightarrow \pi^0 \gamma e^+e^-$, together with the related
decays $K^+ \rightarrow \pi^+ \gamma \gamma$
and $K^+ \rightarrow \pi^+ e^+e^-$.

Since the kaon mass is a very low energy scale, the theoretical
analysis of nonleptonic kaon decays is highly non-trivial. While the
underlying flavor-changing weak transitions among the constituent
quarks are associated with the $W$-mass scale, the corresponding
hadronic amplitudes are governed by the long-distance behavior of the
strong interactions, \ie\ the confinement regime of QCD.

The standard short-distance approach to weak transitions makes
use of the asymptotic freedom property of QCD
to successively integrate out the fields with heavy masses down to
scales $\mu < m_c$.
Using the operator product expansion (OPE) and
renormalization-group techniques, one gets an effective $\Delta S=1$
hamiltonian,
\bel{eq:sd_hamiltonian}
{\cal H}_{\mbox{\rms eff}}^{\Delta S=1} \, = \, {G_F\over\sqrt{2}}
V_{ud}^{\phantom{*}} V_{us}^*\,
\sum_i C_i(\mu) Q_i \, + \, \mbox{\rm h.c.},
\ee
which is a sum of local four-fermion operators $Q_i$,
constructed with the light degrees of freedom
($u,d,s; e,\mu,\nu_l$), modulated by
Wilson coefficients $C_i(\mu)$
which are functions of the heavy ($W,t,b,c,\tau$) masses.
The overall renormalization scale $\mu$
separates the short- ($M>\mu$) and long- ($m<\mu$) distance
contributions, which are contained in $C_i(\mu)$
and $Q_i$, respectively.
The physical amplitudes are of course independent of $\mu$;
thus, the explicit scale (and scheme) dependence of the Wilson
coefficients, should cancel exactly with the corresponding dependence
of the $Q_i$ matrix elements between on-shell states.

Our knowledge of the $\Delta S=1$ effective hamiltonian has improved
considerably in recent years, thanks to the completion of the
next-to-leading logarithmic order calculation of the Wilson
coefficients \cite{BU}.
All gluonic corrections of ${\cal O}(\alpha_s^n t^n)$ and
${\cal O}(\alpha_s^{n+1} t^n)$ are already known,
where $t\equiv\log{(M/m)}$ refers to the logarithm of any ratio of
heavy-mass scales ($M,m\geq\mu$). Moreover, the full $m_t/M_W$ dependence
(at lowest order in $\alpha_s$) has been taken into account.

Unfortunately, in order to predict the
physical amplitudes one is still confronted with the calculation of
the hadronic matrix elements of the quark operators.
This is a very difficult problem, which so far remains unsolved.
%The present technology to calculate low-energy matrix elements is not
%yet developed to the degree of sophistication of perturbative QCD.
We have only been able to obtain rough estimates using
different approximations (vacuum saturation, $N_C\ri\infty$ limit,
QCD low-energy effective action, \ldots)
or applying QCD techniques (lattice, QCD sum rules) which suffer from
their own technical limitations.

Below the resonance region ($\mu < m_\rho$) the strong interaction
dynamics can be better understood with global symmetry considerations.
We can take advantage of the fact that the
pseudoscalar mesons are the lowest energy modes of the hadronic
spectrum: They correspond to the octet of Goldstone bosons associated
with the dynamical chiral symmetry breaking of QCD, $SU(3)_L\otimes
SU(3)_R \rightarrow SU(3)_V$.
The low-energy implications of the QCD symmetries can then be worked out
through an effective lagrangian containing only the Goldstone modes.
The effective Chiral Perturbation
Theory \cite{CPT} (ChPTh)
formulation of the Standard Model is an ideal
framework to describe kaon decays \cite{DR}.
This is because in $K$ decays the
only physical states that appear are pseudoscalar mesons, photons and
leptons, and because the characteristic momenta involved are small
compared to the natural scale of chiral symmetry breaking
($\Lambda_\chi\sim 1$~GeV).

\section{$K \ri \pi \nu {\overline \nu}$}

The decays $K\to\pi\nu\ov{\nu}$ proceed through flavor-changing
neutral current effects. These arise in the Standard Model only at
second (one-loop) order in the electroweak interaction (Z-penguin and
W-box diagrams, Fig. 1) and are additionally GIM suppressed.

The branching fractions are thus very small, at the level of
$10^{-10}$, which makes these modes rather challenging to detect.
However, $K\to\pi\nu\ov{\nu}$ have long been known to be reliably
calculable, in contrast to most other decay modes of interest.
A measurement of $K^+\to\pi^+\nu\ov{\nu}$ and $K_L\to\pi^0\nu\ov{\nu}$
will therefore be an extremely useful test of flavor physics.
Over the recent years important refinements have been added to the
theoretical treatment of $K\to\pi\nu\ov{\nu}$.
Let us briefly summarize the main aspects of why $K\to\pi\nu\ov{\nu}$
is theoretically so favorable and what recent developments have
contributed to emphasize this point.
\begin{figure}[t]
\centering
\leavevmode
\centerline{
\epsfbox{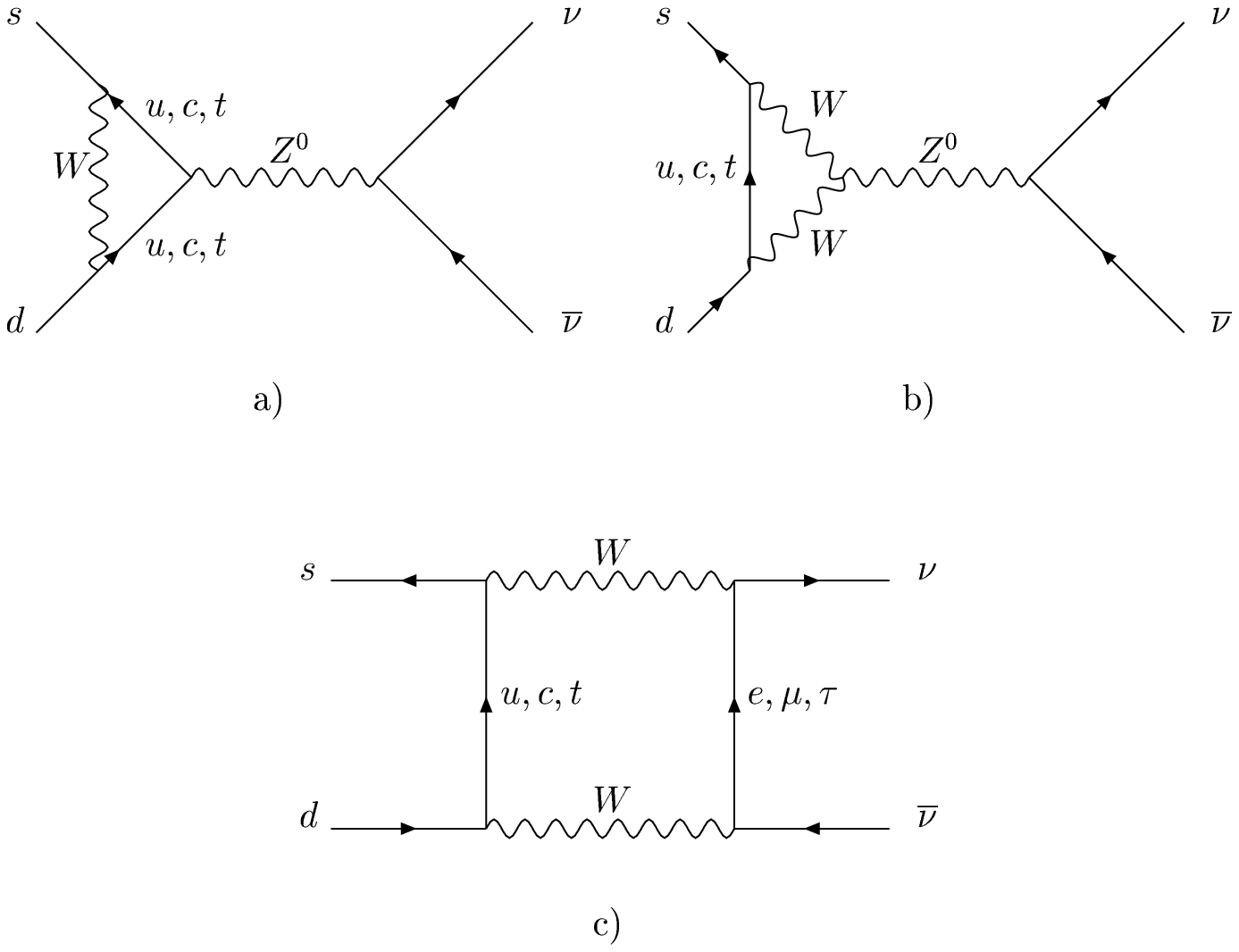}}
\caption{Diagrams inducing the decay $s \ri d \nu {\overline \nu}$
in the Standard Model.}
\end{figure}
\begin{description}
\item[a)]
First, $K\to\pi\nu\ov{\nu}$ is semileptonic. The relevant hadronic
matrix e\-lem\-ents, such as $\langle\pi|(\ov{s}d)_{V-A}|K\rangle$, are just
ma\-trix e\-lem\-ents of a current operator between ha\-dro\-nic states,
which are
already considerably simpler objects than the matrix elements of
four-quark operators encountered in many other observables
($K-\ov{K}$ mixing, $\epsilon^{\prime}/\epsilon$).
Moreover, they are related to the matrix element
\begin{equation}\label{sume}
\langle\pi^0|(\ov{s} u)_{V-A}|K^+\rangle
\end{equation}
by isospin symmetry.
The latter quantity can be extracted from the well measured leading
semileptonic decay $K^+\to\pi^0 l\nu$. Although isospin is a fairly
good symmetry, it is still broken by the small up-down quark mass
difference and by electromagnetic effects. These manifest themselves
in differences of the neutral versus charged kaon (pion) masses
(affecting phase space), corrections to the isospin limit in the
form factors and electromagnetic radiative effects.
Marciano and Parsa \cite{MP} have analyzed these corrections and
found an overall reduction in the branching ratio by $10\%$ for
$K^+\to\pi^+\nu\ov{\nu}$ and $5.6\%$ for $K_L\to\pi^0\nu\ov{\nu}$.
\item[b)] \hskip -0.4pt
Long distance contributions are systematically suppressed
as ${\cal O}(\Lambda^2_{QCD}/$$m^2_c)$ compared to the
charm contribution (which is part of the short distance amplitude).
This feature is related to the hard ($\sim m^2_c$) GIM suppression
pattern shown by the Z-penguin and W-box diagrams, and the absence
of virtual photon amplitudes. Long distance contributions have
been examined quantitatively \cite{LDC} and shown to be
numerically negligible (below $\approx 5\%$ of the charm amplitude).
\item[c)]
The preceding discussion implies that $K\to\pi\nu\ov{\nu}$ are
short distance dominated by top- and charm-loops in general.
The relevant short distance QCD effects can be treated in
perturbation theory and have been calculated at next-to-leading
order \cite{BB2}. This allowed to substantially reduce
(for $K^+$) or even practically eliminate ($K_L$) the leading
order scale ambiguities, which are the dominant uncertainties in
the leading order result.
\end{description}

The decay amplitude for $K\ri\pi\nu\ov{\nu}$,
\bel{eq:pnn} \mbox{}\!\!
{\cal M}(K\ri\pi\nu\ov{\nu})\,\sim\, \sum_{i=c,t}
F(V_{id}^{\phantom{*}} V_{is}^*;x_i)
\left(\ov{\nu}^{\phantom{l}}_L\gamma_\mu\nu^{\phantom{l}}_L\right)
\langle\pi |\ov{s}_L\gamma^\mu d_L|K\rangle
\, , \qquad x_i\equiv m_i^2/M_W^2 ,
\ee
involves the hadronic matrix element of the $\Delta S=1$ vector current,
which (assuming isospin symmetry) can be obtained from $K_{l3}$ decays.
In the ChPTh framework, the needed hadronic matrix element is known
at ${\cal O}(E^4)$; this allows one to estimate the relevant
isospin-violating corrections reliably \cite{LR,MP}.

Summing over the three neutrino flavors and expressing the
quark-mixing factors
through the Wolfenstein parameters \cite{W} $\lambda$, $A$, $\rho$ and
$\eta$, one can write the approximate formula \cite{BU}:

\bel{eq:br_pnn}
BR(K^+\ri\pi^+\nu\ov{\nu}) \,\approx\, 1.93\times 10^{-11}\,
A^4 \, x_t^{1.15}\,\left[ \eta^2 + (\rho_0-\rho)^2\right] \, ,
\qquad \rho_0\approx 1.4 \, .
\ee

\n The departure of $\rho_0$ from unity measures the impact of the
charm contribution.

With the presently favored values for the quark-mixing parameters,
the branching ratio is predicted to be in the range \cite{BU}

\bel{eq:pred_br}
BR(K^+\ri\pi^+\nu\ov{\nu}) \,=\,
(9.1\pm 3.2)\times 10^{-11} \, ,
\ee

\n to be compared with the present experi\-mental upper bound \cite{AD}
$BR(K^+\ri\pi^+\nu\ov{\nu})$ $<$ 2.4$\times 10^{-9}$ (90\% CL).

What is actually measured is the transition $K^+ \ri \pi^+ +$ nothing;
therefore,
the experimental search for this process can also be used
to set limits on possible exotic decay modes like $K^+\ri\pi^+ X^0$,
where $X^0$ denotes an undetected
light Higgs or Goldstone boson (axion, familon, majoron, \ldots).

The CP-violating decay $K_L\ri\pi^0\nu\ov{\nu}$ has been suggested
\cite{LI} as a good candidate to look for pure
direct CP-violating transitions.
The contribution coming from indirect (mass-matrix)
CP violation via $K^0$-${\ov K}^0$ mixing is very small \cite{LI}:
$BR(K_L\ri\pi^0\nu\ov{\nu})_{\rm MM}$ $\sim$ 5 $\times 10^{-15}$.
The decay proceeds almost entirely through direct CP violation, and
is completely dominated by short-distance loop diagrams with top
quark exchanges \cite{BU}:

\bel{eq:br_pnn0}
BR(K_L\ri\pi^0\nu\ov{\nu}) \,\approx\, 8.07\times 10^{-11}\,
A^4 \, \eta^2 \, x_t^{1.15} \, .
\ee

\n The present experimental upper bound \cite{WEA},
$BR(K_L\ri\pi^0\nu\ov{\nu}) < 5.8\times 10^{-5}$
(90\% CL),
is still far away from the expected range \cite{BU}
\bel{eq:pred_br0}
BR(K_L\ri\pi^0\nu\ov{\nu}) \,=\,
(2.8\pm 1.7)\times 10^{-11} \, .
\ee

\n Nevertheless, the experimental prospects to reach the required sensitivity
in the near future look rather promising.
The clean observation of just a single
unambiguous event would indicate the existence of CP-violating
$\Delta S = 1$ transitions.

In Table 1 we have summarized some of the main
features of $K^+\to\pi^+\nu\ov{\nu}$ and $K_L\to\pi^0\nu\ov{\nu}$.
\begin{table}
\centering
\caption{Important properties and results for $K\to\pi\nu\ov{\nu}$.
}
\vskip 0.1 in
\begin{tabular}{|c|c|c|} \hline
& $K^+\to\pi^+\nu\ov{\nu}$ & $K_L\to\pi^0\nu\ov{\nu}$ \\
\hline
\hline
& CP-conserving & CP-violating \\
\hline
CKM & $V_{td}$ & $\mbox{Im} V^*_{ts}V^{\phantom{*}}_{td}\sim J_{\rm
CP}\sim\eta$ \\
\hline
scale uncert. (BR) & $\pm 20\%$ (LO) $\to \pm 5\%$ (NLO) &
                     $\pm 10\%$ (LO) $\to \pm 1\%$ (NLO) \\
\hline
BR (SM) & $(0.9\pm 0.3)\times 10^{-10}$&$(2.8\pm 1.7)\times 10^{-11}$ \\
\hline
exp. limit & $< 2.4\times 10^{-9}$ BNL 787 \cite{AD}
           & $< 5.8\times 10^{-5}$ FNAL 799 \cite{WEA} \\
\hline
\end{tabular}
\label{kpnntab}
\end{table}

While already $K^+\to\pi^+\nu\ov{\nu}$ can be reliably calculated,
the situation is even better for $K_L\to\pi^0\nu\ov{\nu}$.
The charm sector, in $K^+\to\pi^+\nu\ov{\nu}$
the dominant source of uncertainty, is completely negligible for
$K_L\to\pi^0\nu\ov{\nu}$ ($0.1\%$ effect on the branching ratio).
Long distance contributions
($\;\raisebox{-.4ex}{\rlap{$\sim$}} \raisebox{.4ex}{$<$}\; 0.1\%$)
and also the indirect CP violation effect
($\;\raisebox{-.4ex}{\rlap{$\sim$}} \raisebox{.4ex}{$<$}\; 1\%$)
are likewise negligible. In summary, the total theoretical
uncertainties, from perturbation theory in the top sector
and in the isospin breaking corrections, are safely below
$2-3\%$ for $BR(K_L\to\pi^0\nu\ov{\nu})$. This makes this decay
mode truly unique and very promising for phenomenological
applications. Note that the range given as the Standard Model
prediction in Table \ref{kpnntab} arises from our, at present,
limited knowledge of Standard Model parameters (CKM), and not
from intrinsic uncertainties in calculating $BR(K_L\to\pi^0\nu\ov{\nu})$.

With a measurement of $BR(K^+\to\pi^+\nu\ov{\nu})$ and
$BR(K_L\to\pi^0\nu\ov{\nu})$ available very interesting phenomenological
studies could be performed.
For instance, $BR(K^+\to\pi^+\nu\ov{\nu})$ and
$BR(K_L\to\pi^0\nu\ov{\nu})$ together determine the unitarity triangle
(Wolfenstein parameters $\rho$ and $\eta$)
completely.

The expected accuracy with $\pm 10\%$ branching ratio measurements is
comparable to the one that can be achieved by CP violation studies
at $B$ factories before the LHC era \cite{BB96}.
The quantity $BR(K_L\to\pi^0\nu\ov{\nu})$ by itself offers probably the
best precision in determining $\mbox{Im} V^*_{ts}V_{td}$ or,
equivalently, the Jarlskog parameter \cite{JA}
\begin{equation}\label{jcp}
J_{\rm CP}=\mbox{Im}(V^*_{ts}\Vtd \Vus V^*_{ud})=
\lambda\left(1-\frac{\lambda^2}{2}\right)\mbox{Im}\lambda_t.
\end{equation}

The charged mode $K^+\to\pi^+\nu\ov{\nu}$ is being currently
pursued by Brookhaven experiment E787. The latest published result
\cite{AD} gives an upper limit which is about a factor 25 above
the Standard Model range. Several improvements have been implemented
since then and the SM sensitivity is expected to be reached in the
near future \cite{AGS2}.
Recently an experiment has been proposed to measure
$K^+\to\pi^+\nu\ov{\nu}$ at the Fermilab Main Injector \cite{CCTR}.
Concerning $K_L\to\pi^0\nu\ov{\nu}$, a proposal exists at
Brookhaven (BNL E926) to measure this decay at the AGS
with a sensitivity of ${\cal O}(10^{-12})$ (see \cite{AGS2}).
There are furthermore plans to pursue this mode with comparable
sensitivity at Fermilab \cite{KAMI} and KEK \cite{ISS}.

\section{$K_S\ri\gamma\gamma$}

The symmetry constraints do not allow any direct tree-level
$K_1^0\gamma\gamma$ coupling at ${\cal O}(E^4)$ ($K^0_{1,2}$ refer to the
CP-even and CP-odd eigenstates, respectively). This decay proceeds
then through a loop of charged pions and kaons as shown in Fig. 2.

\begin{figure}[t]
\centering
\leavevmode
\centerline{
\epsfbox{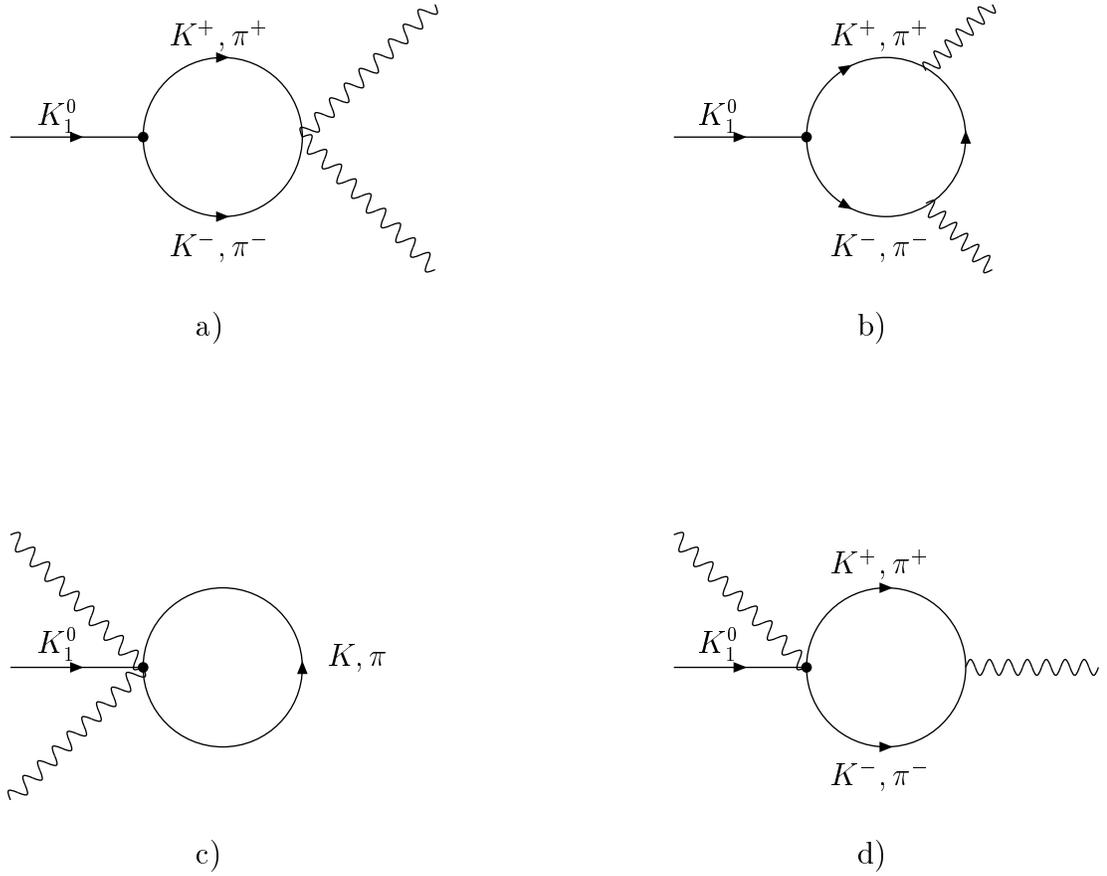}}
\caption{Feynman diagrams for $K^0_1 \ri \gamma\gamma$.}
\end{figure}

\n Since there are no possible counter-terms to renormalize
divergences, the one-loop amplitude is necessarily finite. Although
each of the four diagrams in Fig. 2 is quadratically
divergent, these divergences cancel in the sum. The resulting
prediction \cite{DAE}, $BR(K_S\ri\gamma\gamma) =
2.0 \times 10^{-6}$, is in very good agreement with the experimental
measurement \cite{BA}:

\bel{eq:ksgg}
BR(K_S\ri\gamma\gamma) \, = \,
(2.4 \pm 0.9) \times 10^{-6} \, .
\ee

\section{$K_{L,S}\ri\mu^+\mu^-$}

There are well-known short-distance contributions \cite{BU}
(electroweak penguins and box diagrams)
to the decay $K_L\ri\mu^+\mu^-$.
This part of the amplitude is sensitive to the Wolfenstein parameter
$\rho$.
However, this transition is dominated by long-distance
physics. The main contribution proceeds through a two-photon
intermediate state: $K_2^0\ri\gamma^*\gamma^*\ri\mu^+\mu^-$.
Contrary to $K_1^0\ri\gamma\gamma$,
the prediction for the $K_2^0\ri\gamma\gamma$ decay is
very uncertain, because the first non-zero contribution
occurs\footnote{
%%%%%%%%%%%%%
At ${\cal O}(E^4)$, this decay proceeds through
a tree-level $K_2^0\ri\pi^0,\eta$ transition, followed by
$\pi^0,\eta\ri\gamma\gamma$ vertices.
Because of the Gell-Mann-Okubo relation,
the sum of the $\pi^0$ and $\eta$ contributions
cancels exactly to lowest order.
The decay amplitude is then very sensitive to $SU(3)$ breaking.}
%%%%%%%%%%%%%
at ${\cal O}(E^6)$.
That makes very difficult any attempt to
predict the $K_{L}\ri\mu^+\mu^-$ amplitude.

The long distance
amplitude consists of a dispersive ($A_{\rm dis}$) and an absorptive
contribution ($A_{\rm abs}$). The branching fraction can thus be written
\begin{equation}\label{bklma}
BR(K_L\to\mu^+\mu^-)=|A_{\rm SD}+A_{\rm dis}|^2 + |A_{\rm abs}|^2.
\end{equation}
Using the well-known unitarity bound
\begin{equation}
{{\Gamma(K_L\to\mu^+\mu^-)} \over {\Gamma(K_L\to \gamma\gamma)}} \geq
{{\alpha^2 m^2_{\mu}} \over {2\beta m^2_K}}\left(\log{{1+\beta} \over
{1-\beta}}\right)^2, \qquad \beta = \sqrt{1-4 m^2_{\mu}/m^2_K},
\end{equation}
associated with the $\gamma\gamma$ intermediate state, and knowing
$BR(K_L\to\gamma\gamma)$ it is possible to extract
$|A_{\rm abs}|^2=(6.8\pm 0.3)\times 10^{-9}$ \cite{LV}.
$A_{\rm dis}$ on the other hand cannot be calculated accurately at
present and the estimates are strongly model dependent
\cite{VAR}.
This is rather unfortunate, in particular since
$BR(K_L\to\mu^+\mu^-)$, unlike most other rare decays, has already
been measured, and this with very good precision:
\begin{equation}\label{bklmex}
BR(K_L\to\mu^+\mu^-)=
\left\{ \begin{array}{ll}
        (6.9\pm 0.4)\times 10^{-9}, & \mbox{BNL 791 \cite{HEIN}},\\
        (7.9\pm 0.7)\times 10^{-9}, & \mbox{KEK 137 \cite{AKA}}.
        \end{array} \right.
\end{equation}

\n For comparison we note that $BR(K_L\to\mu^+\mu^-)_{\rm SD}=(1.3\pm
0.6)\times 10^{-9}$ is the expected branching ratio in the Standard
Model based on the short-distance contribution alone. Due to the fact
that $A_{\rm dis}$ is largely unknown, $K_L\to\mu^+\mu^-$ is at present
not a very useful constraint on CKM parameters. Some improvement might
be expected from measuring the decay $K_L \ri
\mu^+\mu^-e^+e^-$, which
could lead to a better understanding of the $K_L \ri \gamma^*\gamma^*$
vertex. First results obtained at Fermilab (E799) give $BR(K_L \ri
\mu^+\mu^-e^+e^-)$ = $(2.9^{+6.7}_{-2.4})\times 10^{-9}$.

The situation is completely different for the $K_S$ decay. A
straightforward chiral analysis \cite{EP} shows that, at lowest
order in momenta, the only allowed tree-level $K^0\mu^+\mu^-$ coupling
corresponds to the CP-odd state $K_2^0$. Therefore, the
$K_1^0\ri\mu^+\mu^-$ transition can only be generated by a finite
non-local two-loop contribution, illustrated in Fig. 3.

\begin{figure}[t]
\centering
\leavevmode
\centerline{
\epsfbox{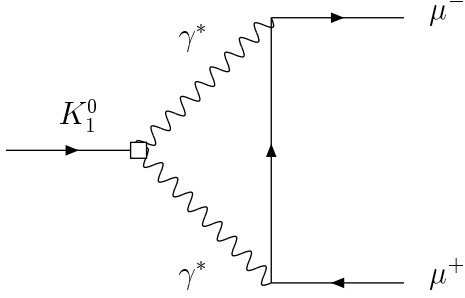}}
\caption{Feynman diagram for the decay $K^0_1 \ri \mu^+\mu^-$. The
$K^0_1\gamma^*\gamma^*$ vertex is generated through the one-loop
diagrams shown in Fig. 2.}
\end{figure}
\newpage

The explicit calculation \cite{EP} gives:

\bel{eq:ksmm_ratios}
{\Gamma(K_S\ri\mu^+\mu^-)\over\Gamma(K_S\ri\gamma\gamma)}
= 2\times 10^{-6}, \qquad\qquad
{\Gamma(K_S\ri e^+ e^-)\over\Gamma(K_S\ri\gamma\gamma)}
= 8\times 10^{-9},
\ee
well below the present (90\% CL) experimental upper limits
\cite{KSE}:    %{PDG:96}:
$BR(K_S\ri\mu^+\mu^-)$ $<$ $3.2\times 10^{-7}$,
$BR(K_S\ri e^+e^-)$ $<$ $2.8\times 10^{-6}$.
Although, in view of the smallness of the predicted ratios, this
calculation seems quite academic, it has important implications for
CP-violation studies.

The longitudinal muon polarization ${\cal P}_L$
in the decay $K_L\ri\mu^+\mu^-$ is an interesting measure of
CP violation.
As for every CP-violating observable in the neutral kaon system,
there are in general two different kinds of contributions to ${\cal P}_L$:
Indirect mass-matrix CP violation through the small
$K_1^0$ admixture of the $K_L$
($\epsilon$ effect), and direct CP violation in the
$K_2^0\ri\mu^+\mu^-$ decay amplitude.

In the Standard Model, the direct CP-violating amplitude is
induced by Higgs exchange with an effective one-loop flavor-changing
$\ov{s} d H$ coupling \cite{BL1}.
The present lower bound
on the Higgs mass, $M_H>66$ GeV (95\% CL), implies a
conservative upper limit
$|{\cal P}_{L,\mbox{\rms dir}}| < 10^{-4}$.
Much larger values, ${\cal P}_L \sim {\cal O}(10^{-2})$, appear quite naturally
in various extensions of the Standard Model \cite{GNMO}.
It is worth emphasizing that ${\cal P}_L$ is especially
sensitive to the presence of light scalars with CP-violating
Yukawa couplings. Thus, ${\cal P}_L$ seems to be a good signature to look
for new physics beyond the Standard Model; for this to be the case,
however, it is very important to have a good quantitative
understanding of the Standard Model prediction to allow us to infer,
from a measurement of ${\cal P}_L$, the existence of a new CP-violation
mechanism.
\newpage

The chiral calculation of the $K_1^0\ri\mu^+\mu^-$ amplitude
allows us to make a reliable estimate
of the contribution to ${\cal P}_L$ due to $K^0$-$\ov{K}^0$ mixing
\cite{EP}:
\bel{eq:p_l}
1.9 \, < \, |{\cal P}_{L,{\rm MM}}| \times 10^3 \Biggl[
{2 \times 10^{-6} \over
BR(K_S\ri\gamma\gamma)} \Biggr]^{1/2} \, < 2.5 \, .
\ee
Taking into account
the present experimental errors in $BR(K_S\ri\gamma\gamma)$ and
the inherent theoretical uncertainties due to uncalculated
higher-order corrections,
one can conclude that experimental indications for
$|{\cal P}_L|>5\times 10^{-3}$ would constitute clear evidence
for additional
mechanisms of CP violation beyond the Standard Model.

\section{$K^+\to\pi^+\mu^+\mu^-$}

The rare decay $K^+\to\pi^+\mu^+\mu^-$ has recently been measured
at Brookhaven (BNL 787) with a branching ratio \cite{B787}
\begin{equation}\label{bkpmm}
BR(K^+\to\pi^+\mu^+\mu^-)=(5.0\pm 0.4\pm 0.6)\times 10^{-8}.
\end{equation}
This compares well with the estimate from ChPTh
$BR(K^+\to\pi^+\mu^+\mu^-)$ = $(6.2^{+0.8}_{-0.6})$ $\times$
$10^{-8}$ \cite{PIC}.
The branching ratio is completely determined by the long-distance
contribution arising from the one-photon exchange amplitude.
A short-distance amplitude from Z-penguin and W-box diagrams also
exists, but is smaller than the long-distance
part by three orders of magnitude and does therefore not play any role
in the total rate. However, while the muon pair couples via a vector
current $(\ov{\mu}\mu)^{\phantom{l}}_V$ in the photon amplitude, the
electroweak short-distance mechanism also contains an axial vector piece
$(\ov{\mu}\mu)_A$. The interference term between these contributions
is odd under parity and gives rise to a parity-violating longitudinal
$\mu^+$ polarization, which can be observed as an asymmetry
$\Delta_{LR}=(\Gamma_R-\Gamma_L)/(\Gamma_R+\Gamma_L)$
\cite{SW,LWS}.
$\Gamma_{R(L)}$ denotes the rate of producing a right- (left-) handed
$\mu^+$ in $K^+\to\pi^+\mu^+\mu^-$ decay. The effect occurs for
a $\mu^-$ instead of $\mu^+$ as well, but the polarization measurement
is much harder in this case.

$\Delta_{LR}$ is sensitive to the Wolfenstein parameter $\rho$.
It is a cleaner observable than $K_L\to\mu^+\mu^-$, although some
contamination through long-distance contributions cannot be excluded
\cite{LWS}. At any rate, $\Delta_{LR}$ will be an interesting observable
to study if a sensitive polarization measurement becomes feasible.
The Standard Model expectation is typically around
$\Delta_{LR}\sim 0.5\%$.

\section{Summary}

The field of rare kaon decays raises a broad range of interesting
topics, ranging from ChPTh, over Standard-Model
flavordynamics to exotic phenomena, thereby covering scales from
$\Lambda_{QCD}$ to the weak scale ($M_W$, $m_t$) and beyond to maybe
several hundred TeV's. In the present chapter we have focused on the
flavor physics of the Standard Model and those processes that can be
used to test it. Several promising examples of short-distance sensitive
decay modes exist, whose experimental study will provide important clues
on flavordynamics.
On the theoretical side, progress has been achieved over recent years
in the calculation of effective Hamiltonians, which by now include
the complete NLO QCD effects in essentially all cases of practical
interest.

The decay $K_L\to\mu^+\mu^-$ is already measured quite accurately;
unfortunately a quantitative use of this result for the determination
of CKM parameters is strongly limited by large hadronic uncertainties.

The situation looks very bright for $K\to\pi\nu\ov{\nu}$. The charged
mode is experimentally already close and its very
clean status promises useful results on $V_{td}$. Finally the decay
$K_L\to\pi^0\nu\ov{\nu}$ is a particular highlight in this field.
It could serve for instance as the ideal measure of the Jarlskog parameter
$J_{\rm CP}$. Measuring the branching ratio is a real experimental
challenge.

It is to be expected that rare kaon decay phenomena will in the
future continue to contribute substantially to our understanding
of the fundamental interactions.

\chapter{THEORETICAL FRAMEWORK}                     % MUST be CAPITALIZED

\section{Motivation}

There are three rare decay modes of the long-lived kaon which have
interrelated theoretical issues: $K_L \rightarrow \pi^0 \gamma
\gamma$, $K_L \rightarrow \pi^0 e^+e^-$ and $K_L \rightarrow \pi^0
\gamma e^+e^-$. The first two have been extensively studied; the
latter has not been previously calculated. It is the purpose of this
thesis to provide a calculation of the latter two processes and describe how
they are related to the phenomenology of the first one.

There is a curious and important inverted hierarchy of these decay
modes. The rate for the radiative decay $K_L \rightarrow \pi^0
\gamma e^+e^-$ is a power of $\alpha$ {\it larger} than the
nonradiative transition $K_L \rightarrow \pi^0 e^+e^-$. This is
because the $K_L \rightarrow \pi^0 e^+e^-$ transition occurs only
through a two-photon intermediate state, or alternatively through
a one-photon exchange combined with CP violation (which numerically
appears to be roughly of the same size as the two-photon
contribution) \cite{DG}. The $K_L \rightarrow \pi^0 e^+e^-$ rate is then of
order $\alpha^4$. However, in $K_L \rightarrow \pi^0 \gamma e^+e^-$ we
need only a one-photon exchange to the $e^+e^-$, leading to a rate of
order $\alpha^3$. Our attention was first called to this inverted
hierarchy by an observation that there are infrared divergences in a
detailed study of the $K_L \rightarrow \pi^0 e^+e^-$ two-photon effect
\cite{DG} which need to be canceled by the one-loop corrections to
the radiative mode $K_L \rightarrow \pi^0 \gamma e^+e^-$ through the
contributions of the soft radiative photons. This implies that the
theoretical {\it and experimental} analyses of $K_L \rightarrow \pi^0
e^+e^-$ and $K_L \rightarrow \gamma \pi^0 e^+e^-$ are tied together.
The soft and collinear photon regions of $K_L \rightarrow \gamma \pi^0
e^+e^-$ form potential backgrounds to the studies of CP violation in
the $K_L \rightarrow \pi^0 e^+e^-$ mode.

The $K_L \rightarrow \pi^0 \gamma e^+e^-$ mode also has an interest of
its own. In recent years there have been important phenomenological
studies of $K_L \rightarrow \pi^0 \gamma\gamma$ in connection with
ChPTh. This decay is calculable at one-loop (\ie, order E$^4$) ChPTh
with no free parameters, yielding a very distinctive spectrum and a
definite rate \cite{EPR1,EPR2}. Surprisingly, when the experiment was
performed the spectrum was confirmed while the measured rate was more
than a factor of 2 larger than predicted. The way out of this
problem appears to have been provided by Cohen, Ecker and Pich (CEP)
\cite{CEP}. By adding an adjustable new effect at order E$^6$, as well
as including known corrections to the $K_L \ri \pi\pi\pi$ vertex, they
found that the predicted rate can be increased dramatically without
modifying the shape of the spectrum much. This is also a surprising
result, yet as far as we know it is the unique solution to the
experimental puzzle. The ingredients of the mode studied in this
thesis, $K_L \rightarrow \pi^0 \gamma e^+e^-$, are the same as for
$K_L \rightarrow \pi^0 \gamma\gamma$, except that one of the photons
is off shell. Within the framework of the CEP calculation, the
ingredients enter with different relative weights for off-shell
photons. This will allow us to test the consistency of the theoretical
resolution proposed for $K_L \rightarrow \pi^0
\gamma\gamma$.

One of the goals of the next generation of rare kaon decay experiments
is to attempt to observe CP violation in the decay $K_L \rightarrow
\pi^0 e^+ e^-$. This reaction is special because we expect that
direct CP violation (as opposed to the ``mass matrix" CP violation
already observed in the parameter $\epsilon$) may be the dominant
component of the amplitude. This is in contrast with $K_L \rightarrow
\pi \pi$, where the direct effect is at most a few parts in a thousand
of $\epsilon$. Direct CP violation distinguishes the Standard Model
from ``superweak"-type models \cite{WL}. Moreover, the magnitude of
the direct CP violation for this reaction is a precise prediction
of the Standard Model,
with very little hadronic uncertainty. In this thesis, we will
update the analysis of the reaction $K_L \rightarrow \pi^0 e^+ e^-$,
attempting to understand if we can be certain that an experimental
measurement is in fact a signal of direct CP violation.

One difficulty is that there are three possible components to the decay
amplitude: 1) A CP-conserving process which proceeds through two-photon
exchanges, 2) a mass matrix CP-violating effect proportional to the known
parameter $\epsilon$, and 3) the direct $\Delta S = 1$ CP-violating effect
that we would like to observe. The existence of the first of these
shows that simply observing the total decay rate is not sufficient to
unambiguously indicate the existence of CP violation. We need to either
observe a truly CP-odd decay asymmetry, or else be confident on the basis
of a theoretical calculation that the CP-conserving effect is safely smaller
than the experimental signal. Unfortunately, the predictions in the literature
for each of the components listed above exhibits a range of values,
including some estimates where all three are similar in magnitude.
However, the quality of the theoretical treatment can improve with time,
effort and further experimental input. We will try to assess the present and
future uncertainties in the theoretical analysis.

There remain significant experimental difficulties before it is possible to
mount a search sensitive to a branching ratio of a few times $10^{-12}$.
We will assume that such a sensitivity is reached. At the same time, it is
reasonable to assume that we will have improved experimental information
on the related rate $K_L \rightarrow \pi^0 \gamma \gamma$, and that
theoretical methods have provided a consistent phenomenology of this
reaction. The CKM parameters will be somewhat more fully constrained in
the future, but hadronic matrix element uncertainties will prevent a precise
determination of the parameters relevant for CP violation, at least until B
meson CP violation has been extensively explored. With these expectations
as our framework, will we be able to prove that the future experimental
observation indicates the presence of direct CP violation?

Our analysis shows that one will {\it not} be able to prove the
existence of direct CP violation from the branching ratio for $K_L
\rightarrow \pi^0 e^+ e^-$ unless the decay rate for the related decay
$K_S \rightarrow \pi^0 e^+ e^-$ is also observed experimentally. This
is yet more difficult than measuring the $K_L$ decay, and poses a
problem for the program of finding direct CP violation. It is possible
but not certain, that the electron charge asymmetry can resolve this
issue and, when combined with the rate, signal direct CP violation.

\section{Overview of the Analysis}

\vspace{2mm}
There is an extensive analysis associated with each of the three components
of the decay amplitude for $K_L \ri \pi^0 e^+ e^-$ that were listed
in the introduction, but only one work has appeared so far to treat
the decay $K_L \ri \pi^0 \gamma e^+ e^-$ \cite{DG1}. We will
devote separate chapters of this thesis to each of these major issues. The
purpose of the present chapter is to highlight the main issues that are to be
discussed more fully later, and to indicate how they fit together in an overall
description of the decay processes.

For the first process, the direct CP component is, in a way, the
simplest. The uncertainties are only in the basic parameters of the
Standard Model, \ie, the mass of the top quark and the CKM parameters.
The relevant hadronic matrix element is reliably known. Unfortunately
the extraction of CKM elements has significant uncertainties, so that
only a range of possible values can be given. This range corresponds
to $K_L \rightarrow \pi^0 e^+ e^-$ branching ratios of a few times
$10^{-12}$. We discuss this range in chapter 3.

The contribution of mass-matrix CP violation is more uncertain. The
rate due to this source is given by the parameter $\epsilon$ times the
rate for $K_S \rightarrow \pi^0 e^+ e^-$, so that the issue is the
prediction of the CP-conserving $K_S$ partial rate. Here the primary
tool is ChPTh, with the pioneering treatment
given by Ecker, Pich and de Rafael (EPR) \cite{EPR1,EPR2}. In chapter
4, we update their analysis, under essentially the same assumptions.
The main new ingredient is the inclusion of the results of the one
loop analysis of nonleptonic decays, which decreases the overall
strength of the weak $K \rightarrow \pi$ transition. This yields a
change in the weak counterterms and a decrease of the rate. However,
more important is an assessment of the uncertainties of such a
calculation, which we describe in chapter 5. Any such calculation has a
range of uncertainties, most of which we are able to estimate based on
past experience with chiral calculations. We systematically discuss
these. Unfortunately we find that one issue in particular has a
devastating sensitivity on this mode. In their analysis, EPR made an
assumption that lies outside of ChPTh, that a
certain weak counterterm satisfies $w_2 = 4 L_9$ where $L_9$ is a
known coefficient in the QCD effective chiral lagrangian. This
assumption has no rigor, and the decay rate is very sensitive to the
deviation $w_2 - 4 L_9$. For any reasonable value of direct CP
violation, there is an equally reasonable value of $w_2$ that can
reproduce the corresponding $K_L \rightarrow \pi^0 e^+ e^-$ decay
rate. Given a measurement, we will then be intrinsically unable to
decide if it is evidence of a nonzero value of direct CP violation or
merely measures a value for $w_2$. It is this what shows a need
to measure the rate $K_S \rightarrow \pi^0 e^+ e^-$.

The third component is the CP-conserving amplitude that proceeds
through the two-photon intermediate state $K_L \rightarrow \pi^0 \gamma
\gamma , \gamma \gamma \rightarrow e^+ e^-$, described in chapter 6.
Here we must first understand the process $K_L \rightarrow \pi^0 \gamma
\gamma$. This has been calculated in ChPTh at one loop
order and has been measured experimentally. While the shape agrees with
the chiral calculation, the theoretical rate misses by a factor of three. This
has prompted some reanalyses of the theory of $K_L \rightarrow \pi^0
\gamma \gamma$, which we will take account of. However, the field has
not reached a satisfactory conclusion on this mode, and it is clear
that in the future
the experimental and phenomenological status of this reaction will
undoubtedly improve. We study how possible resolutions of these analyses
will influence the $K_L \rightarrow \pi^0 \gamma \gamma$ decay rate.
Ultimately this component should be satisfactorily understood.

The ultimate problem is then our inability to distinguish, in a measurement
of the $K_L \rightarrow \pi^0 e^+ e^-$ decay rate, the direct CP violation
from the mass matrix effect. It is possible that the electron energy
asymmetry may allow us to make this separation. The electron asymmetry
comes from the interference of the CP-conserving two-photon process (even
under the interchange of $e^+ e^-$) and the CP-violating one-photon
process (odd under the $e^+ e^-$ interchange). For many values of the
presently favored parameter range, this asymmetry is very large \ie, of
order 50\%. In this case its measurement is not far more difficult than a
good measurement of the rate. If we in fact are able to reach an
understanding of the two-photon process, through future phenomenology
and experiments on $K_L \rightarrow \pi^0 \gamma \gamma$, then the
asymmetry depends most critically on the CP-violating amplitude. If there
is no direct CP violation, there is then a correlation between the $K_L
\rightarrow \pi^0 e^+ e^-$ decay rate and the electron asymmetry,
parameterized by the unknown coefficient $w_2$. As we detail in chapter
7, the presence of direct CP violation would upset this correlation, and in
many cases would lead to a drastically different relative size of the
asymmetry vs. decay rate, often even changing the sign of the asymmetry.
Thus the asymmetry may be used to signal direct CP violation.
Unfortunately this method is not foolproof. There exist combinations of
values of $w_2$ and CKM angles for which the distinction between direct
and mass matrix CP violation is not so great. In this case the
analysis will be muddied by the
inherent uncertainties in the theory. In chapter 7 we also explore
the use of $K_L - K_S$ interference to sort out the direct CP
violating amplitude. We outline the computation for the
$\cal{O}$(E$^4$) ChPTh contributions to the process $K_L \rightarrow \pi^0
\gamma e^+e^-$ in chapter 8, and then we extend it to $\cal{O}$(E$^6$)
in chapter 9. Finally, we recapitulate all our conclusions in chapter 10.

\chapter{DIRECT CP VIOLATION}        % MUST be CAPITALIZED

Direct $\Delta S = 1$ CP violation is manifested in the ``penguin"
reactions pictured in Fig. 4. The QCD short distance corrections to
this mode have been thoroughly analyzed to next-to-leading order by
Buras \etal\ \cite{B} (see also Ciuchini \etal\ \cite{C}), and we will use
their results. The primary weak operator responsible for the
transition has the form

\begin{equation}
{\cal H}^{\Delta S = 1}_W = {G_F \over \sqrt{2}}
\left[ C_{7V} (\mu) Q_{7V} + C_{7A} Q_{7A} \right],
\end{equation}

\noindent where

\begin{eqnarray}
Q_{7V} & = & ( \ov{s} d)_{V-A} (\ov{e} e)_V, \nonumber \\
Q_{7A} & = & ( \ov{s} d)_{V-A} (\ov{e} e)_A.
\end{eqnarray}

\noindent The dominant contribution to the imaginary part of the coefficient
$C_{7i}$ comes from the top quark, so that this is truly a short distance
process. The coefficients have a CP-violating component

\begin{equation}
{\rm Im} \; C_{7i} = -{\rm Im} \left( {V^{\phantom{*}}_{td} V^*_{ts} \over
V^{\phantom{*}}_{ud} V^*_{us}}
\right) y^{\phantom{\dagger}}_{7i},
\end{equation}

\noindent with the results of Ref. \cite{B} yielding

\begin{equation}
{y_{7V} \over \alpha} = 0.743,
\qquad {y_{7A} \over \alpha} = -0.736, \qquad m_t = 175,
\end{equation}

\noindent with very little dependence on $\Lambda_{\overline{MS}}$ (the
above is for $\Lambda_{\overline{MS}}$ = 0.3 GeV) and a negligible
dependence on the low energy scale $\mu$.

\begin{figure}[t]
\centering
\leavevmode
\centerline{
\epsfbox{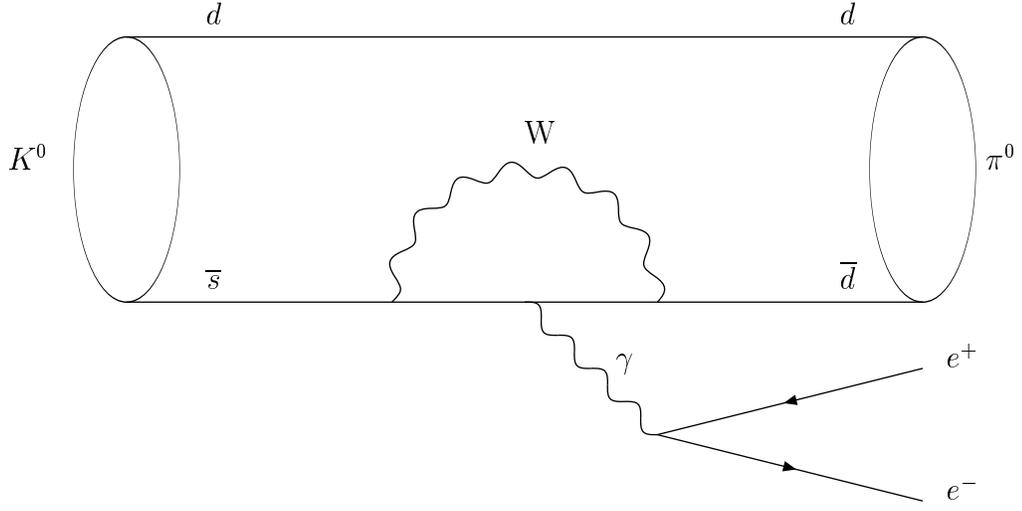}}
\caption{``Penguin'' diagram inducing $\Delta$S = 1 CP
violation.}
\end{figure}

The matrix element involved is well known via isospin symmetry from the
charged current kaon decay, i.e.,

\begin{equation}
\langle \pi^0 (p^{\phantom{l}}_{\pi}) \vert \ov{d} \gamma_{\mu} s
\vert K^0(p^{\phantom{l}}_K) \rangle =
{f_+ (q^2) \over \sqrt{2}} (p^{\phantom{l}}_K +
p^{\phantom{l}}_{\pi})_{\mu} + {f_- (q^2) \over
\sqrt{2}}(p^{\phantom{l}}_K - p^{\phantom{l}}_{\pi})_{\mu},
\end{equation}

\noindent with

\begin{eqnarray}
f_+ (q^2) & = & 1 + \lambda q^2, \nonumber \\
\lambda & = & (0.65 \pm 0.005) \; {\rm fm}^2.
\end{eqnarray}

\noindent The form factor $f_-$ does not contribute significantly to the
decay because its effect is proportional to $m_e$. The decay rate is

\begin{eqnarray}
BR(K_L \rightarrow \pi^0 e^+ e^- )_{\rm dir} = 4.16 ({\rm Im} \lambda_t )^2
(y^2_{7A} + y^2_{7V}), \nonumber \\
{\rm Im} \lambda_t = {\rm Im} V^{\phantom{*}}_{td}
V^*_{ts} = \vert V^{\phantom{*}}_{ub} \vert \vert V^{\phantom{*}}_{cb} \vert
\sin \delta = A^2 \lambda^5 \eta,
\end{eqnarray}

\noindent where $V^{\phantom{*}}_{td} = \vert V^{\phantom{*}}_{ub}
\vert \sin \delta$, and $A, \lambda,
\eta$ refers to the Wolfenstein parametrization of the CKM matrix
\cite{W}. This results in

\begin{equation}
BR (K_L \rightarrow \pi^0 e^+ e^-)_{\rm dir} =
2.75 \times 10^{-12} \times \left( {{\rm Im} \lambda_t \over
10^{-4}} \right)^2 \; {\rm for} \; m_t = 175 \; {\rm GeV}.
\nonumber \\
\phantom{l}
\end{equation}

\noindent The dependence on the top quark mass will of course be removed
by a convincing precise measurement of $m_t$.

The CKM parameter $V_{cb}$ has the most favored values (including the
recent CLEO data) \cite{S}

\begin{equation}
V^{\phantom{*}}_{cb} = \left\{ \begin{array}{ll}
0.0375 \pm 0.0015 \pm 0.0012 \quad & {\rm HQET} \quad\cite{LK},\\
0.036 \pm 0.002 \pm 0.003 & B \rightarrow D^{\ast} \ell \nu \; {\rm
models}, \\
0.039 \pm 0.001 \pm 0.004 & B \rightarrow X \ell \nu. \end{array} \right.
\end{equation}

\noindent The element $V^{\phantom{*}}_{ub}$ is measured by the
inclusive decay $B
\rightarrow X_u e \nu$ in the electron endpoint region. The two inclusive
calculations available yield

\begin{equation}
{V^{\phantom{*}}_{ub} \over V^{\phantom{*}}_{cb}} =
\left\{\begin{array}{ll}
0.082 \pm 0.006 \quad & \left[ {\rm ACCMM} \right] \quad\cite{ACCMM}, \\
0.074 \pm 0.007 & \left[ {\rm RDB} \right] \quad\cite{RDB}.
\end{array} \right.
\end{equation}

\noindent Models that calculate a set of exclusive decays ($B \rightarrow
Me \nu$) can only be used to provide an upper bound on
$V^{\phantom{*}}_{ub}$ since
there are many final states (such as $B \rightarrow \pi \pi e \nu$ with $\pi
\pi$
nonresonant) that are not calculated. These limits are

\begin{equation}
{V^{\phantom{*}}_{ub} \over V^{\phantom{*}}_{cb}} \leq
\left\{\begin{array}{ll}
0.12 \quad & [{\rm ISGW}] \quad\cite{ISGW}, \\
0.087 \quad & [{\rm BSW}] \quad\cite{WSB}, \\
0.067 \quad & [{\rm KS}] \quad\cite{KS}. \end{array} \right.
\end{equation}

\noindent We will use the former measurements to estimate

\begin{equation}
\left\vert {V^{\phantom{*}}_{ub} \over V^{\phantom{*}}_{cb}} \right\vert = 0.078 \pm 0.007 \pm 0.010,
\end{equation}

\noindent with the first uncertainty experimental and the second theoretical.
In the Wolfenstein parametrization of the CKM matrix, the values of
$V^{\phantom{*}}_{cb}$ and $V^{\phantom{*}}_{ub}$ imply

\begin{eqnarray}
A & = & 0.74 \pm 0.06, \nonumber \\
\sqrt{\rho^2 + \eta^2} & = & 0.355 \pm 0.056.
\end{eqnarray}

\noindent Without any further analysis, these measurements imply an upper
bound on ${\rm Im} \lambda_t$:

\begin{eqnarray}
{\rm Im} \lambda_t & = & \vert V^{\phantom{*}}_{ub}
\vert \vert V^{\phantom{*}}_{cb} \vert \sin \delta
\nonumber \\
& \leq & (1.0 \pm 0.3) \times 10^{-4} \sin \delta.
\end{eqnarray}

\noindent A lower bound on this parameter can be found by consideration of
the analysis of $\epsilon$. In the Wolfenstein parametrizations one has the
approximate form

\begin{equation}
\label{15}
\vert \epsilon \vert = (3.4 \times 10^{-3}) A^2 \eta B_K \left[ 1 + 1.3 A^2
(1 - \rho) \left( {m_t \over m_W} \right)^{1.6} \right],
\end{equation}

\noindent where $B_K$ parametrizes the hadronic matrix element and is
estimated to be in the range $0.33 \leq B_K \leq 1$. Using $(1 - \rho) <
1.4$ and $B_K < 1$ one finds for $m_t$ = 175 GeV

\begin{equation}
A^2 \eta \geq 0.13,
\end{equation}

\noindent so that

\begin{equation}
{\rm Im} \lambda_t = A^2 \lambda^5 \eta \geq 6.8 \times 10^{-5}.
\end{equation}

\noindent This brackets the range

\begin{equation}
0.68 \times 10^{-4} \leq {\rm Im} \lambda_t \leq 1.3 \times 10^{-4}.
\end{equation}

\noindent Note that ${\rm Im} \lambda_t$ is positive. These
constraints yield a decay rate from direct CP violation of magnitude

\begin{equation}
1.25 \times 10^{-12} \leq BR (K_L \rightarrow \pi^0 e^+ e^-)_{\rm dir}
\leq 4.6 \times 10^{-12} \; {\rm for} \; m_t = 175.
\nonumber \\
\phantom{l}
\end{equation}

\noindent Alternately, the ``best" values

\begin{equation}
{\rm Im} \lambda_t = 1.0 \times 10^{-4}, \qquad m_t = 175 \; {\rm GeV},
\end{equation}

\noindent which we will take as our standard reference values, lead to a
rate

\begin{equation}
\label{21}
BR (K_L \rightarrow \pi^0 e^+ e^-)_{\rm dir} = 2.32 \times 10^{-12}.
\end{equation}

\noindent A more detailed analysis including a correlation between $\rho$
and $\eta$ inherent in Eq. \ref{15}, as well as the use of $B^0_d
\overline{B}^0_d$ mixing (which constrains $A \sqrt{(\rho - 1)^2 + \eta^2}$ as
a function of $f_B$) narrows the range only slightly because hadronic
uncertainties dominate.

\chapter{REVISING THE EPR ANALYSIS}

In this chapter, we review the formalism for analyzing mass matrix CP
violation, first set forth by Ecker, Pich, de Rafael (EPR)
\cite{EPR1,EPR2}. This amounts to
the prediction of the decay rate for $K_S \rightarrow \pi^0 e^+ e^-$, since
the mass matrix effect is defined by

\begin{eqnarray}
\label{22}
A (K_L \rightarrow \pi^0 e^+ e^-) \vert_{\rm MM} & \equiv & \epsilon A
(K_S \rightarrow \pi^0 e^+ e^-), \nonumber \\
\epsilon & = & (2.258 \times 10^{-3}) e^{i \pi / 4}.
\end{eqnarray}

\noindent We then redo the results taking into account recent work on the
nonleptonic kaon decays to one-loop order. While this produces a
significant numerical change, it is more important as a prelude to our
subsequent analysis of uncertainties in the analysis.

The prediction of $K_S \rightarrow \pi^0 e^+ e^-$ comes from a
comparison with $K^+ \rightarrow \pi^+ e^+ e^-$, which contains many of
the same ingredients. The reactions are displayed schematically in Figs. 5,
6. In these diagrams the round circles represent the electromagnetic
coupling while the square boxes indicate the action of the weak interaction.
We know the electromagnetic interactions of pions and kaons from direct
measurement. The weak $K \rightarrow \pi$ transition of Fig. 5a,b is known
within some theoretical uncertainty from the use of chiral symmetry to relate
it to $K \rightarrow 2 \pi$ and $K \rightarrow 3 \pi$. However, the weak $K
\pi \gamma$ vertex is not known a priori and needs to be extracted from the
analysis of $K^+ \rightarrow \pi^+ e^+ e^-$.

\begin{figure}[t]
\centering
\leavevmode
\centerline{
\epsfbox{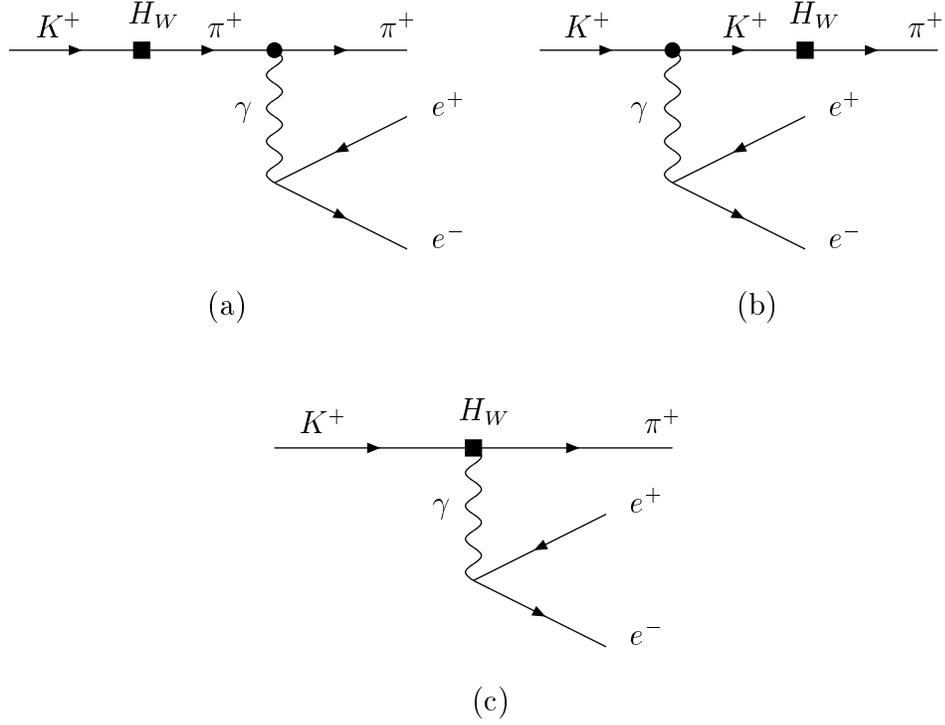}}
\caption{Diagrams contributing to the reaction $K^+ \ri
\pi^+ e^+ e^-$. Round circles represent the electromagnetic coupling
while the square boxes indicate the action of the weak interaction.}
\end{figure}

\begin{figure}[t]
\centering
\leavevmode
\centerline{
\epsfbox{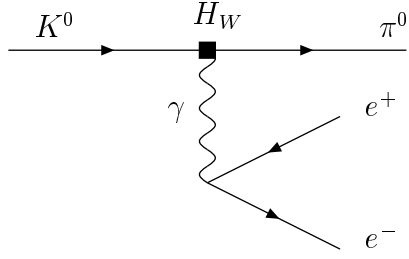}}
\caption{Same as in Fig. 5 for $K_L \ri \pi^0 e^+ e^-$.}
\end{figure}

The nonleptonic weak interactions are described by effective chiral
lagrangians, organized in an expansion in powers of the energy, or
equivalently in numbers of derivatives and masses. At lowest order, called
order $E^2$, the physical transitions are described by a unique lagrangian

\begin{eqnarray}
\label{23}
{\cal L} & = & G_8 Tr (\lambda_6 D_{\mu} U D^{\mu} U^{\dagger}) ,
\nonumber \\
U & \equiv & \exp \left(i {\lambda^A \cdot \phi^A \over
F_{\pi}}\right), \quad A = 1 \ldots 8,
\end{eqnarray}

\noindent where $\phi^A$ are the octet of pseudoscalar mesons ($\pi, K,
\eta$). At next order, order $E^4$, the number of possible forms of
lagrangians is quite large, and has been categorized by Kambor,
Missimer and Wyler \cite{KMW1}. Not all of these contribute to $K
\rightarrow 2 \pi$ and $K\rightarrow \pi$, but certain linear combinations do
influence these amplitudes. The formalism of ChPTh dictates that when
an analysis is carried out to order $E^2$, that one use Eq. \ref{23} at
tree level, in which case one obtains from the $K \rightarrow \pi$ decay rate

\begin{eqnarray}
G_8 & = & {G_F \over \sqrt{2}} \vert V^{\phantom{*}}_{ud} V^*_{us} \vert
g^{\phantom{\dagger}}_8, \nonumber \\
g^{\rm tree}_8 & = & 5.1.
\end{eqnarray}

Note that we neglect the CP violation in the nonleptonic
amplitude, contained in $G_8$, because this is bounded to be tiny by the
smallness of $\epsilon^{\prime}/\epsilon$.

\noindent In contrast, when evaluated at order $E^4$, one must include
one-loop diagrams in addition to the possible order $E^4$ lagrangian. The
loop diagrams involving $\pi \pi$ rescattering in the $I = 0$ channel, $K
\rightarrow (\pi \pi)_{I = 0} \longrightarrow (\pi \pi)_{I = 0}$ as pictured in
Fig. 7, are quite large and are the major part of the order $E^4$ analysis.
While there is some ambiguity in the extraction of
$g^{\phantom{\dagger}}_8$ (see below), the
enhancement from $\pi \pi$ rescattering leads to a smaller value of
$g^{\phantom{\dagger}}_8$, with a good estimate being \cite{KMW2}

\begin{equation}
g^{\rm loop}_8 = 4.3.
\end{equation}

\begin{figure}[t]
\centering
\leavevmode
\centerline{
\epsfbox{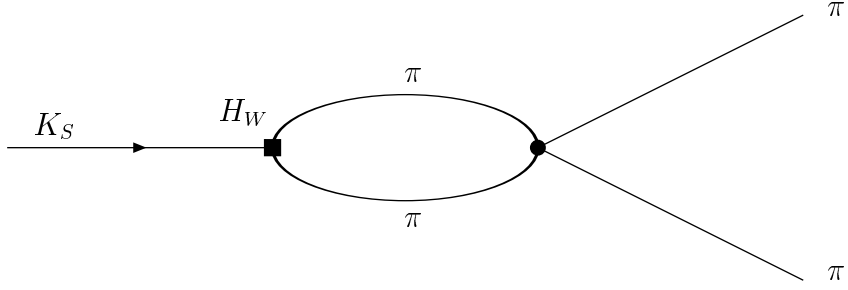}}
\caption{One-loop diagrams involving $ \pi \pi$ rescattering in
the $I$ = 0 channel for $K_S \ri ( \pi \pi )_{I = 0} \longrightarrow
( \pi \pi )_{I = 0}$. The notation for vertices is as in Fig. 5.}
\end{figure}

\noindent The $K \rightarrow \pi$ amplitude used in Figs. 5, 6 does not have
the enhancement from $\pi \pi$ rescattering, and is given in terms of
$g^{\phantom{\dagger}}_8$ by

\begin{eqnarray}
A(K^+ \rightarrow \pi^+) & = & 2 G_8 k^2, \nonumber \\
A(K^0 \rightarrow \pi^0) & = & - \sqrt{2} G_8 k^2.
\end{eqnarray}

\noindent We will explore further uncertainties in the $K \rightarrow \pi$
vertex in the next chapter.

To complete the diagrams of Figs. 5, 6 requires the electromagnetic
vertices of kaons and pions. In ChPTh to order $E^4$
these are given by

\begin{eqnarray}
\label{27}
\langle \pi^+ \vert J^{\mu} \vert \pi^+ \rangle \hskip -8.4pt &=&
\hskip -8.7pt \left\{\hskip -1.212pt 1 +q^2 \hskip -3pt
\left[ {2 L_9 (\mu) \over F^2_{\pi}} - {1 \over 96 \pi^2 F^2_{\pi}}
\hskip -2pt \left(\log
{m^2_{\pi} \over \mu^2} + {1 \over 2} \hskip -0.5pt \log {m^2_K \over
\mu^2} + {3
\over 2} \right) \right] \right\} \hskip -2pt (p^{\phantom{l}}_K +
p^{\phantom{l}}_{\pi})^{\mu}, \nonumber \\
\langle K^+ \vert J^{\mu} \vert K^+ \rangle \hskip -8.4pt &=&
\hskip -8.7pt \left\{\hskip -1.212pt 1 + q^2 \hskip -3pt \left[ {2
L_9 (\mu) \over F^2_{\pi}} - {1 \over 96 \pi^2 F^2_{\pi}}\hskip -2pt
\left(\log
{m^2_K \over \mu^2} + {1 \over 2} \hskip -0.5pt \log {m^2_{\pi} \over
\mu^2} + {3
\over 2} \right) \right] \right\} \hskip -2pt (p^{\phantom{l}}_K +
p^{\phantom{l}}_{\pi})^{\mu}. \nonumber \\
\end{eqnarray}

\noindent The first of these is known more fully from experiment, and has
the form

\begin{eqnarray}
\label{28}
\langle \pi^+ \vert J^{\mu} \vert \pi^+ \rangle & = & {(p^{\phantom{l}}_K +
p^{\phantom{l}}_{\pi})^{\mu} \over {1 - q^2 / m^2 }}, \nonumber \\
m & = & \; 730 \; {\rm MeV}.
\end{eqnarray}

\noindent Taylor expanding the latter form one determines $L_9
(\mu = m_{\eta}) = (7.4 \pm 0.7) \times 10^{-3}$. The experimental
charge radii

\begin{eqnarray}
\langle r^2 \rangle_{\pi^+} & = & (0.44 \pm 0.02) \; {\rm fm}^2, \nonumber \\
\langle r^2 \rangle_{K^+} & = & (0.34 \pm 0.05) \; {\rm fm}^2,
\end{eqnarray}

\noindent are compatible with this value.
The final ingredient required for Figs. 5c, 6 is the weak photonic
coupling. This includes both short distance and long distance physics, as
illustrated in Fig. 8. While the short distance components have a reliable
hadronic matrix element (it is due to the real parts of the coefficients
discussed in the previous chapter), the QCD coefficient depends strongly on
the low energy cutoff $\mu$. In the full matrix element, this dependence is
canceled by a corresponding dependence on $\mu$ of the long distance
physics. However, since the long distance physics is not calculable, we
must attempt to determine this coupling phenomenologically. The
innovation
of EPR was to elucidate the possible forms that this coupling could take.
They found that there were two possible chiral lagrangians that could
contribute to this process:

\begin{eqnarray}
{\cal L}_W & = & {i e G_8 \over 2} F^{\mu \nu} \left[ w_1 Tr \left( Q
\lambda_{6-i7} {\cal L}_{\mu} {\cal L}_{\nu} \right) + w_2 Tr \left( Q
{\cal L}_{\mu} \lambda_{6-i7} {\cal L}_{\nu} \right) \right], \nonumber \\
{\cal L}_{\mu} & \equiv & - \left( \partial_{\mu} U - i e [A_{\mu}, U]
\right) U^{\dagger} = -(D_{\mu} U) U^{\dagger}.
\end{eqnarray}

\noindent In the presence of the short distance electroweak penguin effect
due to Z exchange with an axial electron coupling, we need a third effective
lagrangian, not present in EPR,

\begin{equation}
{\cal L}^{\prime}_W = {i 2 \pi \alpha \over 3} G_8 w_5 \ov{e}
\gamma^{\mu} \gamma_5 e Tr \left( \lambda_{6-i7}{\cal L}_{\mu} \right).
\end{equation}

\noindent The correspondence with the notation of the previous chapter is

\begin{equation}
{\rm Im} \; w_5 = {3 \over 4 \pi} {1 \over {\vert V^{\phantom{*}}_{ud}
V^*_{us} \vert g^{\phantom{\dagger}}_8}} {y^{\phantom{\dagger}}_{7A}
\over \alpha} {\rm Im} \lambda_t.
\end{equation}

\begin{figure}[t]
\centering
\leavevmode
\centerline{
\epsfbox{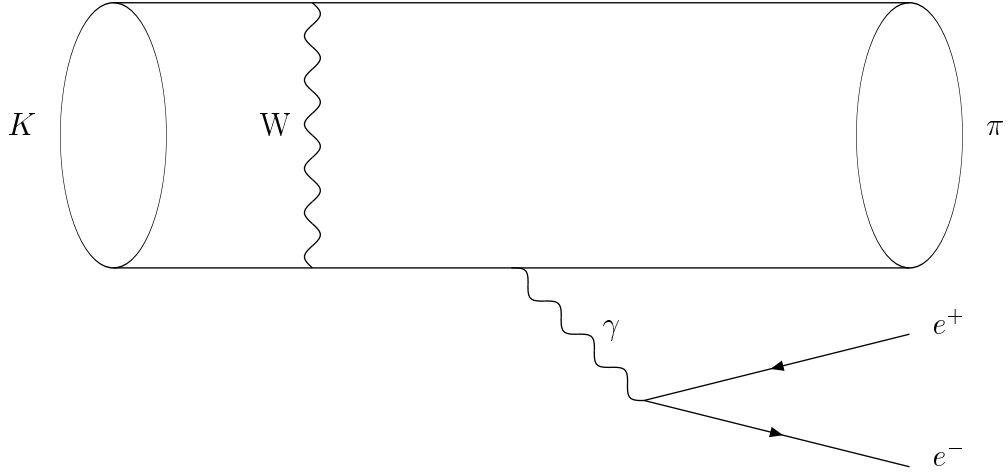}}
\caption{Diagrams contributing to the short distance weak photonic
coupling for $K \ri \pi e^+ e^-$.}
\end{figure}

\noindent Note that the labeling of $w_5$ takes into account the lagrangians
labeled by $w_3$, $w_4$ defined by EPR \cite{EPR1,EPR2,EPR3} that
contributed to other radiative $K$
decays. In fact this form of ${\cal L}^{\prime}_W$ is closely related to $w_1$
term since by using identities of the U matrix it can be shown that

\begin{eqnarray}
F^{\mu \nu} Tr (Q \lambda_{6-i7} {\cal L}_{\mu} {\cal L}_{\nu} ) & = &
-F^{\mu \nu} \partial_{\nu} Tr (Q \lambda_{6-i7} {\cal L}_{\nu} )
\nonumber \\
& = & (\partial_{\mu} F^{\mu \nu} ) Tr(Q \lambda_{6-i7} {\cal L}_{\nu}
) \nonumber \\
& = & (-e) \ov{e} \gamma_{\mu} e Tr(Q \lambda_{6-i7} {\cal L}_{\nu} )
\nonumber \\
& = & \left( {e \over 3} \right) \ov{e} \gamma_{\mu} e Tr (\lambda_{6-
i7} {\cal L}_{\nu}),
\end{eqnarray}

\noindent where in the second line we have integrated by parts thus
subsequently used the equation of motion so that

\begin{equation}
{i e G_8 \over 2} F_{\mu \nu} w_1 Tr (Q \lambda_{6-i7} {\cal L}_{\mu}
{\cal L}_{\nu} ) = {i \pi \alpha G_8 \over 3} w_1 \ov{e} \gamma_{\mu} e
Tr ( \lambda_{6-i7} {\cal L}_{\nu} ).
\end{equation}

\noindent This allows us to identify the short distance CP violating part of
$w_1$:

\begin{equation}
{\rm Im} \; w_1 = {3 \over 4 \pi} {1 \over {\vert V^{\phantom{*}}_{ud}
V^*_{us} \vert g^{\phantom{\dagger}}_8}} {y^{\phantom{\dagger}}_{7V} \over
\alpha} {\rm Im} \lambda_t.
\end{equation}

The real parts of $w_1, w_2$ contain long distance contributions and
hence are not
predictable by present techniques. They need to be extracted from
experimental measurements. The other process available for this procedure
is $K^+ \rightarrow \pi^+ e^+ e^-$ (unless $K_S \rightarrow \pi^0 e^+ e^-
$ is measured in the future). Unfortunately one cannot fix both $w_1,
w_2$ in this way, so that one must add a theoretical assumption in order to
proceed. The relevant amplitudes are

\begin{eqnarray}
{\cal M}(K^+ \rightarrow \pi^+ e^+ e^-) & = &\phantom{-}{G_8\alpha
\over 4 \pi}
d_+ (p^{\phantom{l}}_K + p^{\phantom{l}}_{\pi})^{\mu} \ov{u}
\gamma_{\mu} v, \nonumber \\
{\cal M}(K_S \rightarrow \pi^0 e^+ e^-) & = & - {G_8 \alpha \over 4 \pi}
d_S (p^{\phantom{l}}_K + p^{\phantom{l}}_{\pi})^{\mu} \ov{u}
\gamma_{\mu} v, \nonumber \\
{\cal M}(K_L \rightarrow \pi^0 e^+ e^-) & = & - {G_8 \alpha \over 4
\pi} (p^{\phantom{l}}_K +
p^{\phantom{l}}_{\pi})^{\mu} \ov{u} \left[ d_V \gamma_{\mu} + d_A
\gamma_{\mu}
\gamma_5 \right] v,
\end{eqnarray}

\noindent with

\begin{eqnarray}
\label{37}
d_+ & \equiv & w_+ + \phi_K (q^2) + \phi_{\pi} (q^2), \nonumber \\
d_S & \equiv & {\rm Re} \; w^{\phantom{2}}_S + 2 \phi_K (q^2), \nonumber \\
w_+ & = & - {16 \pi^2 \over 3} \left( w^r_1 + 2 w^r_2 - 12 L^r_9
\right) - {1 \over 6} \log {m^2_K m^2_{\pi} \over m^4_{\eta}}, \nonumber\\
w^{\phantom{2}}_S & = & w_+ + 16 \pi^2 \left( w^r_2 - 4 L^r_9 \right)
+ {1 \over 6} \log {m^2_{\pi} \over m^2_K}, \nonumber \\
\phi_i (q^2) & = & {m^2_i \over q^2} \int^1_0 dx \left[ 1 - {q^2 \over
m^2_i} x (1 - x) \right] \log \left[ 1 - {q^2 \over m^2_i} x (1 - x) \right],
\end{eqnarray}

\noindent and for the CP violating $K_L$ decay

\begin{eqnarray}
d_V & = & \epsilon d_S - {16 \pi^2 \over 3} i {\rm Im} \; w_1, \nonumber \\
d_A & = & {16 \pi^2 \over 3} i {\rm Im} \; w_5.
\end{eqnarray}

\noindent The goal of the search for direct CP violation is to separate the
${\rm Im} \; w_{1,5}$ terms from the mass matrix effect $\epsilon
d_S$. Note that
in these expressions we have neglected the possible direct CP violation in
the $K \rightarrow \pi$ transition (which is bounded to be very small by the
measurement of $\epsilon^{\prime} / \epsilon$) and the contribution of
${\rm Re} \; w_5$
to CP-conserving decays (since ${\rm Re} \; w_5 \approx {\rm Im} \;
w_5 << {\rm Re} \; w_1$).

The EPR analysis of $K^+ \rightarrow \pi^+ e^+ e^-$ uses the tree
level value of $g^{\phantom{2}}_8$, $g^{\rm tree}_8 = 5.1$. The decay
rate is consistent with two values of ${\rm Re} \; w_+$, and a
subsequent analysis of the decay spectrum favored the lower value for
${\rm Re} \; w_1$, i.e. ${\rm Re} \; w_+$ = 1.16 $\pm$ 0.08
\cite{EPR1}. However, given that one is working to one-loop order, it
is more consistent to use the one-loop value for $g^{\phantom{2}}_8$,
$g^{\rm loop}_8$ = 4.3. Because of the presence of the
$L_9$ term, this is not just a rescaling of the value of $w_+$. An
additional change that we make is to use the known full electromagnetic
vertex in the pole diagrams, rather than just the first term in the
expansion of the form factor. Note that because of the factor of
$p^{\phantom{l}}_K \cdot p^{\phantom{l}}_{\pi}$
in the weak matrix element, the only significant form factor
is that of the pion in Fig. 5. This implies the replacement

\begin{equation}
\label{39}
{2 L^r_9 \over F^2_{\pi}} \rightarrow {2L_9^r \over F^2_{\pi} \left( 1 - {q^2
/ m^2_{\rho}} \right)}
\end{equation}

\noindent in the formula for $w_+$. The associated logarithm with $\mu
\approx m_{\eta}$ are much smaller than the $L^r_9$ dependence and are
not influenced much by this replacement. As a technical note, we
comment that some potential modifications using a phenomenological
pion form factor could lead to a lack of gauge invariance. By
modifying the coefficient of a gauge invariant effective lagrangian,
we preserve the gauge invariant nature of the amplitude. With these
changes, we find

\begin{equation}
{\rm Re} \; w_+ = 1.01 \pm 0.10.
\end{equation}

\noindent [Without the second change, we would have had ${\rm Re} \; w_+ = 1.33
\pm 0.065$]. This is illustrated in Fig. 9.

\begin{figure}[t]
\centering
\leavevmode
\epsfxsize=300pt
\epsfysize=300pt
%\rotate[l]
{\centerline{\epsfbox{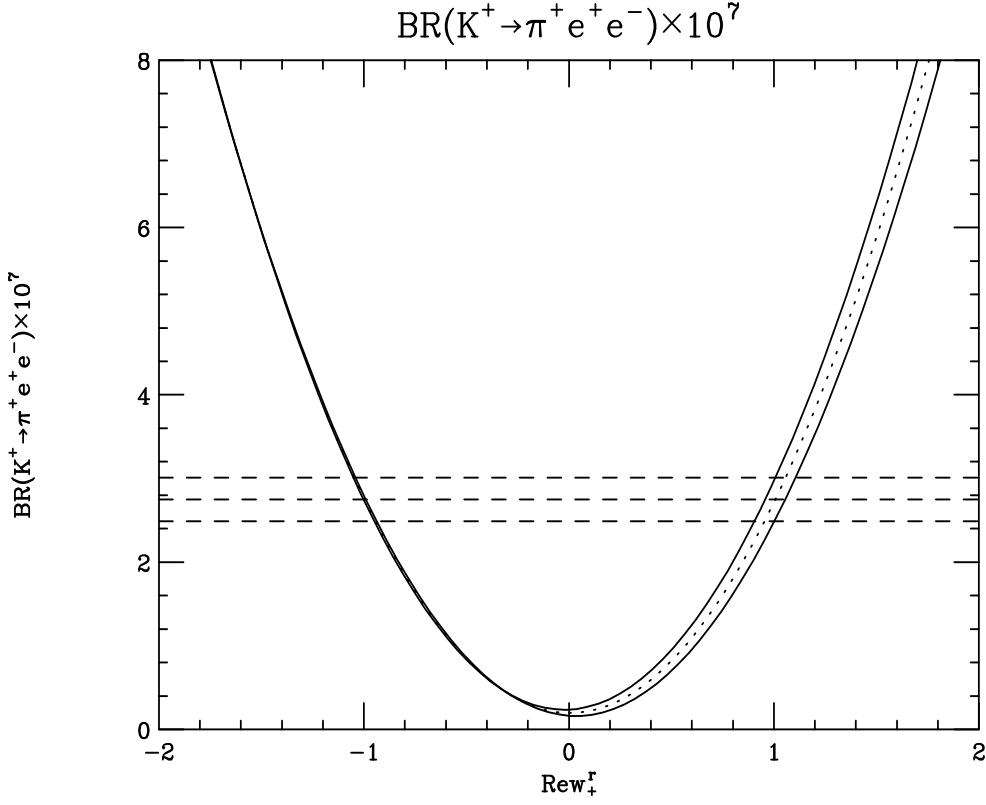}}}
\caption{The branching ratio $BR(K^+ \ri \pi^+ e^+ e^-)$ is
plotted against ${\rm Re} \; w_+$. The solid curves are obtained using
the extreme values of the error intervals of
$\vert V_{ud} \vert$ = 0.9753 $\pm$ 0.0006,
$\vert V_{us} \vert$ = 0.221 $\pm$ 0.003
and $L_9(m_{\eta})$ = (7.4 $\pm$ 0.7)$\times$ $10^{-3}$, while the dashed
curve corresponds to the central values. The experimental value of the
branching ratio $\pm$ its experimental errors are indicated by dashed
horizontal lines.}
\end{figure}

One cannot simply transfer this information to $K_S$ or $K_L$ decays,
because a different linear combination enters

\begin{equation}
w^{\phantom{2}}_S = w_+ + 16 \pi^2 ( w_2 - 4 L_9) + {1 \over 6}
\log {m^2_{\pi} \over m^2_K}.
\end{equation}

\noindent However, EPR deal with this problem by making the assumption
that $w_2 = 4L_9$ resulting in ${\rm Re} \; w^{\phantom{2}}_S = 0.73
\pm 0.08$ \cite{EPR1}.
They note that this assumption is not part of ChPTh, but
do not explore the consequences if it is not
correct. We will discuss this in the next chapter, finding a very
strong sensitivity. At this stage we note that if one makes the
assumption of $w_2 = 4 L_9$, one obtains ${\rm Re} \;
w^{\phantom{2}}_S = 0.58 \pm 0.10$ for our value of ${\rm Re}
\; w_+$.

At this value of $w^{\phantom{2}}_S$, the direct and mass matrix
contributions are comparable:

\begin{eqnarray}
d_V & = & \epsilon d_S - {16 \pi^2 \over 3} i {\rm Im} \; w_1 \nonumber \\
& \approx & e^{i {\pi / 4}} \left( 0.57 \times 10^{-3} \right)
- i 1.0 \times 10^{-3}, \nonumber \\
d_A & = & i 1.0 \times 10^{-3},
\end{eqnarray}

\noindent when evaluated with $m_t = 175$ GeV, ${\rm Im} \lambda_t =
10^{-4}$. This leads to a branching ratio

\begin{equation}
\label{43}
BR ( K_L \rightarrow \pi^0 e^+ e^-)_{\rm MM} = 0.37 \times 10^{-12}
\end{equation}

\noindent if there is no direct CP violation $({\rm Im} \lambda_t = 0)$ [EPR
found $BR_{\rm MM} = 1.5 \times 10^{-12}$ in this case \cite{EPR2}], vs.

\begin{equation}
BR ( K_L \rightarrow \pi^0 e^+ e^-)_{\rm CP} = 1.78 \times 10^{-12}
\end{equation}

\noindent for the full set of parameters given above in Eq. \ref{37}.
Figs. 10 and 11 show the dependence of both the above branching ratios
on ${\rm Re} \; w_+$. The addition
of mass matrix CP violation in this analysis led to a small decrease in the
rate compared to the purely direct CP violation of the previous chapter,
Eq. \ref{21}, because of the cancellation in the imaginary part of
$d_V$. However, this may change if $w_2 \neq 4L_9$.

\begin{figure}[t]
\centering
\leavevmode
\epsfxsize=300pt
\epsfysize=300pt
%\rotate[l]
{\centerline{\epsfbox{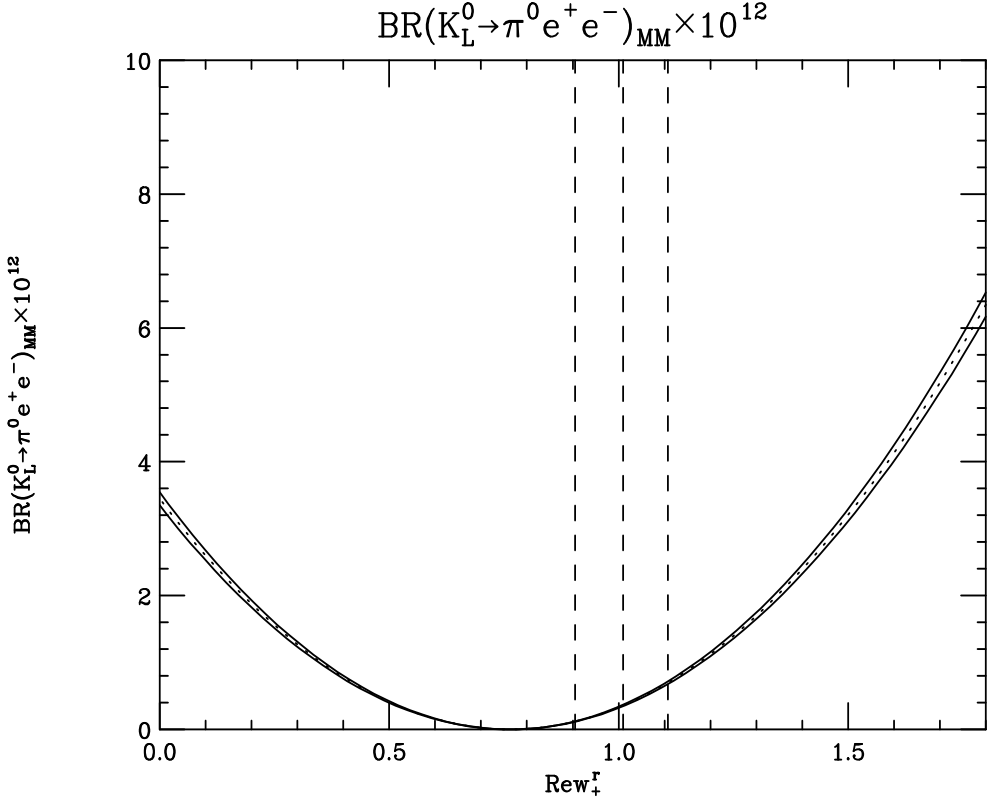}}}
\caption{The branching ratio $BR(K^0_L \ri \pi^0 e^+ e^-)_{\rm MM}$ is
plotted against ${\rm Re} \; w_+$. The solid curves are obtained using
the extreme values of the error intervals of
$\vert V_{ud} \vert$ = 0.9753 $\pm$ 0.0006,
$\vert V_{us} \vert$ = 0.221 $\pm$ 0.003
and $L_9(m_{\eta})$ = (7.4 $\pm$ 0.7)$\times$ $10^{-3}$, while the dashed
curve corresponds to the central values. The experimental limit on
${\rm Re} \; w_+$ $\pm$ its experimental errors are indicated by dashed
vertical lines.}
\end{figure}

\begin{figure}[t]
\centering
\leavevmode
\epsfxsize=300pt
\epsfysize=300pt
%\rotate[l]
{\centerline{\epsfbox{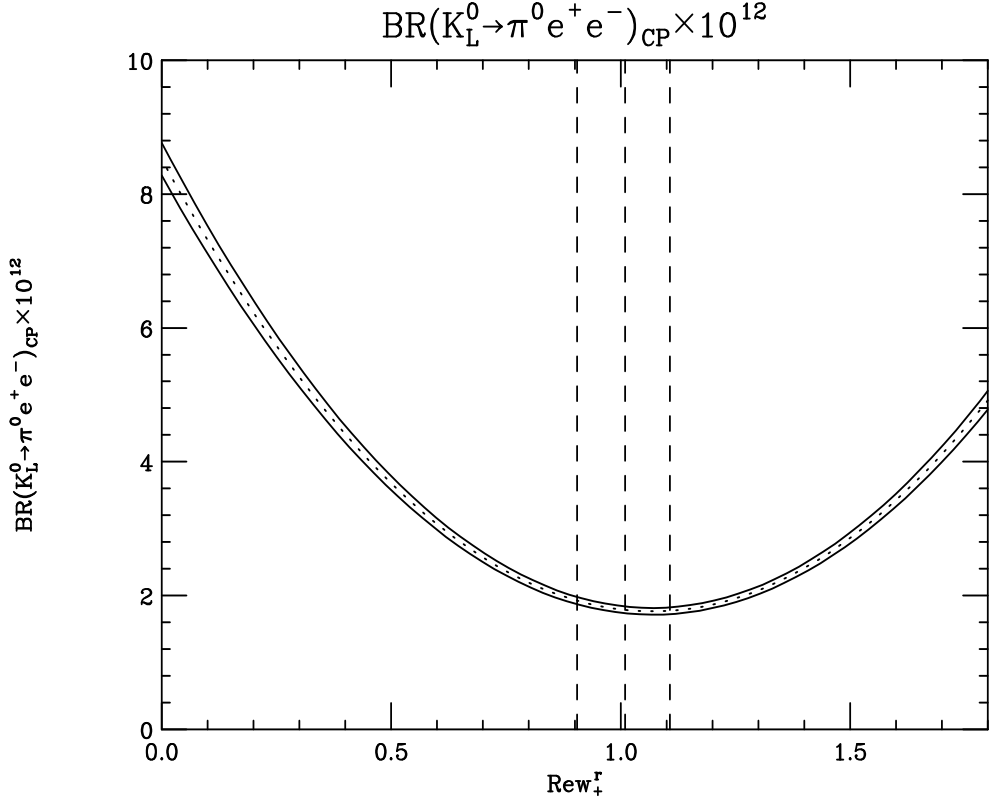}}}
\caption{Same as Fig. 10 for the branching ratio $BR(K^0_L \ri
\pi^0 e^+ e^-)_{\rm CP}$.}
\end{figure}

One of the results that we will see in the next chapter is that the
mass matrix contribution to the branching ratio is near a minimum
value when $w_2$ is close to $4L_9$. For other values of $w_2$, the
rate can easily be an order of magnitude larger. Although our value
for the branching ratio is a factor of four below that of EPR, both
estimates are similar in saying that the mass matrix contribution will
be small as long as $w_2$ = $4L_9$.

Using the tree level value for $g^{\phantom{\dagger}}_8$, $g^{\rm
tree}_8$ = 5.1, the mass matrix contribution to the branching
ratio is

\begin{equation}
\label{45}
BR ( K_L \rightarrow \pi^0 e^+ e^-)_{\rm MM} = 0.55 \times 10^{-12}.
\end{equation}

\chapter{UNCERTAINTIES IN MASS MATRIX CP VIOLATION}

Our goal in this chapter is to assess how well we understand the prediction
for $K_S \rightarrow \pi^0 e^+ e^-$. The most important effect will be
discussed in section 5.3 below, but we proceed systematically to discuss
even contributions that have less uncertainty.

\section{Purely Electromagnetic Vertices}

The electromagnetic vertices enter in diagrams 5a,b, 6. The
uncertainty here is in the choice of whether to use the chiral
expansion of the form factor truncated at order $q^2$, Eq. \ref{27}, or
the full $q^2$ dependence of the monopole form factor, Eqs. \ref{28},
\ref{39}. The first choice is natural when one is working to a given order
in the chiral energy expansion, but the latter choice clearly includes
more of the physics that is known about the electromagnetic vertex.
Note that essentially only the pion form factor is
relevant, because the diagram involving the kaon form factor, Fig.
5b, is suppressed by a factor of $m^2_{\pi}/m^2_K$ with respect to
Fig. 5a because of the momentum dependence of the weak $K \ri \pi$
transition. The use of the full form factor produces a modest
variation in the value of $w_+$ (i.e. $Re\; w_+ = 1.01$ instead of
$Re\; w_+ = 1.33$). Because of cancellations in the $K^0$ amplitude
this provokes a more extreme variation on the decay rate. The lowest order
chiral vertex, Eq. \ref{27}, produces a decay rate

\begin{equation}
BR ( K_L \rightarrow \pi^0 e^+ e^-)_{\rm MM} = 1.96 \times 10^{-12}
\end{equation}

\noindent instead of the result of Eq. \ref{43}. [We note that if we had
also modified $w_2$ in the same way as $4L_9$ as in Eq. \ref{39} we would
have a $BR ( K_L \rightarrow \pi^0 e^+ e^-)_{\rm MM} = 1.33 \times
10^{-12}$.] While we feel that it is good physics to use the full
electromagnetic form factor, one could also interpret these results as
an uncertainty in the analysis due to higher order terms in $q^2$,
with that uncertainty being of order $2 \times 10^{-12}$.

\section{The Weak $K \rightarrow \pi$ Vertex}

We have already given one indication of the sensitivity of the result
to the size of the $K \rightarrow \pi$ transition. Under otherwise
identical assumptions, $g^{\rm tree}_8 = 5.1$ yielded the rate in Eq.
\ref{45}, while $g^{\rm loop}_8 = 4.3$ produced the result of Eq. \ref{43}.
These modifications to $g^{\phantom{l}}_8$ also include corresponding
modifications to the $K\pi\gamma$ vertex required by gauge invariance.
This is automatically maintained, however, by the use of gauge
invariant effective lagrangians. In order to appreciate that this
change in $g^{\phantom{l}}_8$ is not the only uncertainties in the $K
\pi$ amplitude, one needs to understand a bit more about the chiral
phenomenology of $K \rightarrow 2 \pi$ and $K \rightarrow 3 \pi$.

Chiral symmetry relates processes with different numbers of pions, such as
$K \rightarrow \pi$ vs. $K \rightarrow 3 \pi$. The predictions are
compactly
contained in the chiral lagrangians, but can also be obtained using the soft
pion theorems, which was the methodology used in the 1960's. The only
advantage of the latter technique is that it relies only on chiral SU(2) while
modern chiral lagrangian analyses have always involved chiral SU(3)
symmetry. [Presumably the latter could be reformulated in chiral SU(2),
but no one has yet done this.] The soft pion analysis indicates that one
obtains the same relation (up to terms of order $m^2_{\pi}$) between $K
\rightarrow 2 \pi$ and $K \rightarrow 3 \pi$ for any lagrangian that
survives in {\em any} soft pion limit of $K \rightarrow 3 \pi$ (i.e., $p_i
\rightarrow 0$). The only lagrangians that do not survive in any soft
pion limit involve four separate derivatives on the four fields of $K
\rightarrow 3 \pi$, e.g.

\begin{equation}
{\cal L}_{\rm quartic} = {g^{\phantom{2}}_8 \over \Lambda^2_1}
Tr \left( \lambda_6 D_{\mu} U D_{\nu} U^{\dagger} D^{\mu} U
D^{\nu} U^{\dagger} \right).
\end{equation}

\noindent This lagrangian yields a matrix element proportional to $(p_K \cdot
p_1) (p_2 \cdot p_3)$ which clearly vanishes as any $p_i \rightarrow 0$. In
contrast most of the order $E^4$ lagrangians do not vanish in all soft pion
limits. An example is

\begin{equation}
\label{48}
{\cal L}^{\prime} = {g^{\phantom{2}}_8 \over \Lambda^2_2}
Tr \left( U^{\dagger} \lambda_6 D_{\mu} D_{\nu} U
D^{\mu} U^{\dagger} D^{\nu} U \right),
\end{equation}

\noindent which then yields the same relations of $K \rightarrow 3 \pi$ and
$K \rightarrow 2 \pi$ as does the lowest order result ${\cal L}$ given
in Eq. \ref{23}. [This phenomenon is explained in more detail in Ref.
\cite{DGH}.] Since the only inputs to
the chiral phenomenology are the amplitude for $K \rightarrow 2 \pi$ and
$K \rightarrow 3 \pi$, it follows that one cannot distinguish a combination
${\cal L} + {\cal L}^{\prime}$ from a lagrangian involving ${\cal L}$ only.

However, when we discuss the $K \rightarrow \pi$ vertex,
there {\em is} a distinction between these various lagrangians.
For example ${\cal L}^{\prime}$
in Eq. \ref{48} involves a minimum of three meson fields, and hence contributes to
$K \rightarrow 2 \pi$ and $K \rightarrow 3 \pi$ but not at all to $K
\rightarrow \pi$. In contrast, the lowest order lagrangian Eq.
\ref{23} contributes
to all of $K \rightarrow \pi, K \rightarrow 2 \pi, K \rightarrow 3 \pi$. Since
known phenomenology cannot distinguish between linear combinations of
${\cal L} + {\cal L}^{\prime}$, this manifests itself in an
uncertainty in the
$K \rightarrow \pi$ vertex. Since the higher order lagrangians are known
to
make a 25\% difference in relation between $K \rightarrow 2 \pi$ and $K
\rightarrow 3 \pi$, it would be unreasonable to take this uncertainty in $K
\rightarrow \pi$ to be any less than 25-30\%. Using
$g^{\phantom{\dagger}}_8 = 4.3 \times (1 \pm 30\%)$ yields a range

\begin{equation}
BR(K_L \rightarrow \pi^0 e^+ e^-)_{\rm MM} \sim {\cal O}(10^{-16}) \div
1.5 \times 10^{-12}.
\end{equation}

\noindent This again indicates that the analysis has significant
cancellations present and that modest variations in the analysis can
lead to uncertainties of order 1.5$\times 10^{-12}$.

\vspace{-2mm}
\section{The $K \pi \gamma$ Vertex}

The parameter $w_+$ was determined from the analysis of $K^+
\rightarrow \pi^+ e^+ e^-$. This arrangement has some intrinsic
uncertainty because it was performed to a given order in the chiral energy
expansion. It would be reasonable to take this uncertainty at 30\%.

However, this uncertainty is dwarfed by the errors introduced by the
assumption of $w_2 = 4 L_9$. There is nothing about chiral
symmetry that forces such a relation. For example, if the process of Fig.
12 were to contribute to the weak coupling, the relation $w_2 = 4 L_9$
would not occur except for a special value of the $K
\rightarrow a_1$ amplitude. This is highly unlikely, and
we can easily accept $w_2 - 4 L_9 \neq 0$. The crucial distinction
here is between models and rigorous theory. ChPTh is a
rigorous method which expresses true relationships in QCD to a given order
in the energy expansion. However, an assumption such as $w_2 = 4
L_9$ may be true or false in a way that we cannot decide based on QCD. It
may occur within some models, yet we have no guidance as to whether it is
correct in nature. We cannot base something as important as the
observation of direct CP violation on something as flimsy as a model.

\begin{figure}[t]
\centering
\leavevmode
\centerline{
\epsfbox{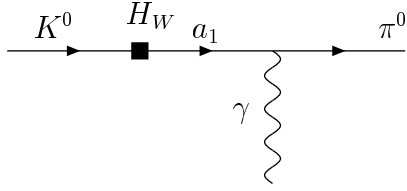}}
\caption{Diagram contributing to the weak photon coupling.
The notation for the vertices is as in Fig. 5.}
\end{figure}

Unfortunately the decay rate for $K_S \rightarrow \pi^0 e^+ e^-$ depends
very strongly on the value of $w_2$. This strong dependence was
observed by Littenberg and Valencia \cite{LV}. If we were to chose $w_2 = 0$,
the rate would be two orders of magnitude larger. In Fig. 13 we plot the
branching ratio for the mass matrix contribution to $K_L \rightarrow \pi^0
e^+ e^-$ vs. $w_2$. We see that reasonable values of $w_2$,
we get a wide range of values of the branching ratio. Conversely, a
measured value of $BR (K_L \rightarrow \pi^0 e^+ e^-)$ in the range of
$10^{-12} \rightarrow {\rm few} \times 10^{-11}$ could be interpreted in terms
of a reasonable value of $w_2$. This is then an enormous uncertainty
in the mass matrix contributions to $K_L \rightarrow \pi^0 e^+ e^-$.

This uncertainty could be removed if one measured the rate of $K_S
\rightarrow \pi^0 e^+ e^-$. The mass matrix contribution would then be
known, see Eq. \ref{22}. However, this task is not easy experimentally.

\begin{figure}[t]
\centering
\leavevmode
\epsfxsize=300pt
\epsfysize=300pt
%\rotate[l]
{\centerline{\epsfbox{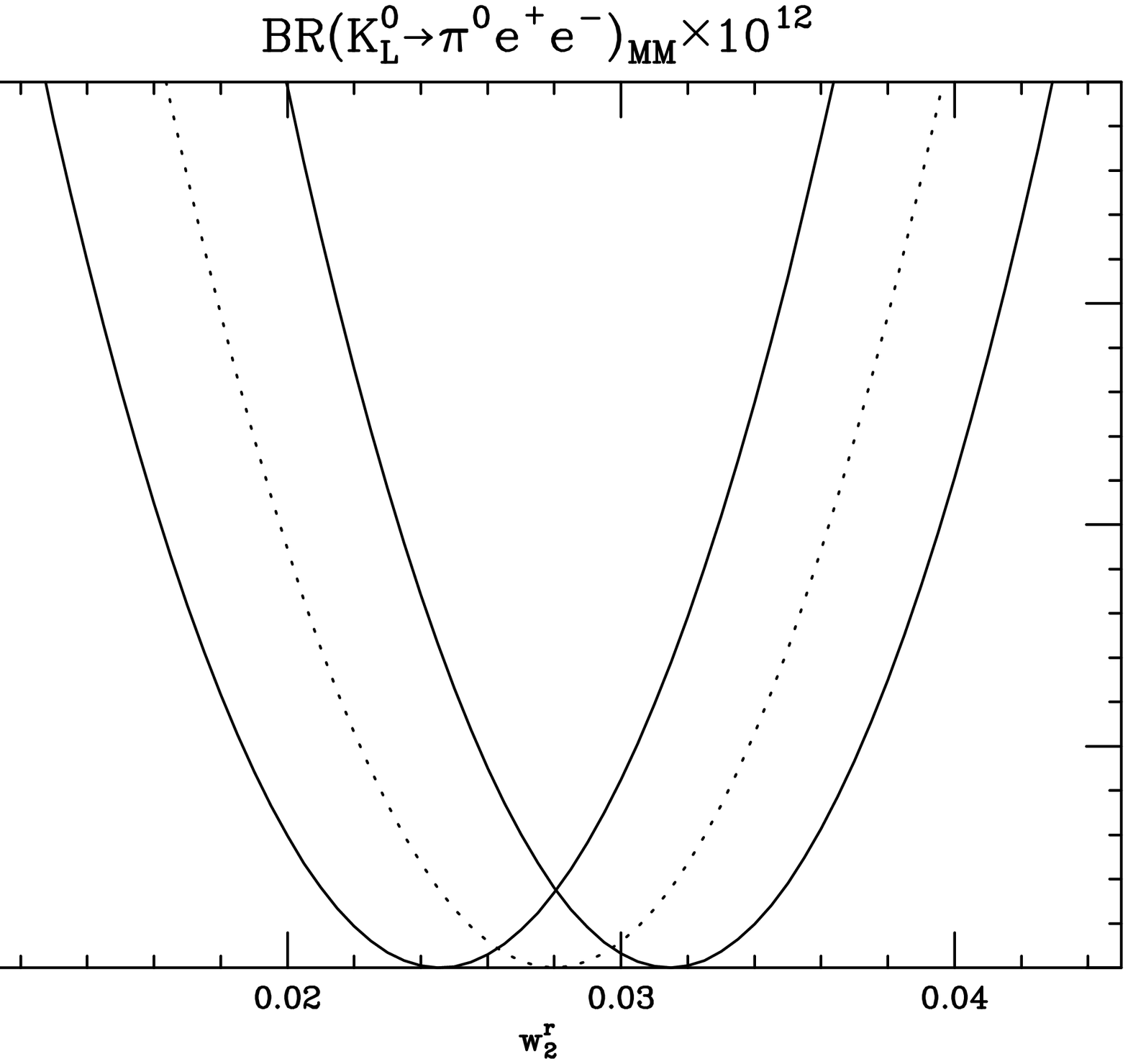}}}
\caption{The mass matrix contribution for the branching
ratio $BR(K_L \ri \pi^0 e^+ e^-)_{\rm MM}$ is plotted against $w_2$. The
convention for the solid and dashed curves is the same as for Fig. 9.}
\end{figure}

\chapter{THE CP-CONSERVING AMPLITUDE}

The $K_L \rightarrow \pi^0 e^+ e^-$ transition can also take place through
a CP-conserving two-photon intermediate state. If we ignore the electron
mass, the form of the amplitude will be

\begin{equation}
\label{50}
{\cal M} (K_L \rightarrow \pi^0 e^+e^-)_{\rm CPC} =
G_8 \alpha^2 K p^{\phantom{l}}_K \cdot (k - k^{\prime}) (p^{\phantom{l}}_K +
p^{\phantom{l}}_{\pi})^{\mu}
\ov{u} \gamma_{\mu} v,
\end{equation}

\noindent where $K$ is a form factor and the extra
antisymmetry under $k \leftrightarrow k^{\prime}$ ($k$ $\equiv$
$k_{e^+}$, $k^{\prime}$ $\equiv$ $k_{e^-}$) is a reflection of the
properties under a CP transformation. In order to calculate this we need to
understand the $K_L \rightarrow \pi^0 \gamma \gamma$ transition first.

We are fortunate that $K_L \rightarrow \pi^0 \gamma \gamma$ is accessible
to present experiments, and significant phenomenology has been performed
on this reaction. We will utilize the work of Cohen, Ecker and Pich
\cite{CEP} as
representative of present work, with the understanding that future
experimental and theoretical work will clarify the analysis considerably.

The most general form of the $K\ri\pi\gamma\gamma$ amplitude
depends on four independent invariant
amplitudes \cite{EPR2} $A$, $B$, $C$ and $D$:

\beqn
\lefteqn{{\cal M}[K(p^{\phantom{l}}_K) \ri \pi^0(p^{\phantom{l}}_\pi)
\gamma(q_1)\gamma(q^{\phantom{l}}_2)]\, =\,
{G_8 {\alpha} \over 4 \pi}\epsilon_\mu(q^{\phantom{l}}_1)
\,\epsilon_\nu(q_2) \, \Biggl\{
    A\,
    \Bigl(q_2^\mu q_1^\nu - q^{\phantom{l}}_1\cdot
q_2 \, g^{\mu\nu}\Bigr) \Biggr. }\no \\
&& \Biggl.
+ 2 {B \over {m^2_K}}\,
\Bigl(p^{\phantom{l}}_K\cdot q^{\phantom{l}}_1 \, q_2^\mu p_K^\nu
+ p^{\phantom{l}}_K\cdot q^{\phantom{l}}_2\, q_1^\nu p_K^\mu
- q^{\phantom{l}}_1\cdot q^{\phantom{l}}_2 \, p_K^\mu p_K^\nu -
p^{\phantom{l}}_K\cdot q^{\phantom{l}}_1\, p^{\phantom{l}}_K\cdot
q^{\phantom{l}}_2 \, g^{\mu\nu}\Bigr) \Biggr. \no\\
&& \Biggl. + C\,\epsilon^{\mu\nu\rho\sigma} q^{\phantom{l}}_{1\rho}
q^{\phantom{l}}_{2\sigma} \Biggr.\\ && \Biggl.
+ {D \over m^2_K}\,\Biggl[
\epsilon^{\mu\nu\rho\sigma}\left(
p^{\phantom{l}}_K\cdot q^{\phantom{l}}_2\, q^{\phantom{l}}_{1\rho} +
p^{\phantom{l}}_K\cdot q^{\phantom{l}}_1\,
q^{\phantom{l}}_{2\rho}\right) p^{\phantom{l}}_{K\sigma}\Biggr. \Biggr. \no \\
&& \Biggl. \Biggl. + \left( p_K^\mu
\epsilon^{\nu\alpha\beta\gamma} +
 p_K^\nu \epsilon^{\mu\alpha\beta\gamma}\right)
p^{\phantom{l}}_{K\alpha} q^{\phantom{l}}_{1\beta}
q^{\phantom{l}}_{2\gamma} \Biggr] \Biggr\}. \no
\eeqn

In the limit where CP is conserved, the amplitudes $A$ and $B$ contribute
to $K_2\ri\pi^0\gamma\gamma$ whereas $K_1\ri\pi^0\gamma\gamma$
involves the other two amplitudes $C$ and $D$. All four amplitudes
contribute to $K^+\ri\pi^+\gamma\gamma$.
Only $A$ and $C$ are non-vanishing to lowest non-trivial order,
${\cal O}(E^4)$, in ChPTh.

%The $K_L \rightarrow \pi^0 \gamma \gamma$ amplitude has two form factors
%$A$, $B$, defined via
%
%\begin{eqnarray}
%{\cal M} \left( K_L \rightarrow \pi^0 \gamma \gamma \right) =
%{G_8 {\alpha} \over 4 \pi} \epsilon_{\mu} (q_1) \epsilon_{\nu} (q_2) \left[
%A \left( q^{\mu}_2 q^{\nu}_1 - q_1 \cdot q_2 \cdot g^{\mu \nu} \right)
%\phantom{{B \over m^2_K}} \right. \nonumber \\
%\left. +2 {B \over m^2_K} \left( p \cdot q_1 q^{\mu}_2 p^{\nu} + p
%\cdot q_2 q^{\nu}_1 p^{\mu} - q_1 \cdot q_2 p^{\mu} p^{\nu} - g^{\mu
%\nu} p \cdot q_1 p \cdot q_2 \right) \right].
%\end{eqnarray}

When the photons couple to $e^+ e^-$, as in Fig. 14, it is well
known that the $A$ amplitude contributes to $K_L \rightarrow \pi^0 e^+ e^-$
only proportionally to $m_e$, which is a small effect that we will drop. It
is the $B$ amplitude that is important for the $e^+ e^-$ final state. The
authors use a representation that fits the known $K_L \rightarrow \pi^+
\pi^- \pi^0$ amplitude in a dispersive treatment of $K_L \rightarrow \pi^0
\gamma \gamma$ and find

\begin{eqnarray}
\label{52}
B(x) & = & c_2 \left\{ {1 \over x} F(x) + {4 \over 3} ( 5 - 2x) \left[ {1 \over
6} + R (x) \right] + {2 \over 3} \log {m^2_{\pi} \over m^2_{\rho}} \right\} +
\beta - 8 a^{\phantom{l}}_V, \nonumber \\
x & = & { \left( k + k^{\prime} \right)^2 \over 4 m^2_{\pi}} ,
\nonumber \\
\beta & = & -0.13, \nonumber \\
c_2 & = & 1.11, \nonumber \\
F_{\rm CEP}(x) & = & 1 - {1 \over x} \left[ \arcsin \left( \sqrt{x}
\right) \right]^2,
\, \qquad x \leq 1, \nonumber \\
& = & 1 + {1 \over 4x} \left( \log {1 - \sqrt{1 - 1/x} \over 1 + \sqrt{1 -
1/x}} + i \pi \right)^2, \, \qquad x \geq 1, \nonumber \\
R_{\rm CEP}(x) & = & - {1 \over 6} + {1 \over 2x} \left[ 1 - \sqrt{1/x
- 1} \arcsin
\left( \sqrt{x} \right) \right], \, \qquad x \leq 1, \nonumber \\
& & - {1 \over 6} + {1 \over 2x} \left[ 1 + \sqrt{1 - 1/x} \left( \log {1 -
\sqrt{1 - 1/x} \over 1 + \sqrt{1 - 1/x}} + i \pi \right) \right], \,
\qquad x \geq 1. \nonumber \\
\phantom{l}
\end{eqnarray}

\begin{figure}[t]
\centering
\leavevmode
\centerline{
\epsfbox{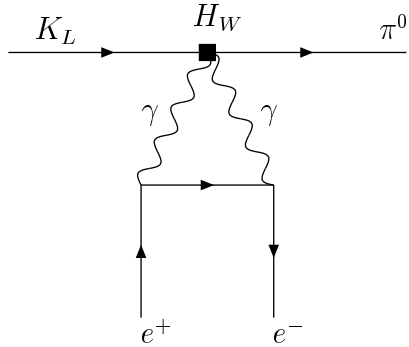}}
\caption{CP-nonviolating diagram involving two photons
coupling to $e^+ e^-$. The notation for the vertices is as in Fig. 5.}
\end{figure}

The most important ingredient above is $a^{\phantom{l}}_V$
which is a parameter representing the vector meson exchange
contributions
to the $A$, $B$ and $D$ amplitudes. A fit (shown in Fig. 15) to the decay rate
[using $BR(K_L \rightarrow \pi^0 \gamma \gamma) = 1.7 \times 10^{-6}$
\cite{NA}] and the $\gamma \gamma$ spectrum in $K_L
\rightarrow \pi^0 \gamma \gamma$ (depicted in Fig. 16) indicates a value
around $a^{\phantom{l}}_V = -0.96$.
This parameter was very important in increasing the chiral prediction of the
decay rate to be in agreement with experiment. We have explored the
sensitivity of this parameter to changes in the analysis and have found that
25\% changes in the dispersive treatment lead to a factor of 2 change in
$a^{\phantom{l}}_V$, so that this value is still quite uncertain.

\begin{figure}[t]
\centering
\leavevmode
\epsfxsize=300pt
\epsfysize=300pt
%\rotate[l]
\centerline{
\epsfbox{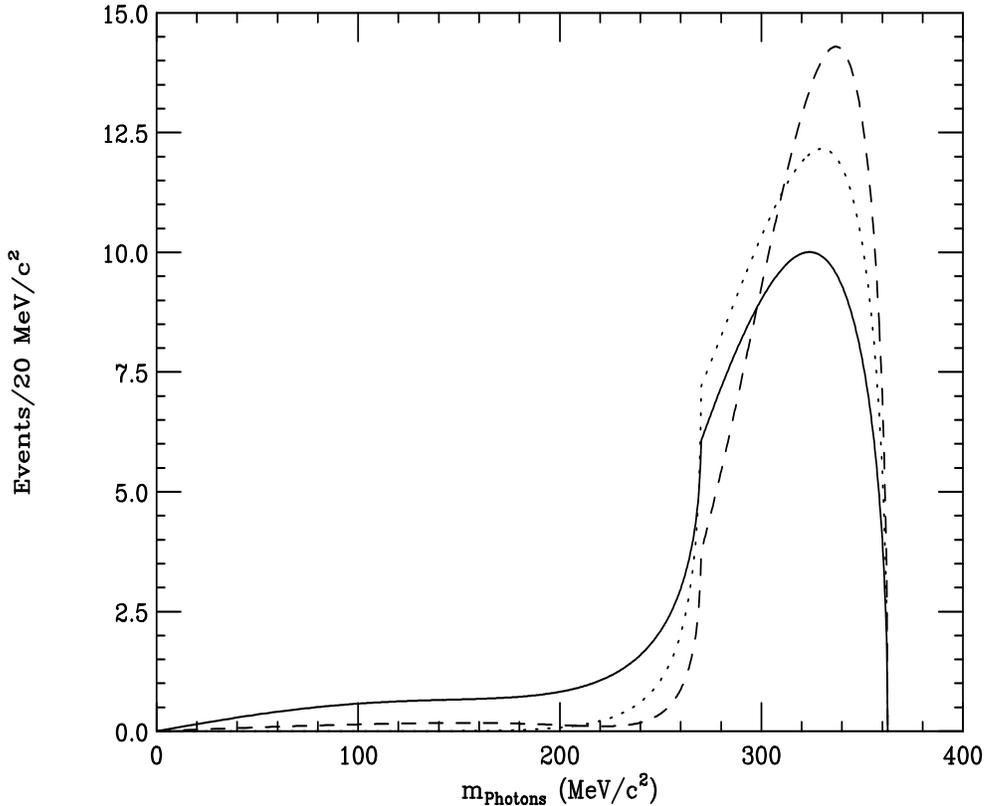}}
\caption{$\gamma\gamma$-invariant distribution for $K_L \ri \pi^0
\gamma\gamma$: ${\cal O}(E^4)$ (dotted curve),
${\cal O}(E^6)$ with $a_V$ = 0 (dashed curve), ${\cal
O}(E^6)$ with $a_V$ = $-0.96$ (full curve). The spectrum
is normalized to the 50 unambiguously events of NA31 \protect\cite{NA}
(without acceptance corrections).}
\end{figure}

\begin{figure}[t]
\centering
\leavevmode
\epsfxsize=300pt
\epsfysize=300pt
\centerline{
\epsfbox{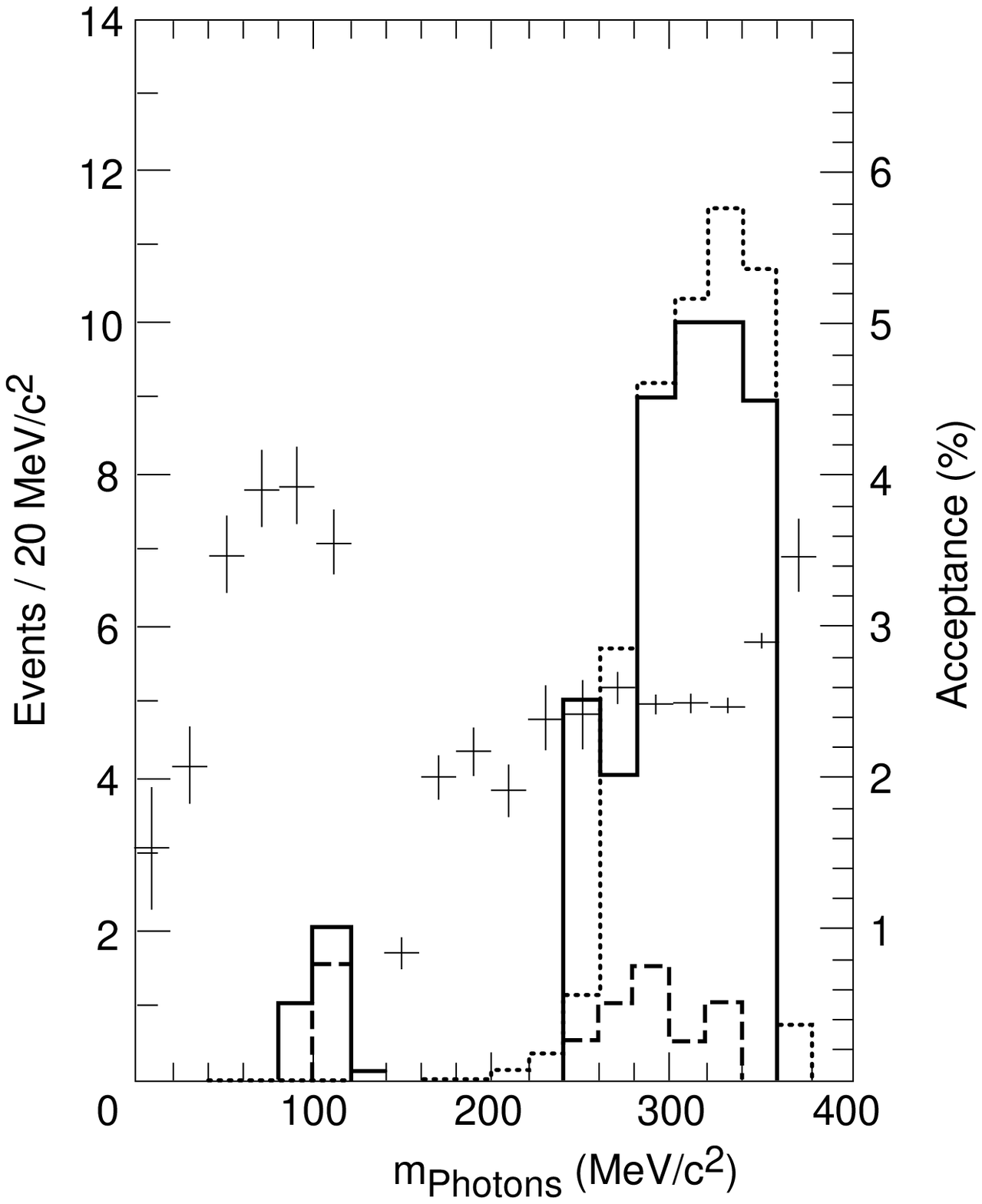}}
\caption{$\gamma\gamma$-invariant mass distributions of the 50 unambiguously
$K_L \rightarrow \pi^0 \gamma\gamma$ events reconstructed
by NA31 \protect\cite{NA} (histograms). The dashed line shows the
estimated background. The dotted line simulates the ${\cal O}(E^4)$
ChPTh prediction. Crosses indicate the experimental acceptance (scale
on the right).}
\end{figure}

Several authors \cite{EPR3,CEP,PH,HS,CDM} have calculated the contribution of
the on-shell two-photon
intermediate state to $K_L \rightarrow \pi^0 e^+ e^-$. Although this is
sometimes referred to as the absorption contribution, it is not the full
absorption part since there is a further cut due to on-shell pions. Besides
this, the full CP-conserving amplitude
also receives contribution form the dispersive part of the amplitude,
with
off-shell photons (and pions). The calculation of this is complicated by the
sensitivity of the loop integral to high momentum, as the Feynman diagram
of Fig. 14 will diverge if we treat the $B$ amplitude as a constant in $q^2_1$
and $q^2_2$. However, the remedy to this is well known; the couplings of
virtual photons to hadrons is governed by form factors which lead to
suppression of the couplings at high $q^2$. We will include an estimate of
these form factors and this will allow us to calculate the dispersive
component of the CP-conserving amplitude.
\newpage

The two-photon loop integral in the limit $m_e \rightarrow 0$ is given by

\begin{eqnarray}
{\cal M} (K_L \rightarrow \pi^0 e^+ e^-)_{\rm CPC}
&=& {G_8 \alpha \over 4 \pi
m^2_K} \int {d^4 \ell \over (2 \pi )^4} {B(k + k^{\prime}) F(\ell + k)
F(\ell - k^{\prime})
\over \ell^2 ( \ell + k) ^2 ( \ell - k^{\prime})^2} \nonumber \\
&\times& \left\{ \rlap/{p^{\phantom{l}}_K} \left[ \ell^2 +
p^{\phantom{l}}_K \cdot (k - k^{\prime})
- p^{\phantom{l}}_K \cdot \ell \; \ell \cdot (k - k^{\prime}) \right.
\right.\nonumber \\
&+& \left. \left. 2p^{\phantom{l}}_K \cdot \ell \; k \cdot k^{\prime}
- p^{\phantom{l}}_K \cdot k \; \ell \cdot k^{\prime} -
p^{\phantom{l}}_K \cdot k^{\prime} \; \ell \cdot k \right] \right.
\nonumber \\
&+& \left. \rlap/{\ell} \left[ ( p^{\phantom{l}}_K \cdot \ell )^2 +
p^{\phantom{l}}_K
\cdot \ell \; p^{\phantom{l}}_K \cdot (k - k^{\prime}) \right]
\right\}. \nonumber \\
\end{eqnarray}

\noindent Here $F(q^2)$ are the form factors for the virtual photon
couplings. The structure above is certainly an approximation, as in
general the virtual photon dependence need not be only an overall
factor of $F(q^2)$. However, the above form would be sufficient to
capture the kinematic variation if only one photon is off-shell [given
an appropriate $F(q^2)$]. Since we only need a minor form factor
suppression to tame the logarithmic divergences, we feel that this
structure will be sufficient for our estimate. We choose

\begin{equation}
F(q^2) = {-m^2_V \over {q^2 - m^2_V}},
\end{equation}

\noindent which is a good representation of almost any mesonic form factor,
with $m^{\phantom{\dagger}}_V \approx m_\rho$. Neglecting terms that
are suppressed by powers of $1/m^2_\rho$, we find the amplitude of Eq.
\ref{50} above, with

\begin{equation}
K = {B(x) \over {16 \pi^2 m^2_K}} \left [
{2 \over 3} \log \left({m^2_{\rho}} \over {-s}\right)
- {1 \over 4} \log \left({-s} \over {m^2_e}\right)
+ {7 \over {18}}
\right ],
\end{equation}

\noindent where $s = \left( k + k^{\prime}\right)^2$.
The log factor is of course expected, since the photon ``absorptive"
part comes from the expansion $\log (-s) = \log s + i\pi$.

This representation of the amplitude leads to a CP-conserving branching
ratio of

\begin{equation}
BR(K_L \rightarrow \pi^0 e^+ e^-)_{\rm CPC} = 4.89 \times 10^{-12}
\end{equation}

\noindent for $a^{\phantom{l}}_V = -0.96$. More generally we show the
CP-conserving branching ratio vs. $a^{\phantom{l}}_V$ in Fig. 17. Note
that while for many values of $a^{\phantom{l}}_V$ the CP-conserving
rate is small compared to the CP-violating rate of previous chapters,
these two rates are comparable for some range of parameters.
Note that there is no interference in the rate between the
CP-conserving and violating components, so that the rates just add, as
shown in Fig. 18.  However, there can be a CP-odd asymmetry in the
electron positron energies, to which we turn in the next chapter.

\begin{figure}[t]
\centering
\leavevmode
\epsfxsize=300pt
\epsfysize=300pt
%\rotate[l]
{\centerline{\epsfbox{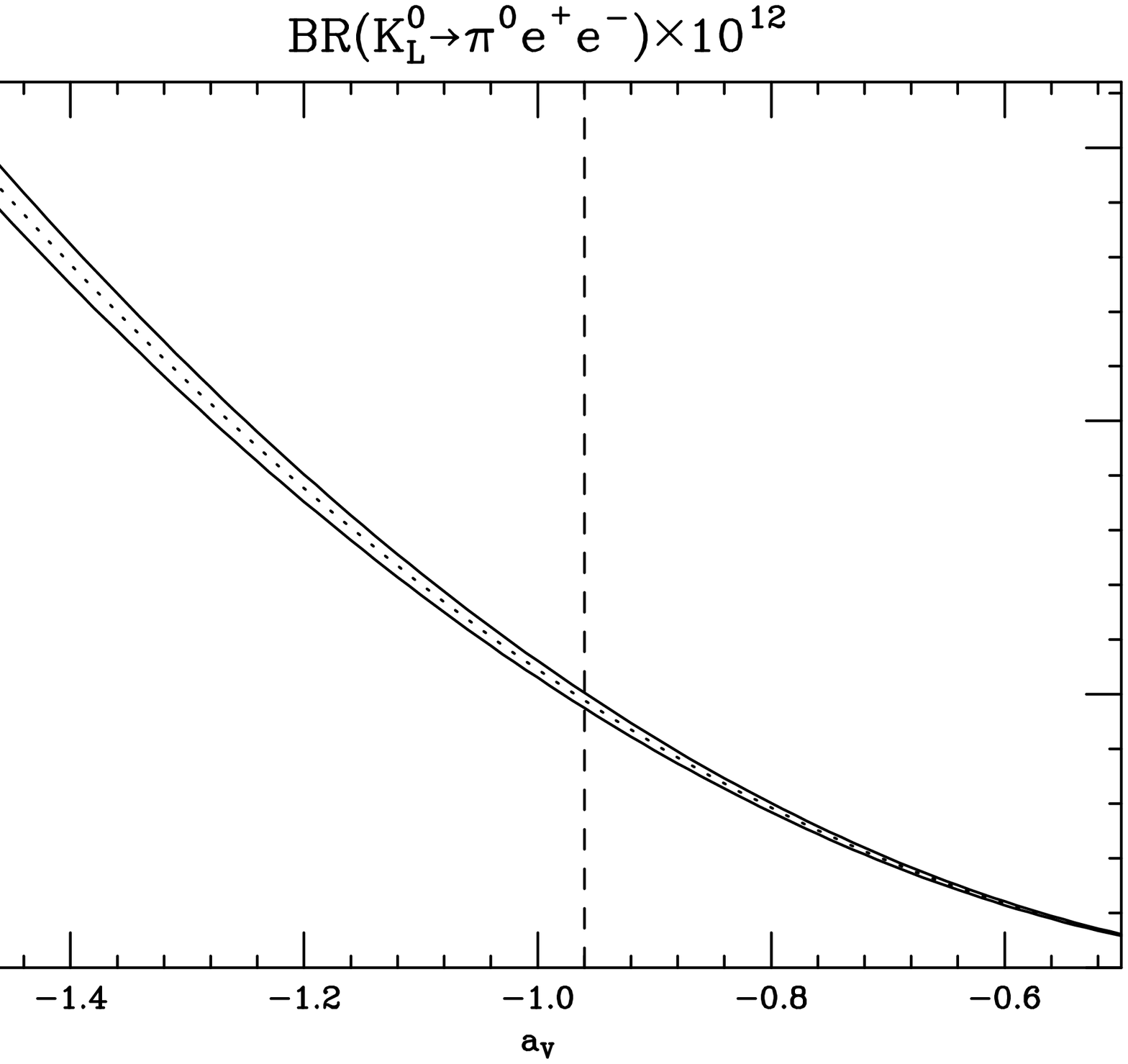}}}
\caption{The CP-conserving branching ratio
$BR(K_L \ri \pi^0 e^+ e^-)_{\rm CPC}$ is plotted against $a_V$. The
convention for the solid and dashed curves is the same as for Fig. 9.
The assumed value for $a_V$ is indicated by the dashed vertical line.}
\end{figure}

\begin{figure}[t]
\centering
\leavevmode
\epsfxsize=300pt
\epsfysize=300pt
%\rotate[l]
{\centerline{\epsfbox{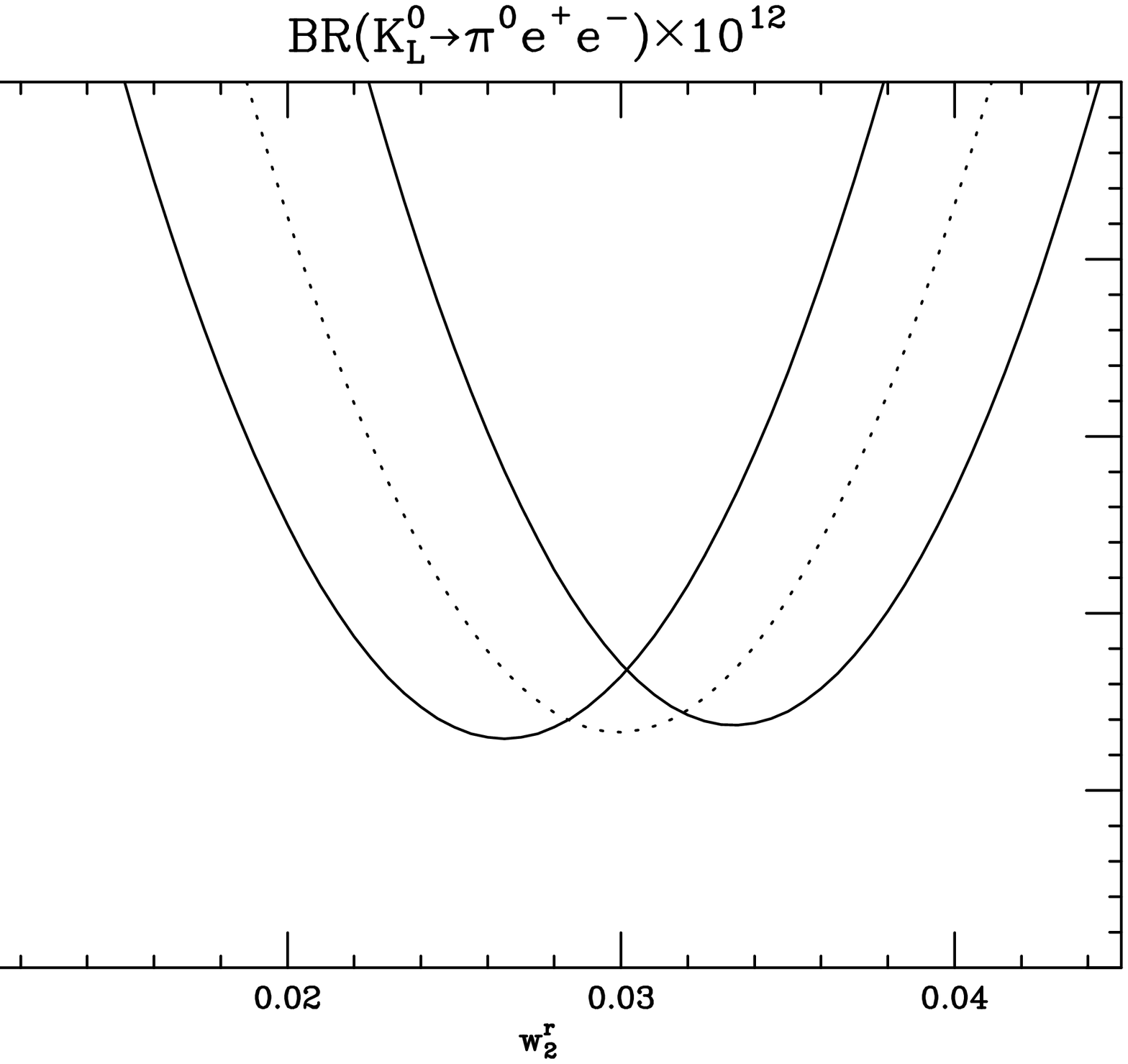}}}
\caption{The complete branching ratio
$BR(K_L \ri \pi^0 e^+ e^-)$ is plotted against $w_2$. The
convention for the solid and dashed curves is the same as for Fig. 9.
$a_V = -0.96$ is assumed.}
\end{figure}

\chapter{THE ELECTRON ENERGY ASYMMETRY AND TIME-DEPEN\-DENT INTERFERENCE}

\section{The Electron Energy Asymmetry}

When both CP-violating and CP-conserving amplitudes contribute to the
decay, there will be an asymmetry in the electron-positron energy
distribution

\begin{equation}
A = {N(E_+ > E_-) - N(E_+ < E_-) \over N(E_+ > E_-) + N(E_+ < E_-)}.
\end{equation}

\noindent This will be quite large if the two amplitudes are comparable. In
contrast to the overall decay rate, this asymmetry is an unambiguous signal
of CP violation. It may be even more useful if it can be used to prove the
existence of direct CP violation. This can occur because the asymmetry is
sensitive to the phase of the CP-violating amplitude, and mass matrix CP
violation has a unique phase (that of $\epsilon$), while direct CP violation
will in general have a different phase.

For the electron energy asymmetry to be useful as a diagnostic of the
form of CP violation, the CP-conserving two-photon amplitudes must be
well known. As we discussed in the previous chapter, it is reasonable
to expect that this will be true in the future after further
phenomenology of $K_L \rightarrow \pi^0 \gamma \gamma$. For
illustrative purposes, we will use $a^{\phantom{l}}_V = -0.96$. The
analysis below will need to be redone in the future if this value of
$a^{\phantom{l}}_V$ changes significantly. However, the pattern of the
analysis and the general conclusions will be valid for a wide range of
values of $a^{\phantom{l}}_V$.

The amplitudes involved have been given in the previous chapter. Note that
the two-photon amplitude has both real and imaginary parts, the mass matrix
CP-violating amplitude has a phase of $45^o$, while the direct component
is purely imaginary. The asymmetry is proportional to the imaginary part of
$B$ times the real part of $d_V$ minus the product of the real part of
$B$ and imaginary part of $d_V$. The asymmetry may be defined in
differential form:

\begin{eqnarray}
{d \Gamma \over dz} (E_+ > E_-) & = & \int_0^{{1 \over 2} \lambda^{{1
\over 2}} (1, z, r^2)} dy {d \Gamma \over dy dz}, \nonumber \\
{d \Gamma \over dz} (E_+ < E_-) & = & \int^0_{-{1 \over 2}
\lambda^{{1 \over 2}} (1, z, r^2)} dy {d \Gamma \over dy dz}, \nonumber
\\
\lambda(1, z, r^2 ) & = & 1 + z^2 + r^4 - 2z - 2r^2 - 2r^2z,
\nonumber \\
r & = & {m_{\pi} \over m_K}, \nonumber \\
A(z) & \equiv & {{d \Gamma (E_+ > E_-)/dz - d \Gamma
 (E_+ < E_-)/dz} \over {d \Gamma (E_+ > E_-) /dz + d \Gamma(E_+ <
E_-)/dz}}, \nonumber \\
z & \equiv & (k + k^{\prime})^2/m^2_K.
\end{eqnarray}

In Fig. 19 we plot the differential asymmetry for several values of
$w_2$ in the case when there is no direct CP violation. Fig. 20 gives the
same information for ${\rm Im} \lambda_t = 10^{-4}$. We see that the
asymmetry is sizable for many values of $w^{\phantom{2}}_2$ and that it depends
significantly on direct CP violation, even possibly changing sign. The
integrated asymmetry is plotted versus $w_2$ in Figs. 21, 22.

\begin{figure}[t]
\centering
\leavevmode
\epsfxsize=300pt
\epsfysize=300pt
%\rotate[l]
{\centerline{\epsfbox{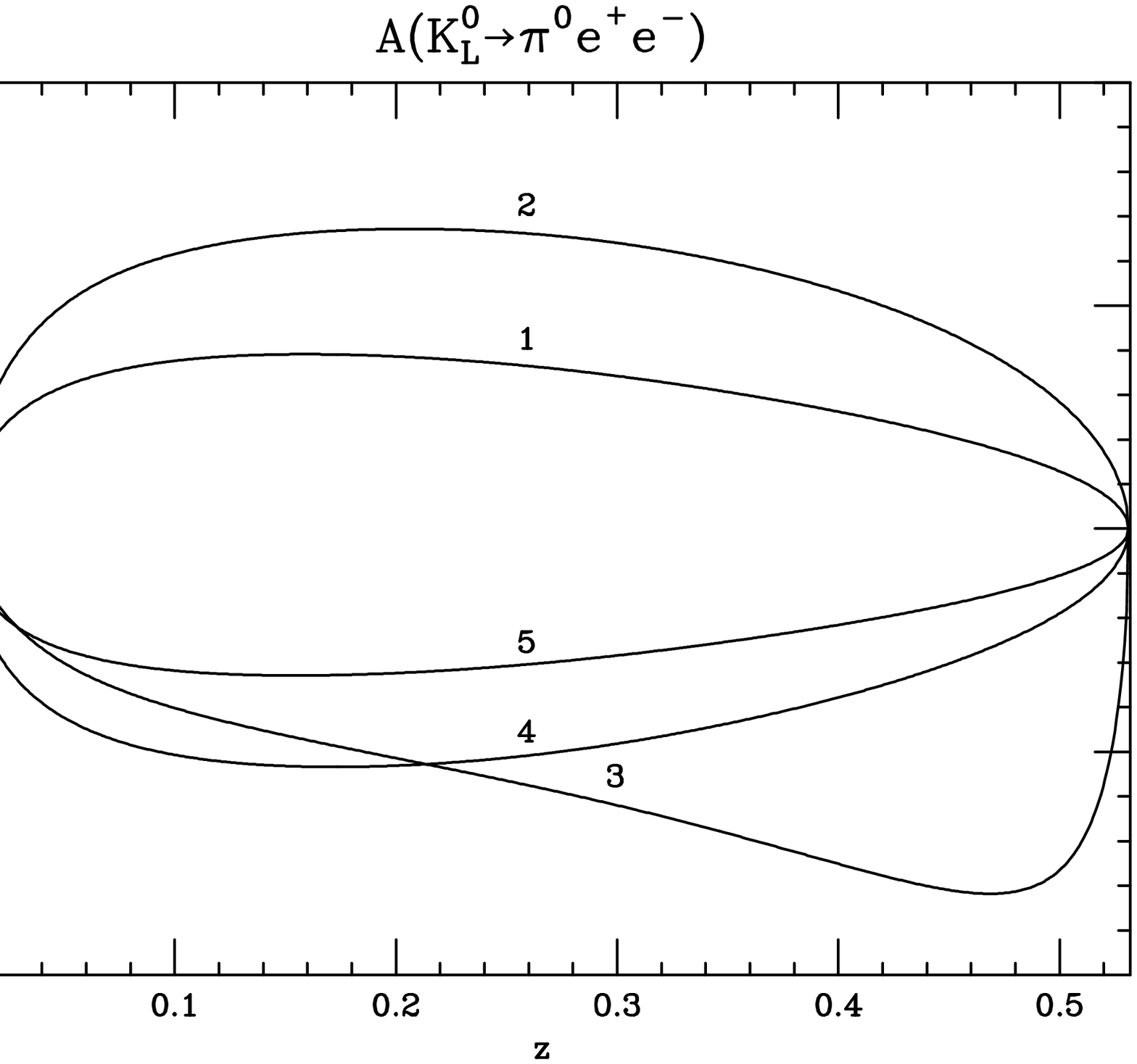}}}
\caption{The differential asymmetry $A(K_L \ri \pi^0 e^+ e^-)$ is plotted
against $z \equiv (k + k^{\prime})^2/m^2_K$ for $w_2$ = 1 $\times
10^{-2}$ (curve 1), 2 $\times 10^{-2}$ (curve 2), $\dots$,
5 $\times 10^{-2}$ (curve 5), in the case when there is no direct
CP violation. $a_V = -0.96$ is assumed.}
\end{figure}

\begin{figure}[t]
\centering
\leavevmode
\epsfxsize=300pt
\epsfysize=300pt
%\rotate[l]
{\centerline{\epsfbox{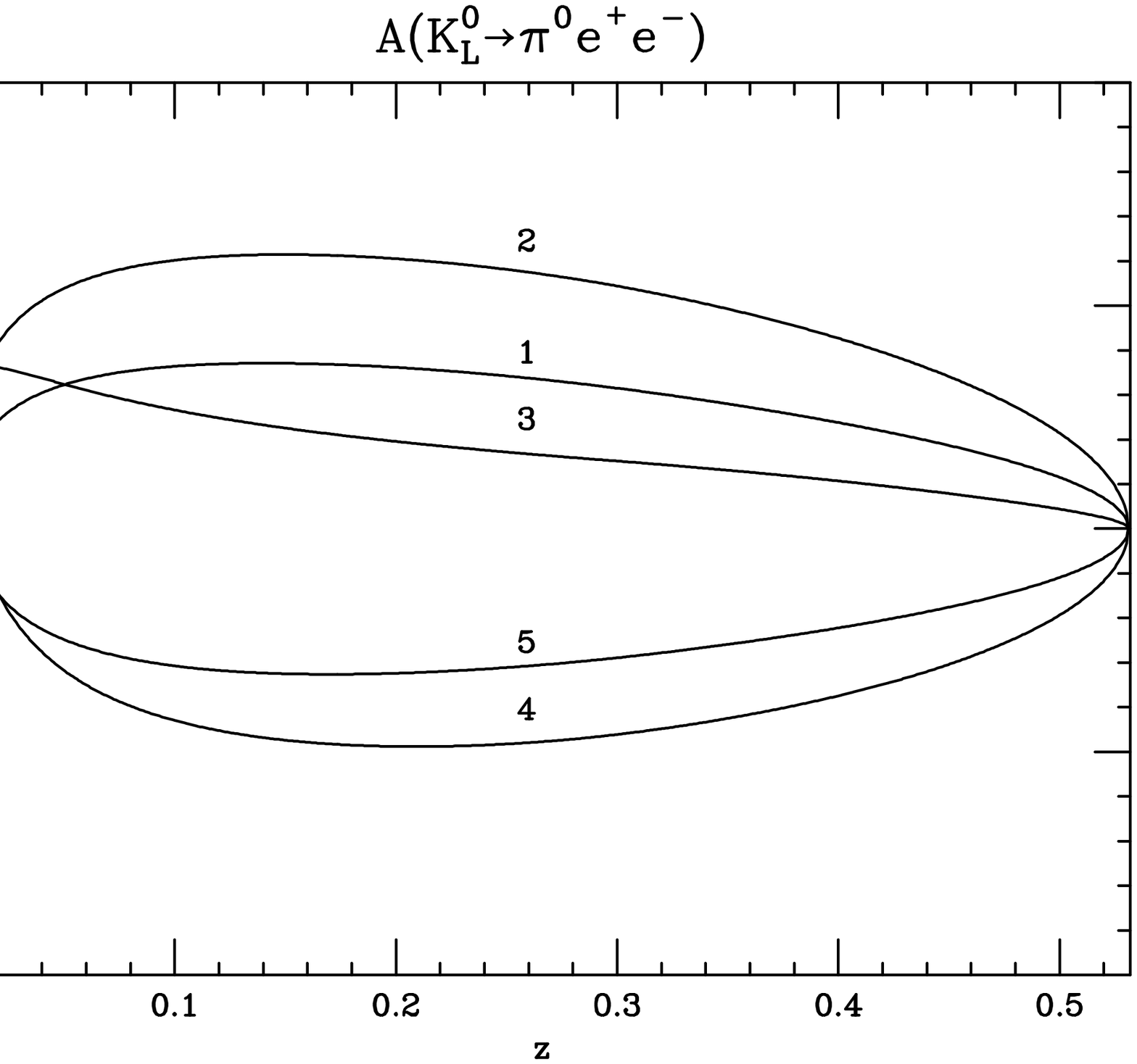}}}
\caption{Same as Fig. 19 for ${\rm Im}\lambda_t$ = $10^{-4}$.}
\end{figure}

\begin{figure}[t]
\centering
\leavevmode
\epsfxsize=300pt
\epsfysize=300pt
%\rotate[l]
{\centerline{\epsfbox{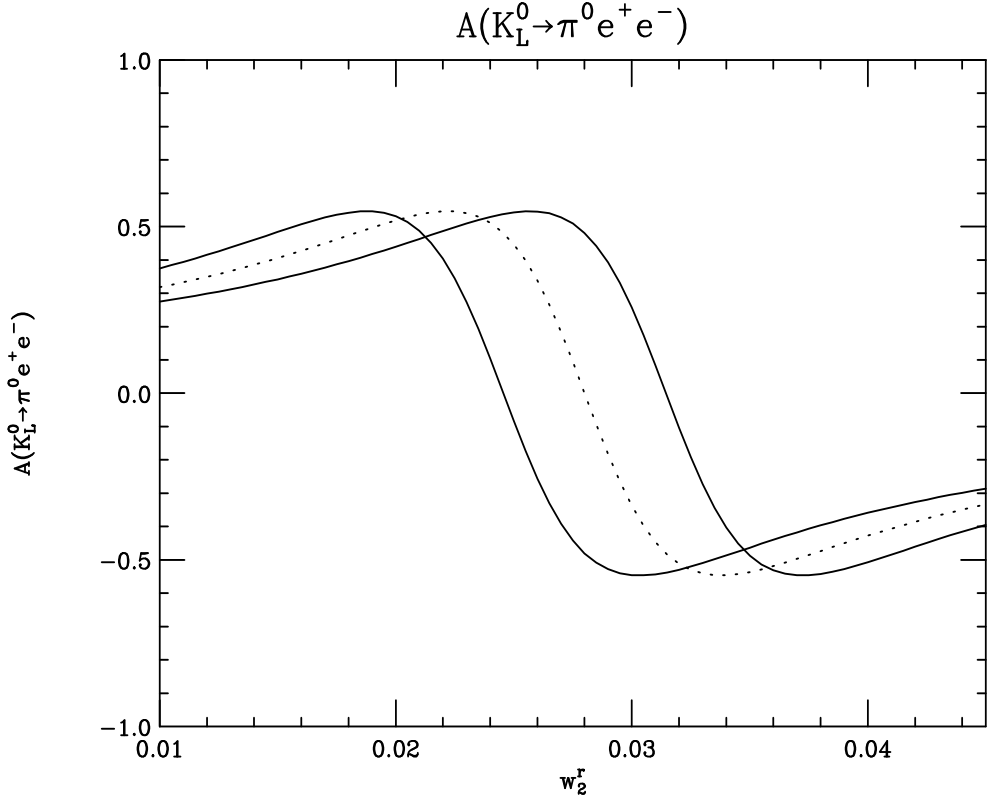}}}
\caption{The integrated asymmetry is plotted vs. $w_2$ in
the case when there is no direct CP violation. The convention
for the solid and dashed curves is the same as for Fig. 9. $a_V =
-0.96$ is assumed.}
\end{figure}

\begin{figure}[t]
\centering
\leavevmode
\epsfxsize=300pt
\epsfysize=300pt
%\rotate[l]
{\centerline{\epsfbox{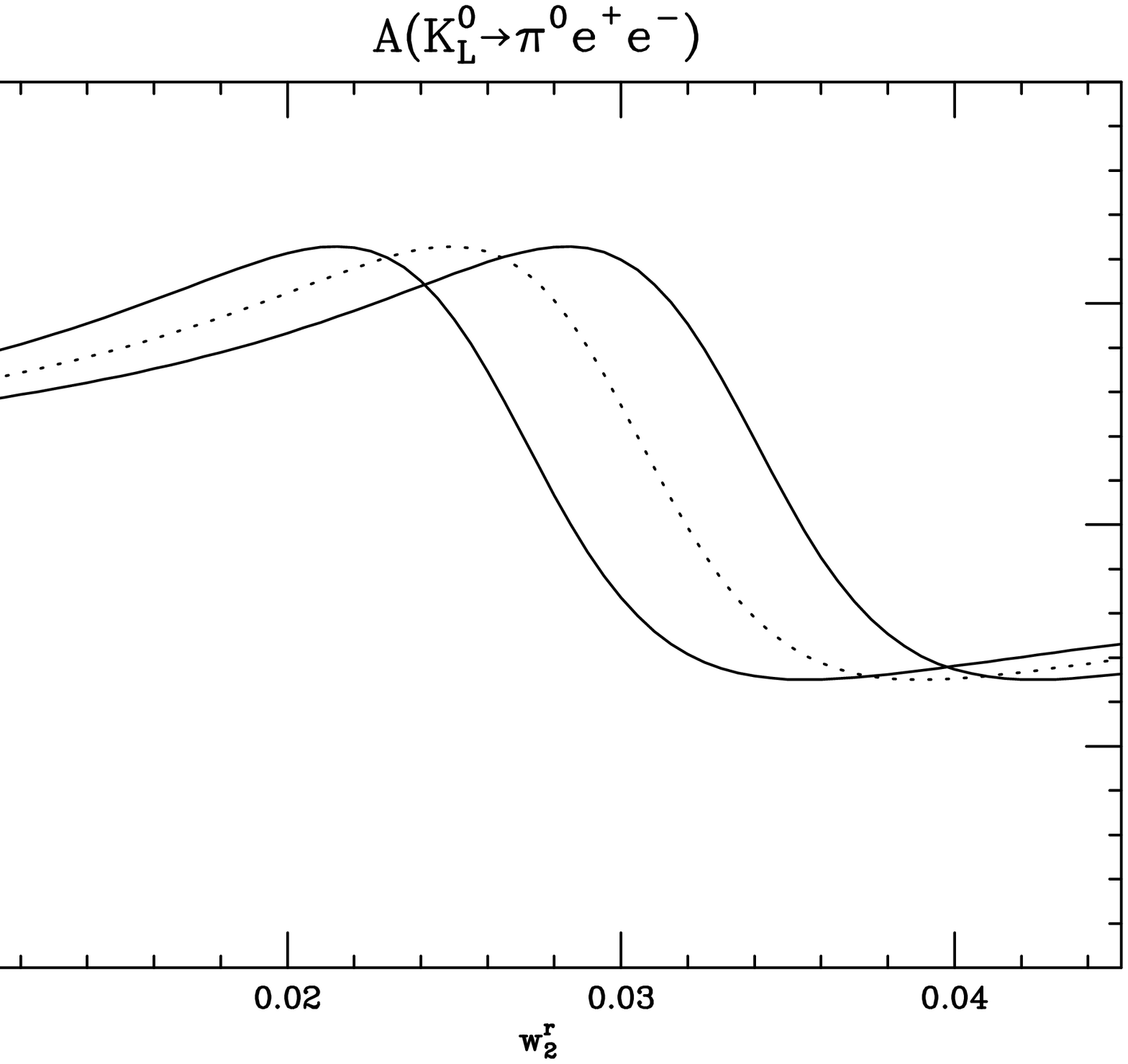}}}
\caption{Same as Fig. 21 for ${\rm Im}\lambda_t$ = $10^{-4}$.}
\end{figure}

Both the decay rate and the asymmetry depend on $w^{\phantom{2}}_S$
and this forms the main uncertainty in the analysis. However, it is
possible to remove this uncertainty by measuring both observables.
For a given value of the CP-conserving amplitude, there is a strict
correlation between the two. In Fig. 23 we plot the values of the
branching ratio and the integrated asymmetry $A$ as one varies $w_2$.
We see that the curves with and without direct CP violation are well
separated for much of the range. To the extent that we understand the
CP-conserving amplitude, this can be used as a diagnostic test for
direct CP violation.

\begin{figure}[t]
\centering
\leavevmode
\epsfxsize=300pt
\epsfysize=300pt
%\rotate[l]
{\centerline{\epsfbox{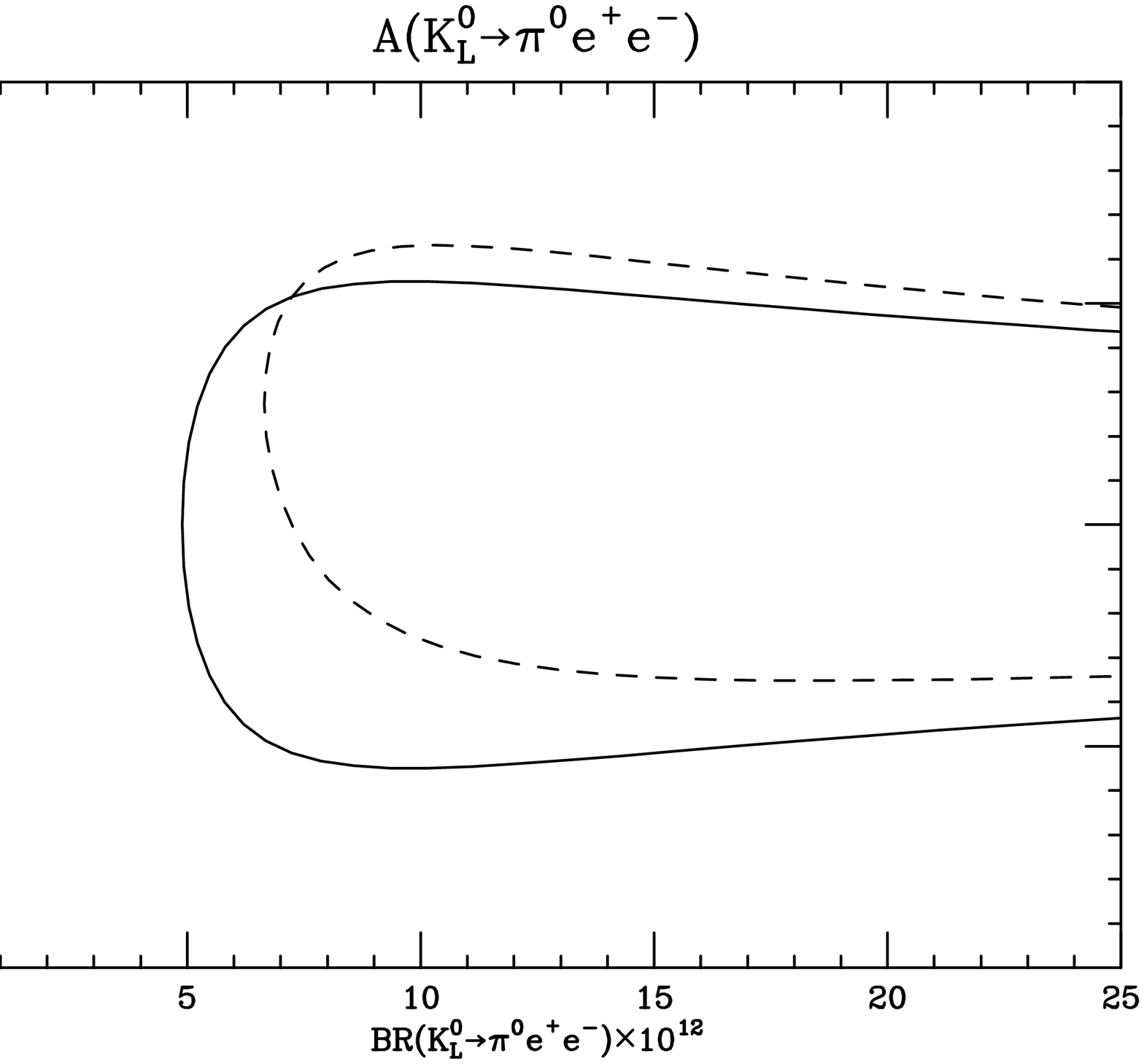}}}
\caption{The integrated asymmetry $A(K_L \ri \pi^0 e^+ e^-)$ is
plotted vs. the
branching ratio $BR(K_L \ri \pi^0 e^+ e^-)$ as one varies $w_2$.
$a_V = -0.96$ is assumed. The solid curve describes the case when there
is no direct CP violation, the dashed curve is for ${\rm Im}\lambda_t$
= $10^{-4}$.}
\end{figure}

\section{Time-Dependent Interference of $K_1$, $K_2$
$\rightarrow \pi^0 e^+ e^-$}

Littenberg \cite{L} has suggested a time-dependent analysis of a state
starting out as a $K^0$ in order to extract maximal information about
the decay mechanism. This is far more demanding experimentally then
simply measuring $K_L$ or $K_S$ decays separately. However, we analyze
this technique in order to assess its usefulness.

A state that at $t=0$ is a $K^0$ will evolve into a mixture of $K_L$
and $K_S$:

\begin{equation}
\vert K^0(t) \rangle = {1 \over {(1 + \epsilon)
\sqrt{2}}}\left\{e^{-iH^{\phantom{l}}_S t} \left[ \vert K_1 \rangle +
\epsilon \vert
K_2 \rangle \right] + e^{-iH^{\phantom{l}}_L t}\left[\vert K_2 \rangle
+ \epsilon \vert K_1 \rangle \right] \right\},
\end{equation}

\noindent where

\begin{equation}
H^{\phantom{l}}_j = m^{\phantom{l}}_j - i{\Gamma_j \over 2}, \qquad
j={\rm S,L}.
\end{equation}

Ignoring small effects such as second order CP violation and
direct CP-violating components in the parameter $\epsilon$, we then
have a time development proportional to

\begin{eqnarray}
\vert \langle \pi^0 e^+ e^- \left \vert {\cal H} \right \vert K^0(t) \rangle
\vert^2 & \approx & {1 \over 2} \left\{ \left \vert A^{\phantom{l}}_S
\right \vert^2 e^{-\Gamma^{\phantom{l}}_S t}+
\left \vert \epsilon A_S + A_{\rm dir} + A_{\rm CPC} \right \vert^2
e^{-\Gamma^{\phantom{l}}_L t} \phantom{e^{-{(\Gamma^{\phantom{l}}_L +
\Gamma^{\phantom{l}}_S) t}\over 2}} \right .\nonumber \\
& + & \left. 2 {\rm Re} \left [ \left (\epsilon A^{\phantom{l}}_S +
A_{\rm dir} + A_{\rm
CPC} \right ) A^*_S e^{-i (m^{\phantom{l}}_L -m^{\phantom{l}}_S) t} \right]
e^{-{{(\Gamma^{\phantom{l}}_L +
\Gamma^{\phantom{l}}_S) t} \over 2}} \right \}. \nonumber \\
\phantom{l}
\end{eqnarray}

Here $A_{\rm dir}$, $\epsilon A_S$ and $A_{\rm CPC}$ are the
amplitudes analyzed in chapters 3, 4 and 6 respectively.
Measurements at
early time $(t \ll 1/\Gamma_S)$ determine $\Gamma(K_S \ri \pi^0 e^+
e^-)$ while at the late time $(t \gg 1/\Gamma_S)$ one observes
$\Gamma(K_L \ri \pi^0 e^+ e^-)$. These contain the information
described in preceding chapters. However, in the interference region
$t$ = ${\cal O} \left( {1/(m^{\phantom{l}}_L - m^{\phantom{l}}_S)}
\right)$ $\sim$ ${\cal O} (1/\tau^{\phantom{\dagger}}_S)$, we obtain
extra information about the separate contributions to the decay
amplitude.

In Fig. 24 we show the time-dependent signal for the cases of pure
mass matrix CP violation and the addition of direct CP violation,
using the analysis of the previous chapters with $w_2 = 4 L_9$. We see
that the shape of the interference region does differentiate these two
cases, but that for the case studied the dependence on direct CP
violation is not so large as to allow an easy experimental
determination.

\begin{figure}[t]
\centering
\leavevmode
\epsfxsize=300pt
\epsfysize=300pt
%\rotate[l]
{\centerline{\epsfbox{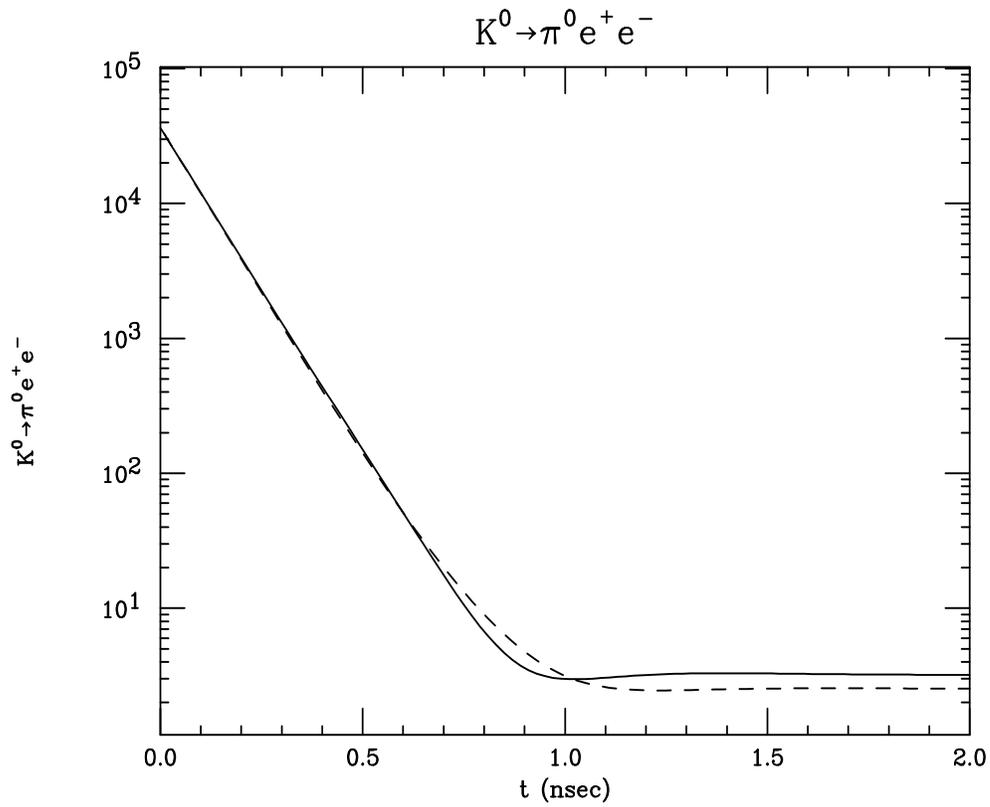}}}
\caption{The normalized time distribution of $K^0 \ri \pi^0 e^+ e^-$
decays is plotted vs. time $t$ in nsecs. $a_V = -0.96$ is assumed.
The solid curve describes the case when there is no direct CP
violation, the dashed curve is for ${\rm Im}\lambda_t$ = $10^{-4}$.}
\end{figure}

\chapter{THE $\cal{O}$(E$^4$) CALCULATION}

First let us provide the straightforward $\cal{O}$(E$^4$) calculation
of ${\cal M}(K_L \rightarrow \pi^0 \gamma e^+e^-)$ within ChPTh. This
is the generalization to $k^2_1 \neq 0$ of the original chiral
calculation of EPR \cite{EPR1,EPR2}. Here $k_1$ is the momentum of the
off-shell photon. This captures all the $k^2_1/m^2_{\pi}$ and
$k^2_1/m^2_K$ variations of the amplitudes at this order in the energy
expansion. There can be further $k^2_1$/(1 GeV)$^2$ corrections which
correspond to $\cal{O}$(E$^6$) and higher. The easiest technique for
this calculation uses the basis where the kaon and pion fields are
transformed so that the propagators have no off-diagonal terms, as
described in Ref. \cite{EPR1,EPR2}. The relevant diagrams are then
shown in Fig. 25. Defining $\ov g$ as

\begin{equation}
{\ov g} = G^{\phantom{l}}_8/3, \qquad\qquad
G^{\phantom{l}}_8 = G^{\phantom{l}}_F \vert \Vud \Vuss \vert
g^{\phantom{l}}_8, \qquad\qquad \vert g^{\phantom{l}}_8 \vert \approx 5.1,
\end{equation}

\noindent the diagrams give the following integrals, respectively:

\begin{equation}
{\cal M}^a_{\mu\nu} = 2 e^2 {\ov g} g_{\mu \nu}
\int{{d^4 l}\over {(2\pi)^4}}
{{3 [(p^{\phantom{l}}_K-p^{\phantom{l}}_0)^2-m^2_{\pi}] - 2
[(l^2-m^2_{\pi})+(l-k_1-k_2)^2-m^2_{\pi}]} \over
{(l^2-m^2_{\pi})[(l-k_1-k_2)^2-m^2_{\pi}]}},
\end{equation}

\begin{eqnarray}
{\cal M}^b_{\mu\nu} &=& -e^2 {\ov g} \int{{d^4 l}\over {(2\pi)^4}}
{{3 [(p^{\phantom{l}}_K-p^{\phantom{l}}_0)^2-m^2_{\pi}] - 2 [(l+k_1)^2 -m^2_{\pi} + (l-k_2)^2
-m^2_{\pi}]} \over
{(l^2-m^2_{\pi})[(l+k_1)^2 - m^2_{\pi}][(l-k_2)^2 - m^2_{\pi}]}}
\nonumber \\
&\times& (2 l + k_1)_{\mu} (2 l -k_2)_{\nu}
+ (k_1, \mu) \leftrightarrow (k_2, \nu),
\end{eqnarray}

\begin{equation}
{\cal M}^c_{\mu\nu} = 8 e^2 {\ov g} g_{\mu \nu} \int{{d^4 l}\over {(2\pi)^4}}
{1 \over {l^2-m^2_{\pi}}},
\end{equation}

\begin{equation}
{\cal M}^d_{\mu\nu} = -4 e^2 {\ov g} \int{{d^4 l}\over {(2\pi)^4}} \left\{
{{(2 l -k_1)_{\mu} (2 l -k_1)_{\nu}} \over {(l^2-m^2_{\pi})[(l-k_1)^2 -
m^2_{\pi}]}} + {{(2 l -k_2)_{\mu} (2 l -k_2)_{\nu}} \over
{(l^2-m^2_{\pi}) [(l-k_2)^2 - m^2_{\pi}]}}\right\}.
\end{equation}

\begin{figure}[t]
\centering
\leavevmode
\centerline{
\epsfbox{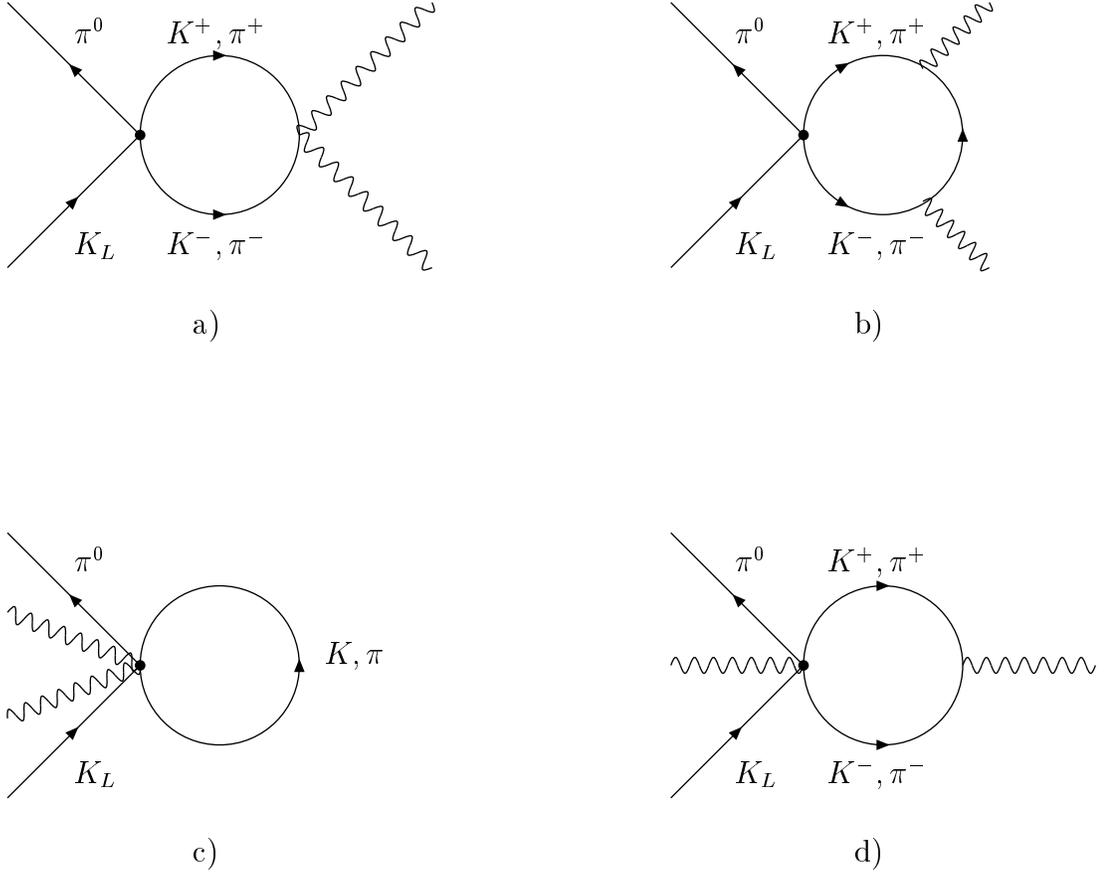}}
\caption{Diagrams relevant to the process $K_L \rightarrow
\pi^0 \gamma e^+ e^-$ at ${\cal O}(E^4)$ and ${\cal O}(E^6)$.}
\end{figure}

\noindent Interestingly when we add these together the $K \ri 3\pi$ amplitude
factors out from the remaining loop integral resulting in

\begin{equation}
{\cal M}^{\pi}_{\mu\nu} = 6 e^2 {\ov g} [(p^{\phantom{l}}_K-
p^{\phantom{l}}_0)^2-m^2_{\pi}]
\int{{d^4 l}\over {(2\pi^4)}} {{[g_{\mu\nu}(l^2-m^2_{\pi})-
(2 l +k_1)_{\mu} (2 l -k_2)_{\nu}]} \over {{(l^2-m^2_{\pi})
[(l+k_1)^2 - m^2_{\pi}] [(l-k_2)^2 - m^2_{\pi}]}}}.
\end{equation}

\noindent It is not hard to verify that this result satisfies the
constraints of gauge invariance $k^{\mu}_1 \cal{M}_{\mu\nu}$ =
$k^{\nu}_2 \cal{M}_{\mu\nu}$ = 0. At this stage, the integral may be
parametrized and integrated using standard Feynman-diagram techniques.
Let us keep photon number one as the off-shell photon and set $k^2_2
= 0$. In this case the amplitude with one photon off-shell is described
by

\begin{equation}
{\cal M}^{\pi}_{\mu\nu} = 6 e^2 {\ov g}
[(p^{\phantom{l}}_K-p^{\phantom{l}}_0)^2-m^2_{\pi}]
\left({-i} \over {16 \pi^2}\right){{(g_{\mu\nu} k_1 \cdot k_2-k_{2\mu}
k_{1\nu})} \over {k_1 \cdot k_2}} [1 + 2 I(m^2_{\pi})],
\end{equation}

\noindent with

\begin{eqnarray}
\label{8}
I(m^2_{\pi}) &=& \int^1_0 dz_1 \int_0^{1-z_1} dz_2
{{m^2_{\pi}-z_1(1-z_1)k^2_1} \over {2 z_1 z_2 k_1 \cdot k_2
+z_1(1-z_1)k^2_1 -m^2_{\pi} +i\epsilon}} \nonumber \\
&=& {m^2_{\pi} \over {s -k^2_1}} [F(s) - F(k^2_1)] - {k^2_1 \over {s
-k^2_1}} [G(s) - G(k^2_1)].
\end{eqnarray}

\noindent The notation is defined by

\begin{equation}
s = (p^{\phantom{l}}_K-p^{\phantom{l}}_0)^2=(k_1 + k_2)^2
\end{equation}

\noindent and

\begin{equation}
F(a) = \int^1_0 {dz_1 \over z_1} \log\left[{{m^2_{\pi} - a(1-z_1)z_1 -
i \epsilon} \over {m^2_{\pi}}}\right],
\end{equation}

\begin{equation}
G(a) = \int^1_0 dz_1 \log\left[{{m^2_{\pi} - a(1-z_1)z_1 -
i \epsilon} \over {m^2_{\pi}}}\right].
\end{equation}

\noindent The above functions are related to those presented by CEP
\cite{CEP}:

\begin{equation}
F(a) = {a \over {2 m^2_{\pi}}} \left[F_{\rm CEP}\left({a \over
4 m^2_{\pi}}\right)-1\right],
\end{equation}

\begin{equation}
G(a) = -{a \over {2 m^2_{\pi}}} \left[R_{\rm CEP}\left({a \over
4 m^2_{\pi}}\right)+{1 \over 6}\right],
\end{equation}

\noindent remembering Eq. \ref{52}:

\begin{eqnarray}
F_{\rm CEP}(x) & = & 1 - {1 \over x} \left[ \arcsin \left(
\sqrt{x} \right) \right]^2, \, \qquad x \leq 1, \nonumber \\
& = & 1 + {1 \over 4x} \left( \log {1 - \sqrt{1 - 1/x} \over 1 + \sqrt{1 -
1/x}} + i \pi \right)^2, \, \qquad x \geq 1, \nonumber \\
R_{\rm CEP}(x) & = & - {1 \over 6} + {1 \over 2x} \left[ 1 - \sqrt{1/x
- 1} \arcsin \left( \sqrt{x} \right) \right], \,
\qquad x \leq 1, \nonumber \\
& & - {1 \over 6} + {1 \over 2x} \left[ 1 + \sqrt{1 - 1/x} \left( \log {1 -
\sqrt{1 - 1/x} \over 1 + \sqrt{1 - 1/x}} + i \pi \right) \right], \,
\qquad x \geq 1. \nonumber \\
\phantom{l}
\end{eqnarray}

\noindent This agrees with the EPR result in the $k^2_1 \ri 0$ limit.

At this order we have also calculated the additional contribution
resulting from the kaons circulating in the loops of Fig. 25. They give
rise to

\begin{equation}
{\cal M}^K_{\mu\nu} = 6 e^2 {\ov g} (m^2_K+m^2_{\pi}-s)
\int{{d^4 l}\over {(2\pi^4)}} {{[g_{\mu\nu}(l^2-m^2_K)-
(2 l +k_1)_{\mu} (2 l-k_2)_{\nu}]} \over {{(l^2-m^2_K)
[(l+k_1)^2 - m^2_K] [(l-k_2)^2 - m^2_K]}}}.
\end{equation}

\noindent The resulting integral is similar to that of Eq. \ref{8},
substituting the mass of the pion with that of the kaon. Attaching an
$e^+e^-$ couple to either photon and adding all
the above contributions together, the result we obtain for the
branching ratio is

\begin{equation}
{\rm BR}(K_L \rightarrow \pi^0 \gamma e^+e^-) = 1.0 \times 10^{-8}.
\end{equation}

\noindent With the definitions

\begin{equation}
z = {s \over {m^2_K}}, \qquad
y = {{p^{\phantom{l}}_K \cdot (k_1 - k_2)} \over {m^2_K}},
\end{equation}

\noindent the decay distributions in $z$ and $y$ provide more detailed
information. We present them in Figs. 26 and 27.

\begin{figure}[t]
\centering
\leavevmode
\epsfxsize=300pt
\epsfysize=300pt
%\rotate[l]
{\centerline{\epsfbox{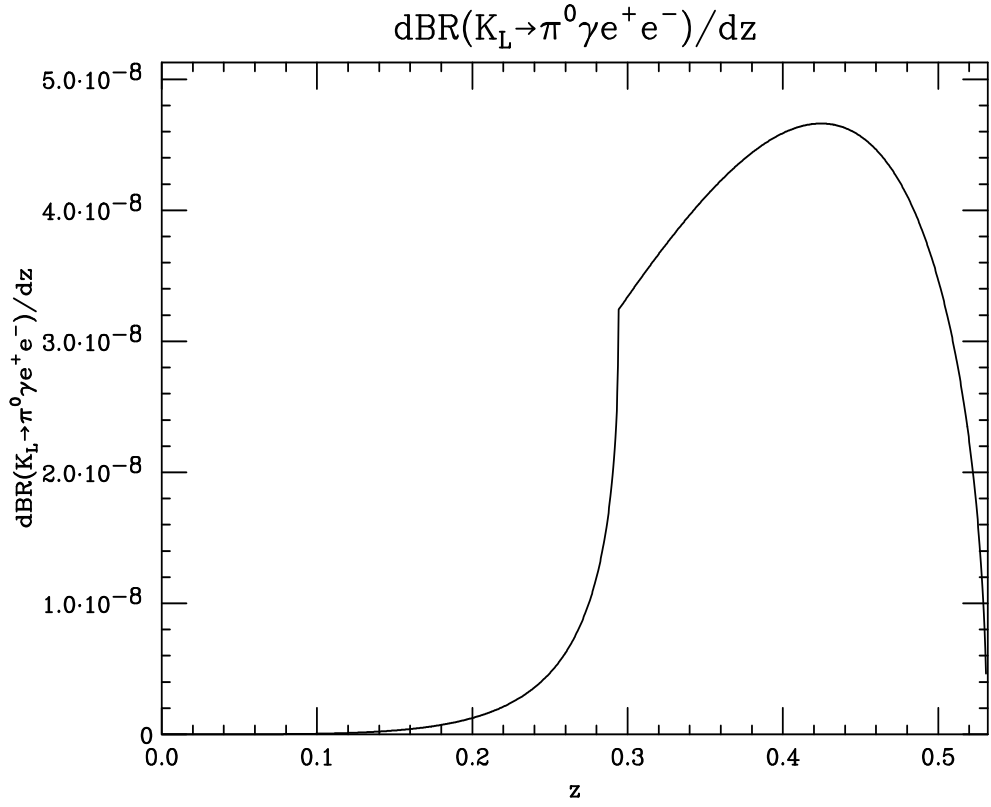}}}
\caption{Differential branching ratio $dBR(K_L \rightarrow
\pi^0 \gamma e^+ e^-)/dz$ to ${\cal O}$(E$^4$).}
\end{figure}

\begin{figure}[t]
\centering
\leavevmode
\epsfxsize=300pt
\epsfysize=300pt
%\rotate[l]
{\centerline{\epsfbox{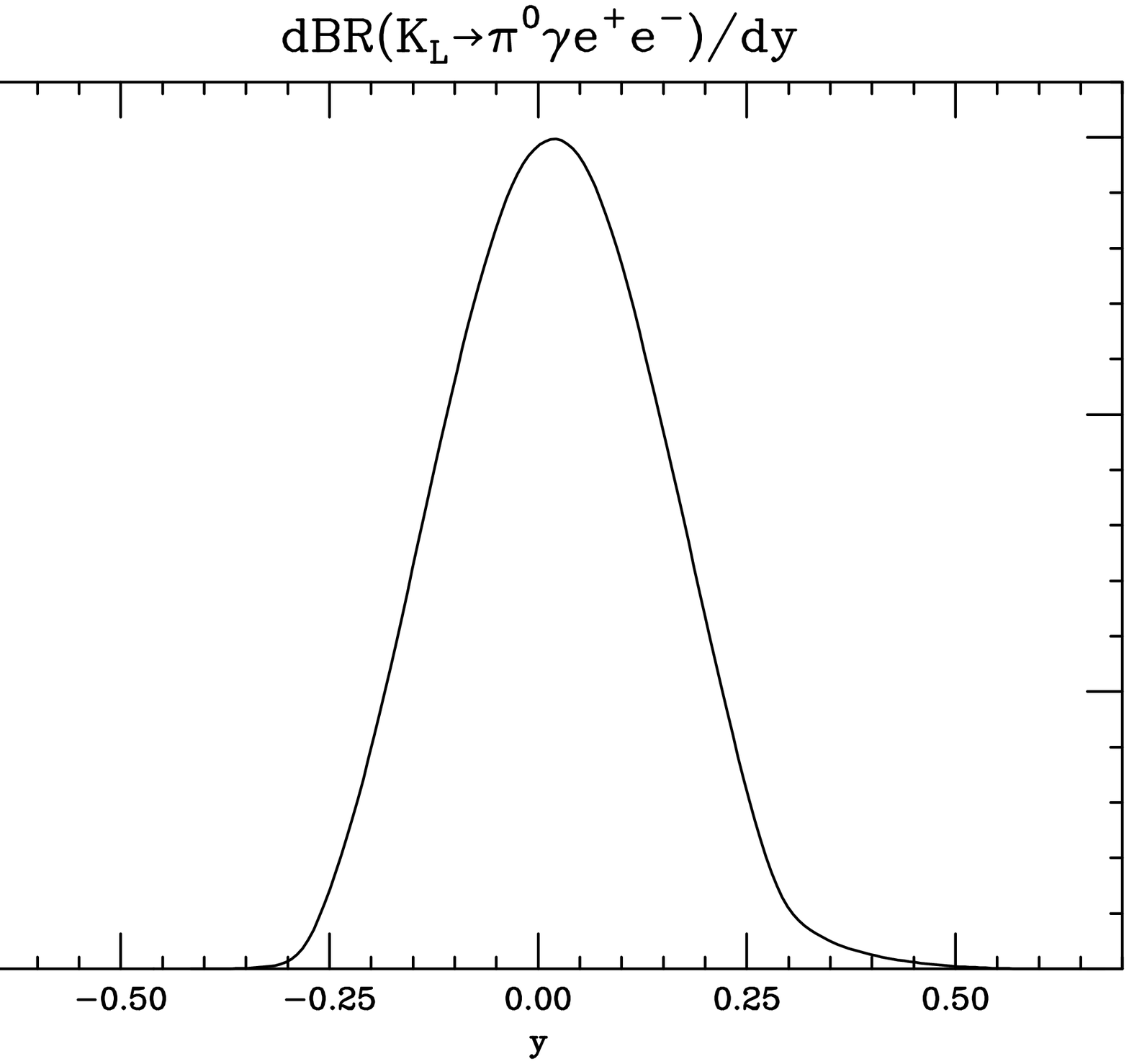}}}
\caption{Differential branching ratio $dBR(K_L \rightarrow
\pi^0 \gamma e^+ e^-)/dy$ to ${\cal O}$(E$^4$).}
\end{figure}

\chapter{THE $\cal{O}$(E$^6$) CALCULATION}

We also wish to extend this calculation along the lines proposed by
CEP \cite{CEP}, who provide a plausible solution to the problem
raised by the experimental rate not agreeing with the $\cal{O}$(E$^4$)
calculation when both photons are on-shell. The two primary new
ingredients involve known physics that surfaces at the next order in
the energy expansion. The first involves the known quadratic energy
variation of the $K \ri 3\pi$ amplitude, which occurs from higher
order terms in the weak nonleptonic lagrangian \cite{DGH,CDM,DA}. While the
full one-loop structure of this is known \cite{KMW1,KMW2,DH}, it
involves complicated
nonanalytic functions and we approximate the result at
$\cal{O}$(E$^4$) by an analytic polynomial which provides a good
description of the data throughout the physical region:

\begin{equation}
\label{18}
{\cal M}(K \ri \pi^+\pi^-\pi^0) = 4 a_1
p^{\phantom{l}}_K \cdot p^{\phantom{l}}_0
p^{\phantom{l}}_+ \cdot p^{\phantom{l}}_- + 4 a_2
(p^{\phantom{l}}_K \cdot p^{\phantom{l}}_+
p^{\phantom{l}}_0 \cdot p^{\phantom{l}}_- +
p^{\phantom{l}}_K \cdot p^{\phantom{l}}_-
p^{\phantom{l}}_0 \cdot p^{\phantom{l}}_+),
\end{equation}

\noindent using

\begin{equation}
a_1 = 3.1 \times 10^{-6} m^{-4}_K \qquad {\rm and}
\qquad a_2 =- 1.26 \times 10^{-6} m^{-4}_K.
\end{equation}

\noindent
$a_1$ and $a_2$ are obtained from a fit to the amplitude for $K_L
\rightarrow \pi^0\pi^+\pi^-$ \cite{DGH} and to the amplitude and
spectrum for $K_L \rightarrow \pi^0 e^+ e^-$ \cite{CEP}, so that their
values are constrained within their theoretical uncertainty of 10 --
20\%. We have numerically verified that such a variation of said
parameters involves a very modest change in the shape of the spectrum
for $K_L \rightarrow \gamma \pi^0 e^+ e^-$ and a change in its final
branching ratio somewhat smaller than the uncertainty on the
parameters.

The other ingredient involves vector meson exchange such as
in Fig. 28. Some of such contributions are known, but there are others
such as those depicted in Fig. 29 which have the same structure but an
unknown strength, leaving the total result unknown. In Ref. \cite{CEP}
the result is parametrized by a ``subtraction constant'' that must be
fit to the data.

\begin{figure}[t]
\centering
\leavevmode
\centerline{
\epsfbox{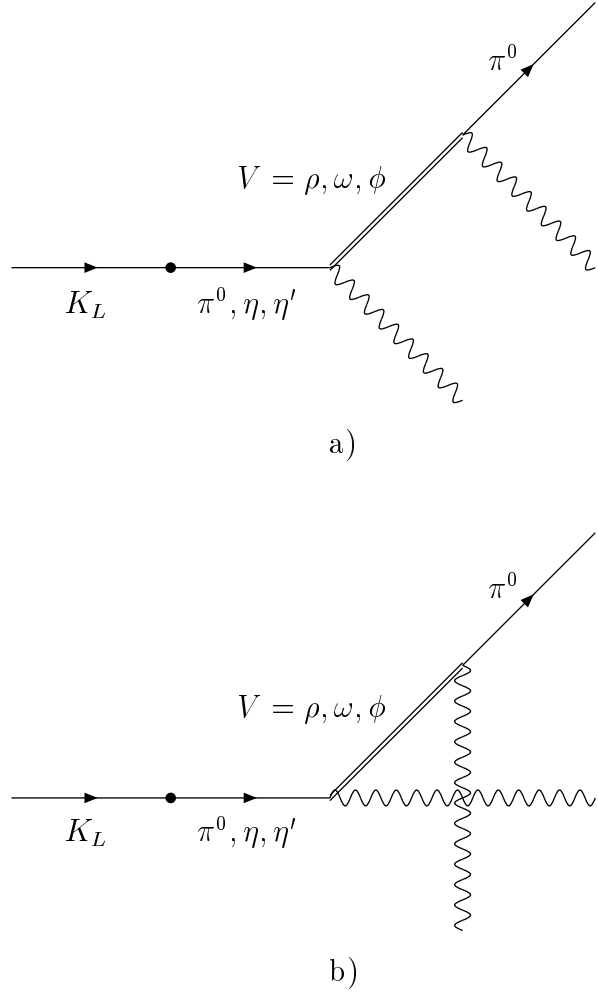}}
\caption{Vector meson exchange diagrams contributing to
$K_L \rightarrow \pi^0 \gamma e^+ e^-$.}
\end{figure}

\begin{figure}[t]
\centering
\leavevmode
\centerline{
\epsfbox{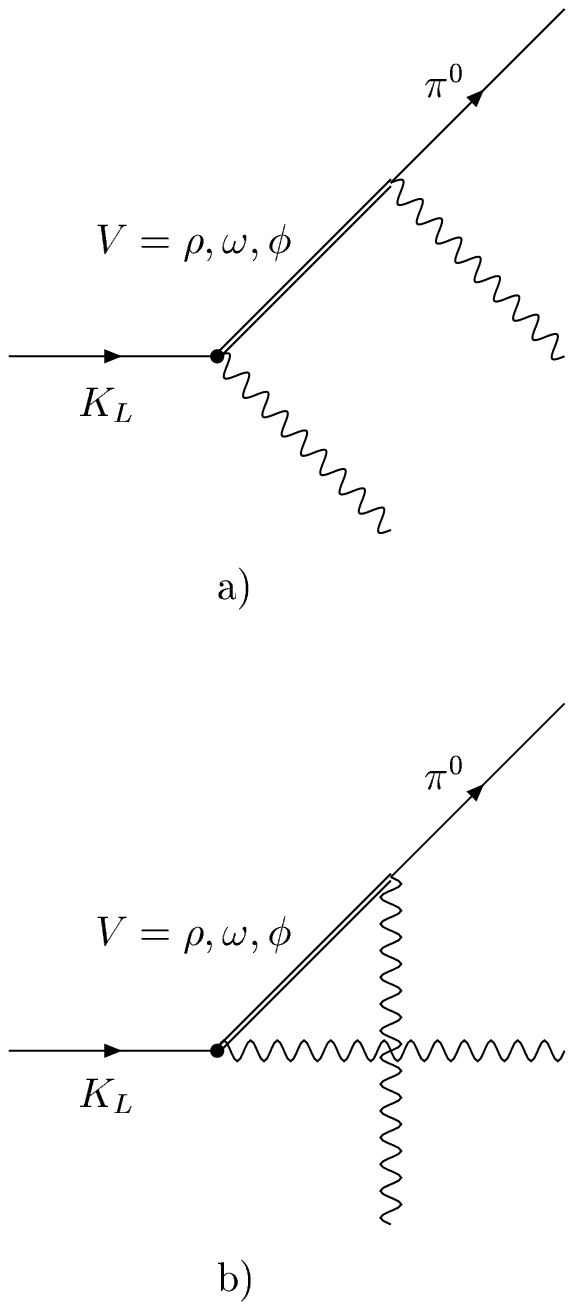}}
\caption{Vector meson exchange diagrams contributing to
$K_L \rightarrow \pi^0 \gamma e^+ e^-$ with unknown strength.}
\end{figure}

In principle one can add the ingredients to the amplitudes and perform
a dispersive calculation of the total transition matrix element. In
practice it is simple to convert the problem into an effective field
theory and and do a Feynman-diagram calculation which will yield the
same result. We follow this latter course.

The Feynman diagrams are the same as shown in Fig. 25, although the
vertices are modified by the presence of $\cal{O}$(E$^4$) terms in the
energy expansion. Not only does the direct $K \ri 3\pi$ vertex change
to the form given in Eq. \ref{18}, but also the weak vertices with one and two
photons have a related change. The easiest way to determine these is
to write a gauge invariant effective lagrangian with coefficients
adjusted to reproduce Eq. \ref{18}. We find

\begin{equation}
{\cal M_{\mu}}(K \ri \pi^+ \pi^- \pi^0 \gamma) = 4 a_1 e
(p^{\phantom{l}}_+ - p^{\phantom{l}}_-)_{\mu}
p^{\phantom{l}}_K \cdot p^{\phantom{l}}_0 + 4 a_2 e
(p^{\phantom{l}}_+ - p^{\phantom{l}}_-)_{\sigma}
(\ps0 \pkm + \psk \p0m),
\end{equation}

\begin{equation}
{\cal M}_{\mu\nu}(K \ri \pi^+ \pi^- \pi^0 \gamma \gamma) =
-8 a_1 e^2 g_{\mu\nu}
p^{\phantom{l}}_K \cdot p^{\phantom{l}}_0 + 8 a_2 e^2 (\pkm \pn0 + \pnk \p0m).
\end{equation}
\newpage

The resulting calculation follows the same steps as
described above, but is more involved and is not easy to present in a
simple form. We have checked that our result is gauge invariant and
reduces to that of CEP in the limit of on-shell photons.

The contribution proportional to $a_1$ can be computed
analogously to those already calculated for the $\cal{O}$(E$^4$) case:

\begin{equation}
{\cal M}_{\mu\nu} = 4 a_1 e^2 (z-2 r^2_{\pi})(1+r^2_{\pi}-z)
{1 \over {z-q}}(g_{\mu\nu} k_1 \cdot k_2-k_{2\mu} k_{1\nu})
[1+2 I(m^2_{\pi})],
\end{equation}

\noindent where

\begin{equation}
r_{\pi} = {m_{\pi} \over m_K}, \qquad z = {s \over {m^2_K}},
\qquad q = {{k^2_1} \over {m^2_K}}.
\end{equation}

The $a_2$ part originates another set of integrals
which can be written as

\begin{equation}
{\cal M}^a_{\mu\nu} = -8 a_2 (\prk \ps0 + \psk \pr0) e^2
g_{\mu\nu}\int {{d^d l}
\over {(2\pi)^d}} {{l_{\rho}(l-k_1-k_2)_{\sigma}} \over
{(l^2-m^2_{\pi})[(l-k_1-k_2)^2 - m^2_{\pi}]}},
\end{equation}

\begin{eqnarray}
{\cal M}^b_{\mu\nu} \hskip -0.4pt &=& \hskip -0.4pt 4 a_2 (\prk \ps0 +
\psk \pr0) e^2\int {{d^d l}
\over {(2\pi)^d}} \left\{ {{(2l+k_1)_{\mu} (2l-k_2)_{\nu}
(l+k_1)_{\rho} (l-k_2)_{\sigma}} \over
{(l^2-m^2_{\pi})[(l+k_1)^2 - m^2_{\pi}][(l-k_2)^2 -
m^2_{\pi}]}} \right. \nonumber \\
\hskip -0.4pt &+& \hskip -0.4pt \left. {{(2l+k_2)_{\nu} (2l-k_1)_{\mu}
(l+k_2)_{\rho} (l-k_1)_{\sigma}} \over
{(l^2-m^2_{\pi})[(l-k_1)^2 - m^2_{\pi}][(l+k_2)^2 -
m^2_{\pi}]}} \right\} \nonumber \\
\hskip -0.4pt &=& \hskip -0.4pt 8 a_2 (\prk \ps0 + \psk \pr0) e^2 \int
{{d^d l} \over {(2\pi)^d}} {{(2l+k_1)_{\mu} (2l-k_2)_{\nu}
(l+k_1)_{\rho} (l-k_2)_{\sigma}} \over
{(l^2-m^2_{\pi})[(l+k_1)^2 - m^2_{\pi}][(l-k_2)^2 -
m^2_{\pi}]}}, \nonumber \\
\phantom{l}
\end{eqnarray}

\begin{equation}
{\cal M}^c_{\mu\nu} = 8 a_2 (\pkm \pn0 + \pnk \p0m) e^2 \int {{d^d l}
\over {(2\pi)^d}} {1 \over {l^2-m^2_{\pi}}},
\end{equation}

\begin{eqnarray}
{\cal M}^d_{\mu\nu} &=& -4 a_2 (\ps0 \pnk + \psk \pn0) e^2 \int {{d^d l}
\over {(2\pi)^d}} {{(2l-k_1)_{\mu} (2l-k_1)_{\sigma}} \over
{(l^2-m^2_{\pi})[(l-k_1)^2 - m^2_{\pi}]}} \nonumber \\
&-& 4 a_2 (\ps0 \pkm + \psk \p0m) e^2 \int {{d^d l}
\over {(2\pi)^d}} {{(2l-k_2)_{\nu} (2l-k_2)_{\sigma}} \over
{(l^2-m^2_{\pi})[(l-k_2)^2 - m^2_{\pi}]}}.
\end{eqnarray}

\noindent From the above formulas we obtain

\begin{eqnarray}
{\cal M}_{\mu\nu} \hskip -4pt &=& \hskip -4pt {1 \over {(4 \pi)^2}}
\left[A(x_1,x_2)(k_{2\mu} k_{1\nu}-k_1 \cdot k_2 g_{\mu\nu})
\phantom{{k^2_1} \over {k_1}} \right. \nonumber \\
\hskip -4pt &+& \hskip -4pt B(x_1,x_2) \left({p^{\phantom{l}}_K \cdot k_1
p^{\phantom{l}}_K \cdot k_2 \over {k_1 \cdot k_2}}g_{\mu\nu}+
p^{\phantom{l}}_{K\mu}p_{K\nu}-{{p^{\phantom{l}}_K \cdot k_1}
\over {k_1 \cdot k_2}} k_{2\mu} p^{\phantom{l}}_{K\nu}-
{{p^{\phantom{l}}_K \cdot k_2} \over {k_1 \cdot k_2}}
k_{1\nu} p^{\phantom{l}}_{K\mu}\right) \nonumber \\
\hskip -4pt &+& \hskip -4pt \left. D(x_1,x_2) \left(k^2_1
{p^{\phantom{l}}_K \cdot k_2
\over {k_1 \cdot k_2}}g_{\mu\nu}-
{{p^{\phantom{l}}_K \cdot k_2} \over {k_1 \cdot k_2}} k_{1\mu} k_{1\nu}+
k_{1\mu}p^{\phantom{l}}_{K\nu}-{{k^2_1} \over {k_1 \cdot k_2}}k_{2\mu}
p^{\phantom{l}}_{K\nu}\right) \right], \nonumber \\
\phantom{l}
\end{eqnarray}

\noindent where

\begin{eqnarray}
{A \over m^2_K} \hskip -4.5pt &=& \hskip -4.5pt 16 a_2 e^2 \{2 [1-2(x_1 +
x_2)] I_1(z_1 z_2)+x_1 I_1(z_2)+x_2 [2 I_1(z^2_2) - I_1(z_2)+
I_1(z_1)]\} \nonumber \\
\hskip -4.5pt &-& \hskip -4.5pt 32 a_2 e^2 \{[2 x^2_1 -x_1(z+q)]
[-I_2(z^3_1 z_2)+I_2(z^2_1 z_2)] \nonumber \\
\hskip -4.5pt &+& \hskip -4.5pt [2 x_1 x_2 - x_1 (z-q)/2 -x_2 (z+q)/2]
[2 I_2(z^2_1 z^2_2)+I_2(z_1 z_2)-I_2(z^2_1 z_2) \nonumber \\
\hskip -4.5pt &-& \hskip -4.5pt I_2(z_1 z^2_2)]+[2 x^2_2-x_2 (z-q)]
[I_2(z_1 z^2_2)-I_2(z_1 z^3_2)]\} \nonumber \\
\hskip -4.5pt &+& \hskip -4.5pt {4 \over 3}
a_2 e^2 \left(1 + \log{m^2_{\pi} \over {m^2_{\rho}}}\right) +
(4 \pi)^2 {\rm VMD}_A,
\end{eqnarray}

\begin{equation}
{B \over m^2_K} = -32 a_2 e^2 I_3 +16 a_2 e^2 I_4 +
{4 \over 3} a_2 e^2 (z-q) \log{m^2_{\pi} \over {m^2_{\rho}}} +
(4 \pi)^2 {\rm VMD}_B,
\end{equation}

\begin{eqnarray}
{D \over m^2_K} &=& 16 a_2 e^2 I_3 -8 a_2 e^2 I_4
- {2 \over 3} a_2 e^2 (z-q) \log{m^2_{\pi} \over {m^2_{\rho}}}
\nonumber \\
&+& 16 a_2 e^2 [2 x_2 -(z-q)/2]
[2 I_1(z_1 z_2)-I_1(z_2)] \nonumber \\
&+& 16 a_2 e^2 (2 y -q)[I_1(z_1)-I_1(1)/2]+
4 a_2 e^2 [2 x_1-(z+q)/2] I_5 \nonumber \\
&+& (4 \pi)^2 {\rm VMD}_D,
\end{eqnarray}

\noindent with the integrals given in the Appendix:

\begin{equation}
I_1(z^n_1 z^m_2) = \int^1_0 dz_1 \int^{1-z_1}_0 dz_2 z^n_1 z^m_2
\log{D_1 \over {m^2_{\pi}}},
\end{equation}

\begin{equation}
{I_2(z^n_1 z^m_2) \over m^2_K} = \int^1_0 dz_1
\int^{1-z_1}_0 dz_2 {{z^n_1 z^m_2} \over D_1},
\end{equation}

\begin{equation}
I_3 m^2_K= \int^1_0 dz_1 \int^{1-z_1}_0 dz_2 D_1 \log{D_1 \over {m^2_{\pi}}},
\end{equation}

\begin{equation}
I_4 m^2_K = \int^1_0 dz_1 D_2 \log{D_2 \over {m^2_{\pi}}},
\end{equation}

\begin{equation}
I_5 = \int^1_0 dz_1 (4 z^2_1-4 z_1+1) \log{D_2 \over {m^2_{\pi}}},
\end{equation}

\noindent and

\begin{eqnarray}
D_1 &=& m^2_{\pi} - 2 k_1 \cdot k_2 z_1 z_2 - k^2_1 z_1 (1-z_1), \nonumber \\
D_2 &=& m^2_{\pi} - k^2_1 z_1 (1-z_1), \nonumber \\
x_1 &=& {{p^{\phantom{l}}_K \cdot k_1} \over {m^2_K}},
\qquad x_2 = {{p^{\phantom{l}}_K \cdot k_2} \over {m^2_K}},
\end{eqnarray}

\begin{equation}
\label{38}
{\rm VMD}_A(x_1,x_2) = - \sum_{V= \omega,\rho} G_V
\left[{{p^{\phantom{l}}_K \cdot (p^{\phantom{l}}_K-k_2)} \over
{(p^{\phantom{l}}_K-k_2)^2-m^2_V}}+
{{p^{\phantom{l}}_K \cdot (p^{\phantom{l}}_K-k_1)} \over
{(p^{\phantom{l}}_K-k_1)^2-m^2_V}}\right],
\end{equation}

\begin{equation}
{\rm VMD}_B(x_1,x_2) = - \sum_{V= \omega,\rho} G_V
k_1 \cdot k_2 \left[{1 \over {(p^{\phantom{l}}_K-k_2)^2-m^2_V}}+
{1 \over {(p^{\phantom{l}}_K-k_1)^2-m^2_V}}\right],
\end{equation}

\begin{equation}
\label{40}
{\rm VMD}_D(x_1,x_2) = \sum_{V= \omega,\rho} G_V
{{k_1 \cdot k_2} \over {(p^{\phantom{l}}_K-k_1)^2-m^2_V}},
\end{equation}

\noindent assuming the numerical values \cite{HS}

\begin{equation}
G_{\rho}m^2_K = 0.68 \times 10^{-8}, \qquad
G_{\omega}m^2_K = -0.28 \times 10^{-7}.
\end{equation}

The loop calculation that we have just described provides all of the
off-shell dependence scaled by the pion mass, and is of the form
$k^2_1/m^2_{\pi}$. There can be an additional dependence of the form
$k^2_1/\Lambda^2$, where $\Lambda$ $\approx$ 1 GeV. We cannot provide a
model independent analysis of the latter. However, experience has
shown that most of the higher order momentum dependence is well
accounted for by vector meson exchange. Therefore we include the
$k^2_1/ \Lambda^2$ dependence which is predicted by the diagrams of
Fig. 28. One can recover the parametrization in $a^{\phantom{l}}_V$
neglecting the dependence on $(p^{\phantom{l}}_K-k_1)^2$ and
$(p^{\phantom{l}}_K-k_2)^2$ in formulas \ref{38}--\ref{40}, and
performing the replacement \cite{HS}

\begin{equation}
{{\pi G_{\rm eff} m^2_K}
\over {2 G_8 \alpha m^2_V}} \ri a^{\phantom{l}}_V ,
\end{equation}

\noindent where $G_{\rm eff}$ $\approx$ $G_{\rho}$+$G_{\omega}$. This
completes our treatment of the $K_L \rightarrow
\pi^0 \gamma e^+ e^-$ amplitude.

The calculation we have presented in this chapter
leads to the total branching ratio of

\begin{equation}
{\rm BR}(K_L \rightarrow \pi^0 \gamma e^+e^-) = 2.4 \times 10^{-8}.
\end{equation}

\noindent The decay distributions are presented in Figs. 30 and 31.

\begin{figure}[t]
\centering
\leavevmode
\epsfxsize=300pt
\epsfysize=300pt
%\rotate[l]
{\centerline{\epsfbox{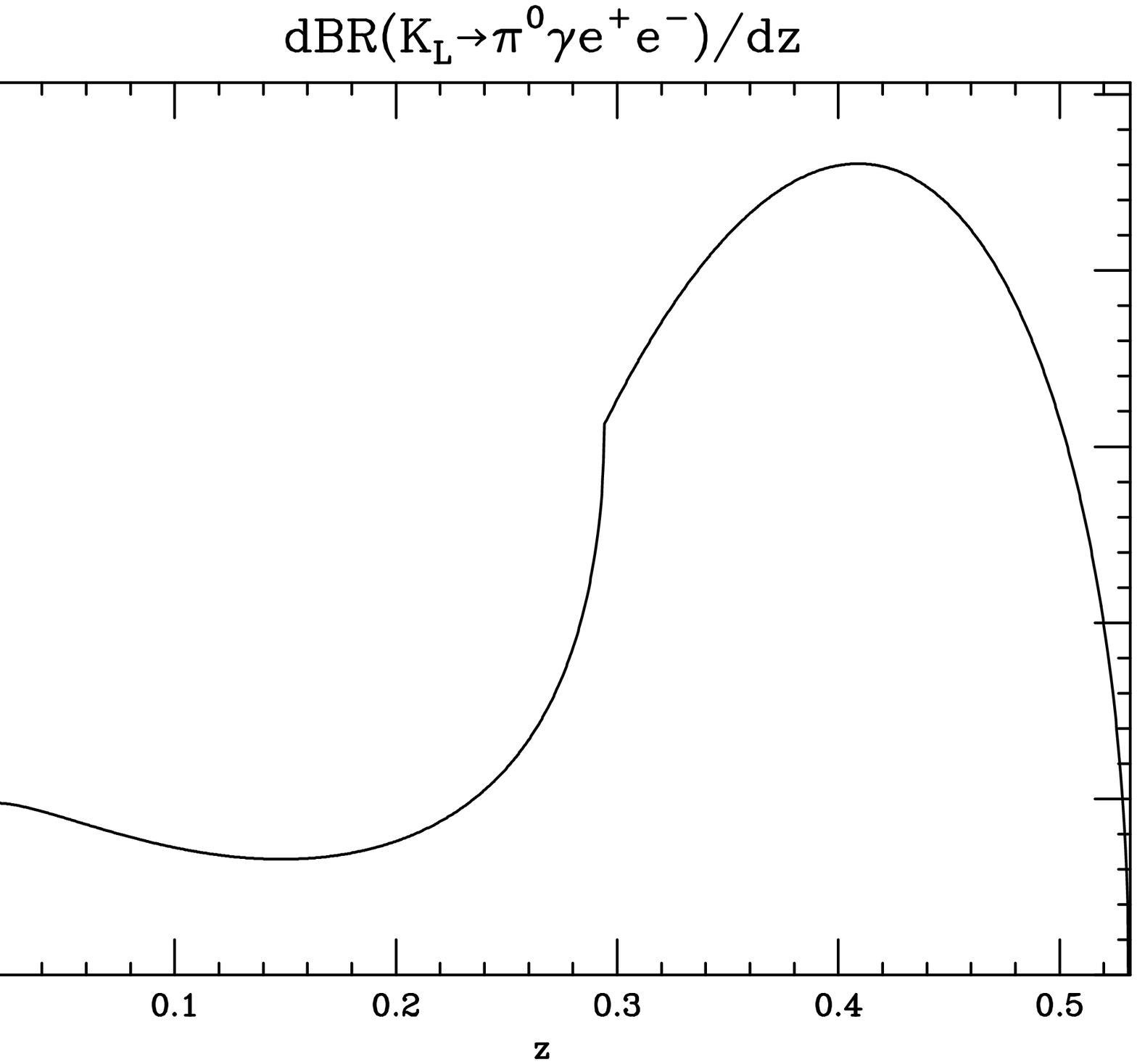}}}
\caption{Differential branching ratio $dBR(K_L \rightarrow
\pi^0 \gamma e^+ e^-)/dz$ to ${\cal O}$(E$^6$).}
\end{figure}

\begin{figure}[t]
\centering
\leavevmode
\epsfxsize=300pt
\epsfysize=300pt
%\rotate[l]
{\centerline{\epsfbox{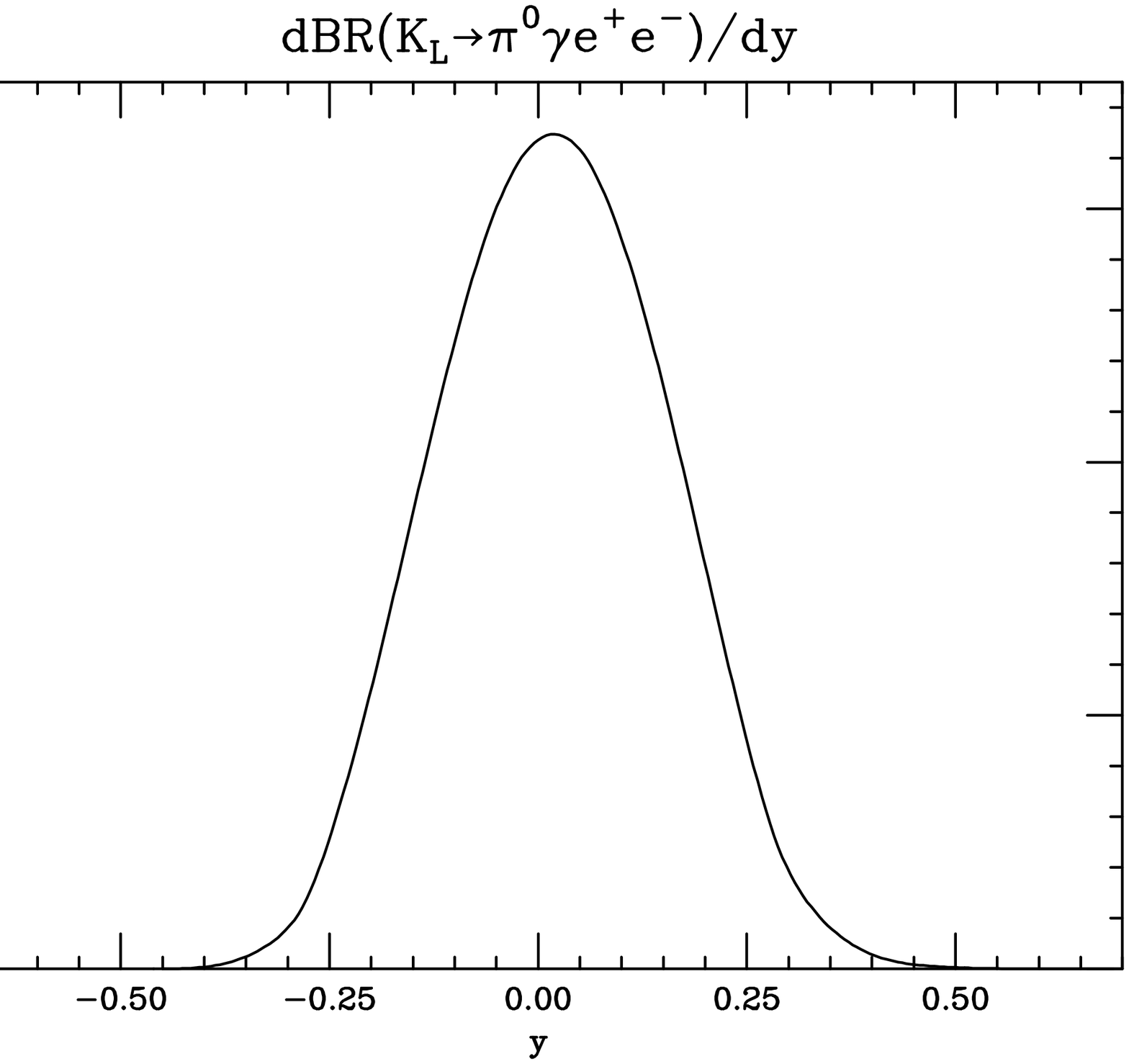}}}
\caption{Differential branching ratio $dBR(K_L \rightarrow
\pi^0 \gamma e^+ e^-)/dy$ to ${\cal O}$(E$^6$).}
\end{figure}

\chapter{CONCLUSIONS}

\vspace{2mm}
Rare kaon decays are an important testing ground of the electroweak
flavor theory. With the improved experimental sensitivity expected in
the near future, they can provide new signals of CP violation
phenomena and an opportunity to explore physics beyond the Standard
Model.

The theoretical analysis of these decays is far from trivial due to
the very low mass of the hadrons involved. The delicate interplay
between the flavor-changing dynamics and the confining QCD
interaction makes very difficult to perform precise dynamical
predictions. Fortunately, the Goldstone nature of the pseudoscalar
mesons implies strong constraints on their low-energy interactions,
which can be analyzed with effective lagrangian methods. The ChPTh
framework incorporates all the constraints implied by the chiral
symmetry of the underlying lagrangian at the quark level, allowing for
a clear distinction between genuine aspects of the Standard Model and
additional assumptions of variable credibility, usually related to the
problem of long-distance dynamics. The low-energy amplitudes are
calculable in ChPTh, except for some coupling constants that are not
restricted by chiral symmetry. These constants reflect our lack of
understanding of the QCD confinement mechanism and must be determined
experimentally for the time being. Further progress in QCD can only
improve our knowledge of these chiral constants, but it cannot modify
the low-energy structure of the amplitudes.

Addressing the two specific rare decays analyzed in this thesis, we can
conclude that because of the three possible contributions to $K_L \ri
\pi^0 e^+ e^-$, the analysis of this process has multiple issues that
theory must address. We have provided an updated analysis of all of
its components. The goal of identifying direct CP violation will not
be easily accomplished. The decay rate by itself suffers from a severe
uncertainty in the analysis of the mass matrix contributions. In the
chiral analysis there is a free parameter, $w_2$, which is not fixed
experimentally, and which has a strong influence on the decay rate.
Measuring the related rate of $K_S \ri \pi^0 e^+ e^-$ would determine
this parameter and will likely allow us to determine whether or not
direct CP violation is present.

Alternatively if the $K_S$ branching ratio is not measured it may be
possible to signal direct CP violation using the asymmetry in the $e^+$,
$e^-$ energies. The direct and mass matrix CP violations have
different phases, and the asymmetry is sensitive to their difference.
There exist some combinations of the parameters where there remains
some ambiguity, but for sizeable portions of the parameter space
direct CP violation can be signaled by a simultaneous measurement of
the decay rate and the energy asymmetry, as illustrated in Fig. 23.

Progress in the theoretical analysis is possible, especially in the
CP-conserving amplitude that proceeds through the two-photon
intermediate state. Here both a better phenomenological understanding
of the related decay $K_L \ri \pi^0 \gamma \gamma$, and a better
theoretical treatment of the dispersive contribution should be
possible in the near future.

Overall, our reanalysis shows that the demands on the experimental
exploration of this reaction are quite severe. The simple observation of a
few events will not be sufficient to indicate direct CP violation. The
measurement of an electron asymmetry requires more events, and may or
may not resolve the issue. Only the simultaneous measurement of $K_S
\rightarrow \pi^0 e^+ e^-$ allows a convincing proof of the existence of
direct CP violation.

The behavior of the $K_L \rightarrow \pi^0 \gamma e^+ e^-$ amplitude
mirrors closely that of the process $K_L \rightarrow \pi^0 \gamma
\gamma$. The more complete calculation at order E$^6$ gives a rate
which is more than twice as large as the one obtained at order E$^4$,
despite the fact that the new parameter introduced at order E$^6$ is
quite reasonable in magnitude. This large change occurs partially
because the order E$^4$ calculation is purely a loop effect, while at
order E$^6$ we have tree level contributions, and loop contributions
are generally smaller than tree effects at a given order. It was more
surprising that the spectrum in $K_L \rightarrow \pi^0 \gamma
\gamma$ was not significantly modified by the order E$^6$
contributions. These new effects are more visible in the low $z$
region of the process $K_L \rightarrow \pi^0 \gamma e^+ e^-$.

This reaction should be reasonably amenable to experimental
investigation in the future. It is 3--4 orders of magnitude larger
than the reaction $K_L \rightarrow \pi^0 e^+ e^-$ discussed above,
which is one of the targets of experimental kaon decay programs, due
to the connections of the latter reaction to CP studies. In fact, the
radiative process $K_L \rightarrow \pi^0 \gamma e^+ e^-$ will need to
be studied carefully before the nonradiative reaction can be isolated.
The regions of the distributions where the experiment misses the
photon of the radiative process can potentially be confused with $K_L
\rightarrow \pi^0 e^+ e^-$ if the resolution is not sufficiently
precise. In addition, since the $\pi^0$ is detected through its decay
to two photons, there is potential confusion related to misidentifying
photons. The study of the reaction $K_L \rightarrow \pi^0 \gamma e^+
e^-$ will be a valuable preliminary to the ultimate CP tests.

\soloappendix
\chapter{RELEVANT INTEGRALS}

In this Appendix we list the explicit expressions for the integrals
used in the calculation of chapter 9. We follow the notation of that
chapter. For $s \leq 4 \mp^2$ and $\k1^2 \leq 4 \mp^2$ we have:

\begin{eqnarray}
\label{first}
I_1(1) &=& {1 \over {s-k^2_1}}\left\{ -{3 \over 2}\left(s-\k1^2\right)-\mp^2
\left[F(s) -F(\k1^2)\right] \phantom{\ak} \right. \nonumber \\
&-& \left. \ka \ak + \sa \as \right\}, \no \\
\end{eqnarray}

\begin{eqnarray}
I_1(z_1) &=& {1 \over {s-k^2_1}}\left[-{{4 s} \over 9} - {(4 m^2_{\pi}- s)
\over {3 \sqrt{s}}} \sa \as\right. \nonumber \\
&+& \left. {{4 k^2_1} \over 9} + {(4 m^2_{\pi}- k^2_1)
\over {3 \sqrt{k^2_1}}} \ka \ak \right],
\end{eqnarray}

\begin{eqnarray}
I_1(z^2_1) &=& {1 \over {s-k^2_1}}\left[-{{2 s} \over 9} - {(4 m^2_{\pi}- s)
\over {6 \sqrt{s}}} \sa \as \right. \nonumber \\
&+& \left. {{2 k^2_1} \over 9} + {(4 m^2_{\pi}- k^2_1)
\over {6 \sqrt{k^2_1}}} \ka \ak \right],
\end{eqnarray}

\begin{eqnarray}
I_1(z_2) &=& -{4 \over 9} + {{3 m^2_{\pi} + k^2_1} \over {6 \ds}} +
{m^2_{\pi} \over {3 s}} + {{(-4 m^4_{\pi} + 5 m^2_{\pi} s -s^2)} \over
{3 s \sqrt{s} \sa }} \as \nonumber \\
&-& {m^4_{\pi} \over {2 \ds^2}} \left[ 2 {\sa \over m^2_{\pi}} \as -
{s \over m^2_{\pi}} \right. \nonumber \\
&+& \left. 2 {\ka \over m^2_{\pi}} \ak +
{\k1^2 \over m^2_{\pi}}\right] \nonumber \\
&-& {k^4_1 \over {3 \ds^2}} \left[{m^2_{\pi} \over s} + {{(-4 m^4_{\pi}
+ 5 m^2_{\pi} s -s^2)} \over {s \sqrt{s} \sa}} \as - {{m^2_{\pi}} \over
k^2_1} \right. \nonumber \\
&+& \left. {{(-4 m^4_{\pi} + 5 m^2_{\pi} k^2_1 -k^4_1)} \over {k^2_1
\sqrt{k^2_1} \ka}} \ak \right] \nonumber \\
&+& {m^2_{\pi} \k1^2 \over \ds^2}\left[F(s) - F(k^2_1)\right] - {{2 m^2_{\pi}
k^2_1} \over \ds^2}\left[{\sa \over \sqrt{s}} \as \right. \nonumber \\
&-& \left. {\ka \over \sqrt{k^2_1}} \ak\right],
\end{eqnarray}

\begin{eqnarray}
I_1(z^2_2) &=& -{2 \over 9} + {m^2_{\pi} \over {4 \ds}} + {k^2_1 \over
{24 \ds}} + {1 \over {3 \ds^2}}\left(m^4_{\pi} - 3 m^2_{\pi} k^2_1 -
{\k1^4 \over 4}\right) \nonumber \\
&+& {\mp^2 \over {3s}}-{\nns \over {6 s \sqrt{s} \sa}} \as \nonumber \\
&-& {\mp^6 \over {3 \ds^3}} \left[ {s \over \mp^2} + {{s \sqrt{s} \sa}
\over \mp^4} \as \right. \nonumber \\
&-& \left. {\k1^2 \over \mp^2} - {{\k1^2 \sqrt{\k1^2} \ka}
\over \mp^4} \ak \right] \nonumber \\
&+& {\k1^6 \over {3 \ds^3}} \left[{\mp^2 \over s} -{\nns \over {2s
\sqrt{s}\sa}} \as \right. \nonumber \\
&-& \left. {\mp^2 \over \k1^2} +{\nnk \over {2 \k1^2
\sqrt{\k1^2}\ka}} \ak \right] \nonumber \\
&+& {{\mp^4 \k1^2} \over {\ds^3}} \left[ -{s \over \mp^2} + {{2 \sa \sqrt{s}}
\over \mp^2} \as \right. \nonumber \\
&+& \left. {\k1^2 \over \mp^2} - {{2 \ka \sqrt{\k1^2}}
\over \mp^2} \ak - F(s) + F(\k1^2) \right] \nonumber \\
&-& {{\mp^2 \k1^4} \over {\ds^3}} \left[ F(s) -{{3 \sa}
\over \sqrt{s}} \as \right. \nonumber \\
&-& \left. F(\k1^2) + {{3 \ka}
\over \sqrt{\k1^2}} \ak \right],
\end{eqnarray}

\begin{eqnarray}
I_1(z_1 z_2) &=& -{13 \over 144}+{{(-6 m^2_{\pi}+k^2_1)} \over {24 \ds}}
+ {\ns \over {12 s \sqrt{s} \sa}} \as \nonumber \\
&+& {m^4_{\pi} \over {2 \ds^2}}\left[F(k^2_1)-F(s)\right] \nonumber \\
&-& {k^4_1 \over {2 \ds^2}} \left[- {{m^2_{\pi}} \over {3 s}} +
{\ns \over {6 s \sqrt{s} \sa}}\as \right. \nonumber \\
&+& \left. {{m^2_{\pi}} \over {3 k^2_1}} - {\nk \over {6 k^2_1
\sqrt{k^2_1} \ka}}\ak \right] \nonumber \\
&+& {{m^2_{\pi} k^2_1} \over \ds^2} \left[ {\sa \over \sqrt{s}}\as
\right. \nonumber \\
&-& \left. {\ka \over \sqrt{\k1^2}}\ak \right],
\end{eqnarray}

\begin{eqnarray}
{I_2(z_1 z_2) \over m^2_K} &=& -{1 \over {2 \ds}} -{\mp^2 \over \ds^2}
\left[ F(s) - F(\k1^2)\right] \nonumber \\ &+& {\k1^2 \over \ds^2}
\left[ {\sa \over \sqrt{s}} \as
\right. \nonumber \\
&-& \left. {\ka \over \sqrt{\k1^2}} \ak \right],
\end{eqnarray}

\begin{eqnarray}
{I_2(z_1 z^2_2) \over m^2_K} &=& -{1 \over {6 \ds}} + {1 \over
\ds^2}\left({\k1^2 \over 3} + \mp^2\right) \nonumber \\ &-& {\mp^2
\over \ds^3}\left[2 \sa \sqrt{s} \as -s \right. \nonumber \\ &-&
\left. 2 \ka \sqrt{\k1^2} \ak + \k1^2 \right] \nonumber \\ &-& {{2
\k1^4} \over {3 \ds^3}}\left[{\mp^2 \over s}+ {{(-4 m^4_{\pi}+5
m^2_{\pi} s-s^2)} \over {s \sqrt{s} \sa}} \as
\right. \nonumber \\
&-& \left. {\mp^2 \over \k1^2} - {{(-4 m^4_{\pi} + 5 m^2_{\pi} k^2_1 -k^4_1)}
\over {k^2_1 \sqrt{k^2_1} \ka}} \ak \right] \nonumber \\
&+& {{2 \mp^2 \k1^2} \over \ds^3} \left[F(s) - F(\k1^2)
- {{2 \sa} \over \sqrt{s}}\as \right. \nonumber \\
&+& \left. {{2 \ka} \over \sqrt{\k1^2}}\ak
\right],
\end{eqnarray}

\begin{eqnarray}
{I_2(z_1 z^3_2) \over m^2_K} &=& -{1 \over {12 \ds}} + {1 \over \ds^2}
\left({{6 \mp^2 +\k1^2} \over 4} \right)\nonumber \\ &-& {1 \over
\ds^3} \left({{4 \mp^4 -12 \mp^2 \k1^2 -\k1^4} \over 4}\right)
\nonumber \\ &-& {\mp^6 \over \ds^4}\left[{s \over \mp^2} - {\k1^2
\over \mp^2} + {{s \sqrt{s}\sa} \over \mp^4} \as \right. \nonumber \\
&-& \left. {{\k1^2 \sqrt{\k1^2} \ka}
\over \mp^4} \ak - {s^2 \over {4 \mp^4}} + {\k1^4 \over
{4 \mp^4}} \right] \nonumber \\
&+& {\k1^6 \over \ds^4}\left[{\mp^2 \over s} -{\nns \over {s
\sqrt{s}\sa}} \right. \as \nonumber \\
&-& \left. {\mp^2 \over \k1^2} + {\nnk \over {\k1^2
\sqrt{\k1^2}\ka}} \ak \right] \nonumber \\
&+& {{3 \mp^4 \k1^2} \over \ds^4} \left[ {{2 \sa \sqrt{s}}\over
{\mp^2}}\as - {s \over \mp^2}\right. \nonumber \\
&+& \left. {\k1^2 \over \mp^2}- {{2 \ka \sqrt{\k1^2}}\over
{\mp^2}}\ak -F(s) + F(\k1^2) \right] \nonumber \\
&-& {{3 \mp^2 \k1^4} \over \ds^4} \left[ F(s) - {{3 \sa} \over
\sqrt{s}}\as \right. \nonumber \\
&-& \left. F(\k1^2) + {{3 \ka} \over \sqrt{\k1^2}}\ak\right],
\end{eqnarray}

\begin{eqnarray}
{I_2(z^2_1 z_2) \over m^2_K} &=& -{1 \over {6 \ds}} + {1 \over
\ds^2}\left[ -{{4 \mp^2}
\over 3} \right. \no \\
&-& \left. {{\sqrt{\k1^2} \ka} \over 3}\ak \right. \no \\
&+& \left. {{4 \mp^2} \over {3\sqrt{\k1^2}}}\ka \ak \right]
- {1 \over \ds^2}\left[-2 \mp^2 + {{2 \k1^2 \mp^2} \over {3 s}}
\right. \no \\
&-& \left. \sa {{(2 \k1^2 \mp^2 +\k1^2 s - 6 \mp^2 s)}
\over {3 s \sqrt{s}}}\as \right], \no \\
\end{eqnarray}

\begin{eqnarray}
{I_2(z^2_1 z^2_2) \over m^2_K} &=& -{1 \over {24 \ds}} -{1 \over
{12\ds^2}}(6 \mp^2 -
\k1^2)-{\mp^4 \over \ds^3}\left[F(s) - F(\k1^2)\right] \nonumber \\
&-& {{\k1^4 \over \ds^3}}\left[-{\mp^2 \over {3s}} + {\ns \over {6
s \sqrt{s} \sa}}\as \right. \nonumber \\
&+& \left. {\mp^2 \over {3\k1^2}} - {\nk \over {6 \k1^2 \sqrt{\k1^2}
\ka}}\ak \right] \nonumber \\
&+& {{2 \mp^2 \k1^2} \over \ds^3}\left[-{\ka \over \sqrt{\k1^2}}\ak
\right. \nonumber \\
&+& \left. {\sa \over \sqrt{s}}\as \right],
\end{eqnarray}

\begin{eqnarray}
{I_2(z^3_1 z_2) \over m^2_K} &=& -{1 \over {12 \ds}} + {1 \over
{2\ds^2}}\left[ -{{4 \mp^2}
\over 3} \phantom{\ak}\right. \no \\
&-& \left. {{\sqrt{\k1^2} \ka} \over 3}\ak \right. \no \\
&+& \hskip -1pt \left. {{4 \mp^2} \over {3\sqrt{\k1^2}}}\ka \ak \right]
\hskip -0.8pt - \hskip -0.8pt {1 \over {2\ds^2}}\left[-2 \mp^2 + {{2
\k1^2 \mp^2} \over {3 s}} \right. \no \\
&-& \left. \sa {{(2 \k1^2 \mp^2 +\k1^2 s - 6 \mp^2 s)} \over {3 s
\sqrt{s}}}\as \right], \no \\
\end{eqnarray}

\begin{eqnarray}
I_3 m^2_K &=& -{13 \over 12}\mp^2 + {13 \over 144}\left(s +\k1^2\right)
- {\mp^4 \over {2 \ds}}\left[F(s) - F(\k1^2)\right] \nonumber \\
&-& {\k1^4 \over {2 \ds}}\left[
- {\nk \over {6\k1^2}\sqrt{\k1^2} \ka} \ak \right. \nonumber \\
&+& \left. {\ns \over {6 s\sqrt{s} \sa}} \as \right] \nonumber \\
&+& {{\k1^2 \mp^2} \over {s-k^2_1}} \left[-{\ka \over
\sqrt{\k1^2}}\ak \right. \nonumber \\
&+& \left. {\sa \over \sqrt{s}}\as\right] \nonumber \\
&-& \sa {{\left(2 \k1^2 \mp^2 +\k1^2 s -10 \mp^2 s +s^2\right)} \over
{12 s \sqrt{s}}}\as, \no \\
\end{eqnarray}

\begin{eqnarray}
I_4 m^2_K = {{5 \k1^2} \over {18}} - {{4 \mp^2} \over 3} + {{(4 \mp^2 -
\k1^2) \ka} \over {3 \sqrt{\k1^2}}}\ak,
\end{eqnarray}

\begin{eqnarray}
\label{last}
I_5 = -{8 \over 9} + {{8 \mp^2} \over {3 \k1^2}} - {{2(4 \mp^2
-\k1^2) \ka} \over {3 \k1^2 \sqrt{\k1^2}}}\ak.
\end{eqnarray}

\n In the cases when $s > 4 \mp^2$ or $\k1^2 > 4 \mp^2$, we have to
perform the substitutions:

\begin{eqnarray}
\as &\ri& -{1 \over {2i}}\left[\log\left({{1-\sqrt{1-4\mp^2/s}} \over
{1+\sqrt{1-4\mp^2/s}}}\right)+i\pi\right], \nonumber \\
\ak &\ri& -{1 \over {2i}}\left[\log\left({{1-\sqrt{1-4\mp^2/\k1^2}} \over
{1+\sqrt{1-4\mp^2/\k1^2}}}\right)+i\pi\right],
\end{eqnarray}

\n and

\begin{eqnarray}
\sa &\ri& i\sqrt{s-4 m^2_{\pi}}, \nonumber \\
\ka &\ri& i\sqrt{k^2_1-4 m^2_{\pi}},
\end{eqnarray}

\n respectively, in formulas \ref{first}--\ref{last}.

%
%%%

%%%% 
%%   These are some gadgets that may make things more readable and dense!
%%   Can be skipped on first reading; make sure to play with this. 
%%   
%% 
% \beginsinglespace
%   Some single spaced text ...... .
% \endsinglespace
%
% In particular,
% \beginsinglespace
%       \include{1-chapter}
%       \include{2-chapter}
%       \appendix  % Use \soloappendix if it has one chapter only.
%       \include{appendix}
%       \include{bibliography}
% \endsinglespace 
%%%%

%%% That's all !

\end{document}